
%
%
%
%
%
%
%
%
%
%
%
\input amsppt.sty
\nopagenumbers
\magnification 1200
\TagsOnRight
\refstyle{A}
\widestnumber\key{AdCKP}

\def \inv
{^{-1}}
\def \demop
{\demo{Proof}}
\def\qed
{$\blacksquare$}
\def\qede
{\newline\line{\hfill\qed} \enddemo}

\def \coker
{\text{coker}}
\def \Cinf
{\Bbb C^\infty}
\def \demop
{\demo{Proof}}
\def \diag
{{\text{diag}}}
\def \ddxl
{ \overleftarrow \partial}

\def \Epsilon
{\Cal E}
\def \glinf
{{gl_\infty}}
\def \glinfmjn
{{gl_{\infty-}^j}}
\def \glinfpjn
{{gl_{\infty+}^j}}
\def \Glinf
{{Gl_\infty}}

\def \gllf
{{gl^{\text{\it lf}}_\infty}}

\def \Gllf
{{Gl^{\text{\it lf}}_\infty}}
\def \Glnlf
{{Gl^{0,\text{\it lf}}_\infty}}
\def \hatgllf
{{\hat{gl}^{\text{\it lf}}_\infty}}
\def \hatGlnlf
{{\hat{Gl}^{0,\text{\it lf}}_\infty}}
\def \hatGllf
{{\hat{Gl}^{\text{\it lf}}_\infty}}
\def \hatvacmbf
{ {\hat w_0^\nbar}  }
\def \Heisnbar
{\Cal H^{\nbar}}
\def \im
{\text{Im\ }}
\def \inv
{^{-1}}
\def \index
{\text{ind}}

\def \nbar
{{\underline n}}

\def \psresa
{{\Cal R_a}}
\def \pslatresai
{{\Cal U_{\alpha_i} }}
\def \Proj
{\Bbb P}
\def\qed
{$\blacksquare$}
\def\qede
{\newline\line{\hfill\qed} \enddemo}
\def\resz
{\text{Res}_z}

\def \semi
{\Lambda^{\infty\over 2}\Bbb C^\infty}
\def \sfwinf
{S_\infty^{fw}}
\def \tauwjn
{\tau_W^{j,\nbar}}
\def \tpsresa
{{\tilde{\Cal R}_a}}
\def \tpslatresai
{{\tilde{\Cal U}_{\alpha_i} }}

\def \tvacmbf
{ {\tilde w_0^\nbar}  }
\def \tWmbf
{{\tilde w_W}}

\def \vacwn
{ {w^\nbar_0}  }
\def \vacw
{ {w_0}  }
\def\wkfin
{ W^{(k)}_{\text{fin}}  }

\def \Wwave
{{w_W}}

\def \datum
{Urbana, December $14^{\text{th}}$, 1992}
\topmatter

\title
Partitions,\\ Vertex Operator Constructions\\ and Multi-component KP
equations
\endtitle

\rightheadtext
{Partitions, Vertex operator constructions,  \datum \dots}

\author
M.J. Bergvelt, A.P.E. ten Kroode,\datum
\endauthor

\affil
Urbana, Illinois and Amsterdam, the Netherlands
\endaffil

\abstract
For every partition of a positive integer $n$ in $k$ parts and every
point of an infinite Grassmannian we obtain a solution of the $k$
component differential-difference KP hierarchy and a corresponding
Baker
function.
A partition of $n$ also
determines a vertex operator construction of the fundamental
representations
of
the infinite matrix algebra $\glinf$ and  hence a $\tau$ function. We
use
these
fundamental representations
to study the Gauss decomposition in the infinite matrix group
$Gl_\infty$
and to express the Baker function in terms of $\tau$-functions.
The reduction to loop algebras is discussed.

\endabstract

\date
\datum
\enddate

\address
(Bergvelt) Department of Mathematics, University of Illinois, Urbana,
IL
61820, USA
\endaddress
\email
bergv\@huygens.math.uiuc.edu
\endemail
\address
(ten Kroode) Nieuwe Prinsengracht $77^1$, 1018 VR Amsterdam, The
Netherlands
\endaddress


\endtopmatter

\document
\subhead {\bf 1.} Introduction\endsubhead

\subsubhead {\bf 1.1 } Infinite Grassmannians and Hirota equations
\endsubsubhead

Sato  discovered that the Kadomtsev--Petviashvili (KP) hierarchy
of soliton equations could be interpreted as the Pl\"ucker equations
for the
embedding of a  certain infinite Grassmannian in  infinite
dimensional projective  space, see e.g. \cite{Sa1, Sa2}.

Let us first recall the finite dimensional situation. The
Grassmannian
$Gr_j(\Bbb C^n)$ consists of all $j$-dimensional subspaces $W$ of the
$n$-dimensional complex linear space $\Bbb C^n$. Let
$\{e_i\mid i=1,2,\dots,n\}$
be a basis for	$\Bbb C^n$ and let $H_j\in Gr_j(\Bbb C^n)$ be the
subspace
spanned by the first $j$ basis vectors $e_1,e_2,\dots,e_{j}$. The
stabilizer
in
$Gl(n,\Bbb C)$ of $H_j$ is the ``parabolic'' subgroup $P_j$
consisting of
invertible matrices $X=\sum X_{ab}	E_{ab}$, with $X_{ab}=0$ if
$a> j$
and $b\le j$. Here $E_{ab}$ is the elementary
matrix with as only non zero entry a 1 on the $(a,b)^{\text{th}}$
place. So
$Gr_j(\Bbb C^n)$ can be identified with the homogeneous space
$Gl(n,\Bbb
C)/P_j$.
Now this homogeneous space  is projective, i.e., admits an embedding
into a
projective space.
Explicitly, let $\Lambda\Bbb C^n$ be the exterior algebra generated
by
the basis elements $e_a$ of $\Bbb C^n$ and $\Lambda^j\Bbb C^n$ the
degree $j$
part, i.e., the linear span of elementary wedges $e_{i_1}\wedge
e_{i_2}\wedge\dots \wedge e_{i_{j}}$.
For $W\in Gr_j(\Bbb C^n)$ with basis $w_1,w_2,\dots,w_{j}$ we
have the element $w_1 \wedge w_2 \wedge\dots \wedge w_{j}$ which is
up to
multiplication by a non zero
scalar independent of the choice of basis. This then defines
an embedding $\phi_j:Gr_j(\Bbb C^n)\to \Proj
\Lambda^j\Bbb C^n$. (If $V$ is a vector space $\Proj V$ denotes the
associated
projective space.)  The image of $\phi_j$ is the projectivization of
the $Gl(n,\Bbb C)$ orbit of the
highest weight vector $e_1 \wedge e_2 \wedge\dots \wedge e_{j}$ and
is
described by the following quadratic
equations: if $\tau\in \Lambda^j \Bbb C^n$, $\tau\ne0$ then $[\tau]$,
the line through $\tau$, belongs to $\im \phi_j$
iff it satisfies the following equation:
$$
\sum_{a=1}^{n} \psi(a)\tau\otimes\psi(a)^* \tau=0.
\tag1.1.1
$$
Here $\psi(a)$ and $\psi(a)^*$, $a=1,\dots, n$ are  fermionic
creation and annihilation
operators on $\Lambda \Bbb C^n$ that act on
elementary wedges by
$$
\aligned
\psi(a) \cdot(e_{a_1}\wedge e_{a_2} \wedge\dots \wedge e_{a_j})
&=e_a \wedge e_{a_1}\wedge e_{a_2}
\wedge\dots \wedge e_{a_j},\\
\psi^*(a) \cdot(e_{a_1}\wedge e_{a_2} \wedge \dots \wedge e_{a_j})
&=\sum_{k=1}^j(-1)^{k+1}\delta_{aa_k}
e_{a_1}\wedge e_{a_2} \wedge
\dots \wedge\hat e_{a_k}\wedge\dots \wedge e_{a_j}.
\endaligned
\tag1.1.2
$$
Here the hat $\hat{}$ denotes deletion.
The equation  \thetag{1.1.1} is one of the forms of  the famous
Pl\"ucker
equations, cf., \cite{GH}.

The infinite dimensional situation relevant for soliton equations of
KP-type is initially
very much the same as in finite dimensions: one considers  a group
$G$ of
certain invertible infinite matrices  indexed by $\Bbb Z$,
a parabolic subgroup $P$  and the
homogeneous space  $Gr=G/P$, an infinite Grassmannian. (We will be
sketchy in
this introduction
about the precise definition of the infinite dimensional objects $G$,
$P$ etc;
there are various choices for them, corresponding to various classes
of
solutions
of the hierarchies).

The group $G$ has a central
extension $0\to \Bbb C^*\to \hat G\to G\to0$. (Such an extension also
occurs
in the finite dimensional situation of $Gl(n,\Bbb C)$,
but is there necessarily trivial and is usually ignored). There is an
integrable highest  weight representation $L_\lambda$
for $\hat G$, with $\lambda$ an integral dominant weight,
such that the lift $\hat P$ of $P$ stabilizes the
highest weight vector $v_\lambda$ of $L_\lambda$. Then the
projectivized group orbit $\Proj(\hat G\cdot v_\lambda)\subset
\Proj L_\lambda$ is isomorphic to $ G/P$ and so this
construction gives a projective embedding of $G/P$. The
representation
				$L_\lambda$
can be realized explicitly as a homogeneous component (with respect
to the
grading by ``charge'') of a
``semi infinite wedge space'' on which fermionic creation and
annihilation
operators $\psi(a)$ and $\psi(a)^*$, $a\in \Bbb Z$ act by formulae
analogous to \thetag {1.1.2}. The image of $G/P$ in
$\Proj L_\lambda$ is described by \thetag{1.1.1}, but with now the
summation
running
over all integers.

To obtain the KP hierarchy one next
considers the principal Heisenberg subalgebra $\hat s^{princ}$
of the Lie algebra $\hat g$ associated to  the group $\hat G$ and one
proves that
$L_\lambda$, now thought of as a module for $\hat g$, remains
irreducible under the action of the subalgebra $\hat
s^{princ}$. By uniqueness of representations of Heisenberg
algebras one 	concludes that $L_\lambda $ is isomorphic to a
polynomial algebra $\Bbb C[x_1,x_2,\dots]$ in an infinite
number of variables. An element $\tau$ of this polynomial
algebra corresponds to a point of the group 	orbit $\hat
G\cdot v_\lambda$ precisely when it satisfies an infinite
collection of differential equations (Hirota equations) of the
form
$$
P(\partial/\partial x)\tau(x)\cdot\tau(x):=P(
\partial/\partial y)\tau(x+y)        \tau(x-y)|_{y=0} = 0,
\tag1.1.3
$$
for certain polynomials $P$. The equations \thetag{1.1.3} are a
``bosonized''
form of the fermionic Pl\"ucker equations \thetag{1.1.1}. This then
is the KP
hierarchy in so  called Hirota form and gives the defining equations
for
the  projectivized group orbit.

One sees that the construction of the KP hierarchy depends
essentially on
the choice of
the principal Heisenberg algebra to obtain   a
concrete, bosonic, realization of the representation
$L_\lambda$. It is therefore natural to investigate what
happens if one focuses one's attention to other
Heisenberg subalgebras $\hat s$ of $\hat g$, that, as is well known,
give rise to other so called vertex
operator constructions (\cite{KaP2, Lep}).
In general the representation $L_\lambda$
will not remain irreducible under   other Heisenberg algebras
but in our situation there is in the
group $\hat G$ a subgroup $\hat T$, the {\it
translation group}, of elements that
commute with $\hat s$ in the representation $L_\lambda$ and such
that $L_\lambda$ remains irreducible under the action of the pair
$(\hat
s, \hat T)$.

Investigating examples (see for example \cite{tKB},
\allowlinebreak\cite{KaW}) one quickly discovers that
one obtains from these other constructions of
$L_\lambda$ in  much the same way as before defining equations for
the group
orbit, but the equations can have a rather different
character; in particular one will find hierarchies
that contain also difference, as opposed to just
differential, equations. For example the Toda lattice
can be obtained in this way. The difference equations are ``caused''
by the occurrence of the translation group in the vertex operator
construction sketched above. Also the differential equations that one
obtains
for other Heisenberg algebras look rather different: in the simplest
case
one obtains the Davey-Stewartson equation in stead of the KP
equation.

In this paper we want
to discuss the hierarchies of soliton equations
related to  certain vertex operator constructions of  the central
extension $
\hat g$ of the infinite matrix algebra.
These constructions  use Heisenberg algebras of $\hat g$
obtained from all possible Heisenberg algebras of the  affine Lie
algebras $\hat{gl}(n,\Bbb C)$, where we think of $\hat{gl}(n,\Bbb C)$
as a
subalgebra of $\hat g$.

\subsubhead {\bf 1.2 } Lax and zero curvature form
\endsubsubhead

Until now in this introduction we have described  soliton
equations in Hirota form, using the representation theory of
a central extension of the infinite linear group.

Other approaches to these equations are the Lax and zero curvature
formalisms. Let us sketch how these approaches
are related to the representation theoretic one. As we discussed
before
the main ingredient in the recipe for the construction of Hirota
equations was the choice of Heisenberg system
$(\hat s, \hat T)$ consisting of a Heisenberg subalgebra of $\hat g$
and a translation subgroup $\hat T$ of $\hat G$. Now using the
sequences
$$
0\to\Bbb C\to \hat g\to g\to 0,\quad 0\to\Bbb C^*\to \hat G\to G\to 0
\tag 1.2.1
$$
we obtain a pair $( s,  T)$ consisting of a commutative
subalgebra in $g$ and a subgroup $T$ in $G$ that commutes with $s$.
The decomposition $\hat s=\hat s_+\oplus\hat s_-\oplus \Bbb Cc$ in
annihilation operators, creation operators and central elements
induces a
decomposition $ s= s_+\oplus s_- $. Then one considers on $G/P$
``continuous and discrete time flows'' from the
pair $(\gamma_+,T)$, where $\gamma_+=\exp(s_+)$. The compatibility or
commutativity conditions of
these  flows will then be the Lax or zero curvature equations
(depending
on how one sets things up).

It is  well known (at least for the KP hierarchy) that the Hirota
type
equations are equivalent to the Lax or zero curvature equations.
The main point of this equivalence is the connection
between   the so called Baker function, or wave function, and the
$\tau$-function,  well known from the Japanese literature (e.g.,
\cite{DJMK}) and also, in the algebro-geometric situation, in Russian
works (e.g., \cite {Kr}). In the case of the principal Heisenberg
algebra this
connection  was given by \cite{SeW, 5.14}, by a ``simple
but mystifying proof'' in the words of \cite{W1}. For the case of the
homogeneous Heisenberg algebra \cite{Di3} uses this
connection to {\it define} the $\tau$-function.

It is the aim of this paper to
give a representation theoretic derivation of the connection between
the
$\tau$-function and the wave function. (We will use the term wave
function
instead of Baker function in the main part of the paper).
In our  set-up this
relation is derived from the
observation that the  wave  function consists, essentially, of  one
or several columns  in a
lower triangular matrix in the defining representation of $G$ (for
the
principal
Heisenberg algebra or in general, respectively). Furthermore one
finds that
one can calculate matrix elements of such matrices in terms of the
fundamental
highest weight  representations of the central extension $\hat G$.
The
explicit use of matrix elements of fundamental representations of
Lie algebras to solve integrable systems goes at least back to the
Kostant's
solution of the finite non-periodic Toda lattice, \cite{Ko}.

Let us describe this last simple, but essential, step in a finite
dimensional situation.
Consider the group $Gl(n,\Bbb C)$ acting on the vector space $H=\Bbb
C^n$,
with, as before,  basis $e_1,\dots,e_n$ and $H_j$ the subspace
of $\Bbb C^n $ spanned by the first $j$ basis vectors. We recall that
any
$g\in Gl(n,\Bbb C)$ admits a {\it Gauss-decomposition}
$g=g_-Pg_+$, with $g_-$ a lower triangular matrix with 1\rq s on
the diagonal, $g_+$ an invertible upper triangular matrix and $P$ a
permutation matrix. The permutation matrix is uniquely determined
by $g$, but the $g_\pm$
are not, unless $P=1_n$. We say that $g\in
Gl(n,\Bbb C)$
{\it belongs to the big cell for $H_j$} if $PH_j=H_j$. In this case
we can
choose the factors in a variant $g=g^j_-Pg^\prime_+$ of the
Gauss-decomposition in a unique way such that $P$ is the permutation
matrix
from the regular Gauss decomposition and $g_-^j$ is of the form
$$
g^j_-=1_{n\times n}+\sum_{{r> j}\atop {s\le j}}g_{rs}E_{rs},
\tag1.2.2
$$
If we put $g_+^j=Pg^\prime_+$ we have $g_+^jH_j=H_j$ and we see that
the $j$-dimensional subspace $W=gH_j$ of $\Bbb C^n$ projects
isomorphically to
$H_j$ by the natural projection. Now it is not too difficult
to see that the matrix elements $g_{rs}$ of $g^j_-$
can all be calculated using the fundamental representation
$\Lambda^j \Bbb C^n$ (cf. \cite {BaR\c a,
ch. 3.11}):
we have
$$
g_{rs}=\langle E_{rs} \cdot \bold{v}_j\mid g^j_- \cdot
\bold{v}_j\rangle=\langle E_{rs} \cdot
\bold{v}_j\mid g\cdot \bold{v}_j\rangle/\langle \bold{v}_j\mid g\cdot
\bold{v}_j\rangle,\quad r=1,2,\dots,n,\quad s\le j,
\tag1.2.3
$$
where $\bold{v}_j=e_1 \wedge e_2 \dots \wedge e_j$ and $\langle\,
\cdot \mid
\cdot\,
\rangle$ is the canonical Hermitian form on $\Lambda^j \Bbb C^n$ such
that $\langle   \bold{v}_j\mid   \bold{v}_j\rangle=1$. The
denominator
$\tau_j=\langle \bold{v}_j\mid g\cdot \bold{v}_j\rangle$ of
\thetag{1.2.3} is
the finite dimensional analogue of the famous $\tau$-function; if we
write
$$
g_+^j=\pmatrix A&B\\0&D\endpmatrix,
\tag1.2.4
$$
with $A$ of size $j\times j$, $B$ of size $j\times n-j$ and $D$ of
size
$n-j\times n-j$, then $\tau_j=\det(A)$. The most important property
is that
$\tau_j$ is nonzero iff $g$
belongs to the big cell for $H_j$ iff the projection from $W$ to
$H_j$ is an
isomorphism.

Translating \thetag{1.2.3} to our
infinite dimensional situation gives Lemma 5.5.1. Recall
that one can associate to every partition $\nbar$ of $n$ into $k$
parts a
vertex operator construction for the infinite matrix algebra,
using the technology of ``bosonized $k$ component fermion fields''
(see, e.g.,
\cite {tKvdL} and references therein).  For each
of these constructions we find solutions to the
$k$-component differential-difference
KP hierarchy and we obtain in
theorem 5.5.2. by a straightforward calculation	the
relation between the wave  function and the $\tau$-function for these
hierarchies.  In the literature $k$-component KP hierarchies were
introduced
in \cite{DJKM2} and studied in \cite{UT, Di1, Di3}. However there
apparently
only the  solutions related to the homogeneous partition
$n=1+1\dots+1$
are considered and also the difference equations in the hierarchies
seem to
be included only implicitly.

It is for various reasons interesting to study
more general partitions. Recall (from \cite{SeW}, say), that one can
associate
a solution of the
KP hierarchy to algebro-geometric data, consisting of a Riemann
surface $X$, a
point $p\in X$
with local coordinate $z\inv$ at $p$, a line bundle, etc. If  $z$
happens to
be the $n^{\text{th}}$ root
of a global meromorphic function on $X$ with only a pole at $p$ we
have
a covering map $X\to\Bbb P^1$ with $p$ as the $n$-tuple inverse image
of
the point $\infty\in \Bbb P^1$ and one obtains a solution of
the $n$-KdV hierarchy. The natural generalization (\cite{AB1, AB2})
of
this construction consists in considering $n$-fold coverings of $\Bbb
P^1$
such that the pull back divisor of $\infty\in \Bbb P^1$ is of the
form
$\sum_{a=1}^k
n_ap_a$, for $p_1,\dots,p_k$ points on $X$ and the $n_a$ positive
integers.
This gives us a partition $\nbar$ of $n$ and by choosing other
appropriate
geometric
data (line bundle, trivializations, etc) one finds a solution of the
$k$-component KP hierarchy. In \cite{LiMu} this construction is used
to
study
the analogue of the Schottky problem for Prym varieties. In
\cite{McI} these
type of soliton equations are studied in terms of flows on
generalized
Jacobians, see also \cite{Pr}. In section 5. we will spend quite some
time
discussing the fermionic translation operators $\hat Q_a$, the
translation group $\hat T$ constructed from it  and the relation with
the
infinite Grassmannian. In the algebro-geometric language the
operators $\hat
Q_a$
correspond to tensoring the line bundle with a bundle with divisor
$p_a$.

Another way in which the hierarchies related to arbitrary partitions
might
be
of interest is the following. Recently there has been much renewed
interest
in the Hamiltonian structure of soliton equations in relation to the
so called
$W$-algebras of conformal field theory. For instance in \cite{FeFr}
the $W_n$
algebras are constructed using  vertex operator algebras and
the (modified) $n$-KdV hierarchy corresponding to the principal
partition of
$n$.
The Hamiltonian structure of the $n$-KdV hierarchy is there obtained
using a remarkable duality of $W$ algebras. It seems
reasonable to expect that there exist for every partition of $n$
(or more generally for every vertex operator construction of affine
Kac-Moody
algebras) a related $W$ algebra and that using duality of $W$
algebras one can
obtain Hamiltonian structures for the corresponding soliton
hierarchies. Much
here
remains to be worked out, but see \cite{BdG, BdGH, dGHM}.

There are many papers on soliton equations, so we list only  a few.
Our
main sources
have been the papers
\cite{DS, DJKM, SeW,  KP}. For background and further references
on soliton theory we refer to the
books \cite{AbS, Ca, N, Di2}. Infinite dimensional Grassmannians and
infinite
dimensional Lie algebras are discussed in the monographs \cite{PrS,
Ka}.
Hierarchies of soliton equations in Hirota bilinear form
related to Heisenberg algebras and vertex operator constructions
have been discussed in \cite{KaW}.
\subhead{\bf 2.}
The infinite Grassmannian
\endsubhead
\subsubhead {\bf 2.1} Infinite matrix algebra and group\endsubsubhead
Let   $\Cinf$ be the vector space over $\Bbb C $ with basis
$\epsilon_i$,
$i\in
\Bbb Z$. Let $\glinf$ be the Lie algebra over $\Bbb C$ with
generators the
{\it elementary   matrices} (of size $\infty\times\infty$)
$\Epsilon_{ij},i,j\in \Bbb Z$ that have as only non zero matrix entry
a
1 on the $i,j$th place. We say, as usual, that $\Epsilon_{ij}$ is
upper
(lower) triangular if $i\le j$ ($i\ge j$). We have a natural action
of
$\glinf$ on $\Cinf$ given by
$$
\Epsilon_{ij}\cdot \epsilon_k=\epsilon_i\delta_{jk}.\tag 2.1.1
$$
The group
corresponding to $\glinf$ is $\Glinf$, consisting of infinite
invertible
matrices
$X=\sum_{i,j\in \Bbb Z} X_{ij}\Epsilon_{ij} $ such that only a finite
number
of  the $X_{ij}-\delta_{ij}$ is nonzero.

We will need in the sequel   infinite linear combinations of the
$\epsilon_i$
and the $\Epsilon_{ij}$. These don't occur in $\Cinf,\glinf$ and
$\Glinf$ and therefore we introduce
$$
\aligned
H&=\{\sum_{i=-\infty}^m c_i\epsilon_i\mid c_i\in \Bbb C,
m\in \Bbb Z\},\\
\gllf&=
\{\sum_{i,j\in \Bbb Z} c_{ij}\Epsilon_{ij}\mid c_{ij}\ne0
\text{ for only a finite number of }m=i-j>0\}, \\
\Gllf&=
\{\sum_{i,j\in \Bbb Z} X_{ij}\Epsilon_{ij}\mid X\text
{ invertible}, X_{ij} \ne0   \text{ for only a finite
number}\\
&\qquad\qquad\qquad\qquad\text{ of } m=i-j>0\}.
\endaligned\tag 2.1.2
$$
So the matrices we consider have only a finite number of
non zero lower triangular diagonals but are for the rest arbitrary.
The
Lie
algebra $\gllf$ and the group $\Gllf$ act on $H$ by extension of the
action \thetag{2.1.1}. This definition ensures
that the  exponential map of a strictly upper triangular
matrix in $\gllf$ is a well defined element of $\Gllf$, which  is the
main use we will make of these infinite sums.
To deal with  matrices with an infinite number of both upper and
lower
triangular diagonals, for instance in applications in algebraic
geometry,
one could use  the analytical setup of \cite{SeW} or of
\cite{ADKP}. We warn the reader
that in the
literature on the Sato Grassmannian (e.g., \cite{Sa2, AdC, KNTY, Mu})
one allows, in effect, sums that are infinite precisely in the
opposite
direction from our definition, e.g., infinite number of lower
triangular
diagonals, but a
finite number of upper triangular diagonals. In this approach one
cannot
define the exponential of an upper triangular matrix. As these papers
show,
one can circumvent this technical problem, and one would, by
following
this
path, obtain a larger infinite Grassmannian than we do
and a wider class of (formal)
solutions of the hierarchies we are going to construct. In this paper
we
prefer to avoid these technicalities, so as to be able to use later
on (in
chapter 5) the results of the representation theory of \cite{tKvdL},
which
is
set up just in the present context.

\subsubhead {\bf 2.2} Infinite Grassmannian and Gauss
decompositions\endsubsubhead
Define in $H$ for every integer $j$ a subspace
$$
H_j=\{\sum^{j}_{k=-\infty}c_k \epsilon_k\}\subset H\tag 2.2.1
$$
We define the {\it infinite Grassmannian} $Gr$ of $H$ as the
collection
of subspaces $W$ of $H$ of the form $W=gH_j$, for $g\in \Gllf$ and
some
$j\in \Bbb Z$.

The Grassmannian that we have defined here corresponds, mutatis
mutandis, to
what
is called the {\it polynomial Grassmannian} in \cite{SeW, PrS}). We
will need just a few facts about our
Grassmannian that can be conveniently derived from a
factorization of elements of the group $\Gllf$ as products of lower
triangular, permutation and upper triangular matrices.
This so called Gauss decomposition will also play an important r\^ole
in
the
construction of the soliton hierarchies in chapter 4.

To  formulate the Gauss decomposition in our infinite dimensional
context let
$\sfwinf$ be the group of infinite permutation matrices of
finite width. So $P\in\sfwinf$ iff $P\in \Gllf$ has a finite number
of non
zero diagonals and each row and column contains precisely one non
zero
entry,
which is equal to 1. A permutation matrix $P\in \sfwinf$ acts on
$\Cinf$
by
$P\epsilon_i=\epsilon_{\sigma_P(i)}$, where $\sigma_P:\Bbb Z\to \Bbb
Z$ is
a
permutation. This gives us a bijection of $\sfwinf$ with the
permutations
$\sigma$  of $\Bbb Z$ such that there exist an integer $N$ such that
$|\sigma(i)-i| <N$, for all $i\in \Bbb Z$.

\proclaim{Lemma 2.2.1 (Gauss decomposition)}
Every $g\in \Gllf$ can be factorized as
$$
g=g_-Pg_+,\quad g_-,P,g_+\in \Gllf\tag 2.2.2
$$
where $g_-\in \Gllf$, respectively $g_+\in \Gllf$, is strictly lower,
respectively
upper triangular, i.e., $g_-=1+\sum_{i>j} g_{ij}\Epsilon_{ij}$, and
$g_+=\sum_{i\le j} g_{ij}\Epsilon_{ij}$, $P\in \sfwinf$. In case $g$
happens to belong to $\Glinf$ also the factors $g_{\pm}$ and $P$ do.
\endproclaim

The proof, which is not essentially different from the finite
dimensional
case,
is left to the reader. Note that, as in the finite dimensional
situation, the permutation matrix $P$ is uniquely determined by $g$,
but that
the
factors
$g_\pm$ are not, unless $P=1_\infty$.
We will need a variant of the Gauss decomposition \thetag{2.2.2}
determined by  the choice of an integer $j$.  Let
$$
\aligned
\glinfpjn&=\{X=\sum X_{rs}\Epsilon_{rs}\mid X_{rs}=0 \text{ if }r>j
\text{
and }s\le j\},\\
\glinfmjn&=\{X=\sum X_{rs}\Epsilon_{rs}\mid X_{rs}=0 \text{ if }r\le
j
\text{ or }s> j\}.
\endaligned
\tag2.2.3
$$
There is a natural projection $pr_{W,j}:W\to H_j$, given by
$pr_{W,j}(f)=\sum_{i\le j}f_i
\epsilon_i$ if $f=\sum_{i=-\infty}^{m}f_i \epsilon_i$.
We will say that an element $W\in Gr$ belongs to the $H_j$ cell when
the
natural projection $W\to H_j$ is an isomorphism.

\proclaim{Lemma 2.2.2 (Gauss decomposition adapted to $H_j$)}
Let $g\in \Gllf$ be such that $W=gH_j$ is in the
$H_j$ cell. Then there is a unique decomposition of $g$ of the
form
$$
g=g_-^jg_+^j
$$ with
$$
g_-^j=1_{\infty}+X,\quad X\in\glinfmjn ;
\quad g_+ \cdot
H_j=H_j.
$$
\endproclaim
\demo{Proof}
By the Gauss decomposition we have $g=g_-Pg_+$ and since $W$
belongs to the
$H_j$ cell we have $Pg_+H_j=H_j$. Now write for the
minus
component   of  the Gauss decomposition
$g_-=1_{\infty}+\sum_{\ell>m} (g_-)_{\ell m}\Epsilon_{\ell m}$.
Then define a matrix
$$
f^j=1_{\infty}+ \sum_{m<\ell\le j }
(g_-)_{\ell m} \Epsilon_{\ell
m}+ \sum_{ \ell>m>j}
(g_-)_{\ell m} \Epsilon_{\ell
m}.\tag2.2.4
$$
This matrix is lower triangular with ones on the diagonal, so
is invertible and we can define a new decomposition
$$
g=g^j_-g^j_+,
\tag2.2.5
$$
where $g^j_-=g_- \cdot(f^j)\inv$, $g^j_+=f^j \cdot P
\cdot g_+$.
Then   $g^j_-$ is of the required form and also $g_+^jH_j=H_j$.
\qede
We say that the decomposition described by this lemma is  {\it
adapted to
$H_j$}, or to $j$ for short.

These decompositions are all related by conjugation. Indeed, if
$\Lambda$ is the shift matrix $\sum_{\ell\in \Bbb
Z}\Epsilon_{\ell\ell+1}$
in
$\Gllf$, then
$$
\Lambda\inv \Epsilon_{ij}\Lambda=\Epsilon_{i+1,j+1},
\tag2.2.6
$$
so we see that
$$
\aligned
\glinfpjn&=
\Lambda^{-j}gl_{\infty,+}^{0}\Lambda^{j},\\
\glinfmjn&=
\Lambda^{-j}gl_{\infty,-}^{0}\Lambda^{j}.
\endaligned
\tag2.2.7
$$
This will be used in section 7.

In general an element $f\in W$ is an infinite sum
$f=\sum_{i=-\infty}^{m}f_i \epsilon_i$. We say that an element $f\in
W$
has
{\it finite order} (in the negative direction) if there is an integer
$s$,
called the order of $f$, such that
$$
f=\sum_{i=s}^{m}f_i \epsilon_i,\quad f_s\ne0.
\tag2.2.8
$$
We denote
by $W^{\text{fin}}$
the collection of finite order elements in $W$.

We use the Gauss decomposition to introduce a canonical basis for
$W^{\text{fin}}$. Let $W=gH_j$ and let $g=g_-Pg_+$ be the Gauss
decomposition
as in the lemma. Then we have $W =g_-PH_j$, since $g_+$ is an
automorphism
of
$H_j$.  Now a basis for the finite order part of $H_j$ is given by
$\{\epsilon_i\mid i\le j\}$ and a basis for $(PH_j)^{\text{fin}}$ is
provided
by $\{\epsilon_{\sigma_P(i)}\mid i\le j\}$, where $\sigma_P$ is the
permutation corresponding to $P$. Then, since $g_-$ is strictly lower
triangular, we see
that by taking linear combinations of the finite order elements $g_-
\cdot
\epsilon_{\sigma_P(i)}$ we can obtain a canonical basis of
$W^{\text{fin}}$
given by
$$
w_s=\epsilon_s+\sum_{i\notin S^j_P\atop  i>s}c_i \epsilon_i,
\tag2.2.9
$$
where $s$ runs
over the set $S^j_P=\{\sigma_P(i)\mid i\le j\}$ of orders that
occur in $W$, and where for each $s$ we have (for our definition of
the Grassmannian) a finite  summation. $S^j_P$ is a set of integers
that
is
obtained from $\Bbb Z_{\le j}$ by deleting a finite number of
elements and
adding a finite number of  integers $>j$. So $S^j_P$ contains all
sufficiently small integers. Note that if $s$ is small enough
$w_s=\epsilon_s$, since there are in $g_-$ only finitely many
diagonals
below
the main one.

The natural projection $pr_{W,j}:W\to H_j$ has
finite dimensional 	kernel and cokernel. This follows,  for
instance,
easily
from the remarks about  the canonical basis of $W^{\text{fin}}$ we
just
made.
So we can define the {\it index} of $pr_{W,j}$ as
$\index(pr_{W,j})=\dim(\ker(pr_{W,j}))-\dim(\coker(pr_{W,j}))$. We
have
also
$\index(pr_{W,j})=\#(S^j_P-\Bbb Z_{\le j})-\#(\Bbb
Z_{\le j}- S^j_P)$,
so that the index depends only on the permutation matrix $P$
occurring in the Gauss decomposition of $g$, where $W=gH_j$.
The index of $pr_{W,0}$ is
also called the {\it virtual dimension}  of $W$, written
$\text{Virtdim}(W)$.
The Grassmannian  decomposes into disjoint components of
fixed virtual dimension: $Gr=\cup_{j\in \Bbb Z} Gr_j$ with
$Gr_j=\{W\in Gr\mid \text{Virtdim}(W)=j\}$. For instance $H_j$
belongs to
$Gr_j$.

An element $g$ of  $\Gllf$ is said to belong to the {\it big cell} if
it has
a Gauss decomposition with the permutation matrix $P$ the identity.
More
generally we say that $g$  belongs to the {\it  $H_j$ cell} if it has
a
Gauss decomposition with a permutation matrix $P$ as middle factor,
such
that
the corresponding permutation $\sigma_P$ is a product of
two (commuting)
permutations
$\sigma_j$, $ \sigma_{j}^\perp$ with $\sigma_j$ ($\sigma_j^\perp$)
leaving the sets of integers $\{i\mid i> j\}$ ($\{i\mid i\le j\}$)
pointwise
fixed. An element $g$ belongs to the
big cell iff it belongs to the  $H_j$ cell for all $j\in \Bbb Z$.

If $g$ belongs to
the  $H_j$ cell the corresponding element $W=gH_j$ can be written  as
$W=g_- H_j$, and projects therefore isomorphically
to $H_j$. In other words $g$ belongs to the  $H_j$ cell iff the
element
$W=gH_j$ does. This argument also
shows that the virtual dimension of $W$ is  in this case $j$.
An element $W\in Gr$ that is in the  $H_j$
cell has a particular simple canonical
basis (for its finite order part $W^{\text{fin}}$):
$$
w_s=\epsilon_s+\sum_{i>j}c_i \epsilon_i,\quad s\le j.
\tag2.2.10
$$
Again, only a finite number of $w_s$ differ from $\epsilon_s$.
\subhead{\bf 3.}
Partitions and associated Heisenberg systems
\endsubhead
\subsubhead {\bf 3.1} Relabeling associated to a
partition\endsubsubhead
Fix an integer $n>1$. Let $\nbar=(n_1\ge n_2\ge\dots\ge n_k>0)$ be a
partition
of $n$ into $k$ parts, so that we have $n=\sum_1^kn_a$.
We relabel the basis for
$\Cinf$
such that we have
$$
\Cinf=\bigoplus_{a=1}^k\bigoplus_{i\in \Bbb Z}
\Bbb C\epsilon_a(i),
\tag3.1.1
$$
with $ \epsilon_a(i)= \epsilon_j $, where $ j=np+n_1+\dots+n_{a-1}+q$
if $i=n_ap+q$ and $1\le q\le n_a$.  We call $\epsilon_a(i)$ the type
$\nbar$
relabeling of $\epsilon_j$. For all positive integers $n$ the {\it
principal
partition} $\nbar=n$, i.e., into one part, leads to same, trivial,
relabeling: $\epsilon_j=\epsilon_1(j)$.

The relabeling of the
basis for $\Cinf$ induces a natural relabeling of the basis for
$\glinf$:
an
infinite matrix is then thought to be build up out of $n\times n$
matrices,
each of which consists of blocks of size $n_a\times n_b$, $1\le
a,b\le
k$.
More explicitly   we put
$$
\Epsilon_{a b}^{n_ap+q \ n_br+s}=
\Epsilon_{np+n_1+\dots+n_{a-1}+q, nr++n_1+\dots+n_{b-1}+s},
\tag 3.1.2
$$
and we have
$$
\Epsilon^{i j}_{a b} \epsilon_c(\ell)=
\epsilon_a(i)\delta_{j\ell}\delta_{bc}.
\tag3.1.3
$$
The multiplication for the generators after relabeling reads:
$$
\Epsilon_{ab}^{ij}\Epsilon_{cd}^{kl}=\Epsilon_{ad}^{il}
\delta_{bc}\delta_{jk}.
\tag3.1.4
$$
We extend this
relabeling process for vectors and matrices in the obvious way to $H$
and
$\gllf$.

\subsubhead {\bf 3.2} The numbers $r_b(j)$\endsubsubhead
Fix an integer $j$ and a partition $\nbar=(n_1,n_2,\dots,n_k)$ of $n$
into
$k$ parts. We then can write $j=np+n_1+n_2+\dots+n_{a-1}+q$, with
$1\le
q\le
n_a$, so that
$\epsilon_j$ corresponds to $\epsilon_a(i)$, $i=n_ap+q$,
in the relabeling of
section 3.1. We associate to
these data the numbers $r_b(j)$ defined by:
$$
r_b(j)=\left\{\aligned n_b(p+1)\\n_ap+q\\n_bp
	    \endaligned\right.\quad
	    \aligned b&<a\\b&=a\\b&>a.
	    \endaligned
\tag 3.2.1
$$
In the sequel we will usually have fixed $j\in \Bbb Z$ and we then
will
write
simply $r_b$ for $r_b(j)$.
These numbers satisfy:
$$
\aligned
    j&=r_1(j)+r_2(j)+\dots +r_k(j),\\
    r_b(j) +n_b\ell &=r_b(j+\ell n),\quad \ell\in \Bbb Z\\
\endaligned
\tag3.2.2
$$
If $\nbar=(n)$ then $r_b(j)=r_1(j)=j$.
Note that, for any partition, the numbers $r_b(0)$ are all zero, so
the reader might wish to keep this simpler case in mind.

The meaning of these numbers is the following: consider the natural
ordering
on
the basis elements
of $\Bbb C^\infty$: $\epsilon_\ell\le\epsilon_j$ iff $\ell\le j$.
Let
$\epsilon_a(i)$ be the type $\nbar$ labeling of
$\epsilon_j$, so that $j=np+n_1+\dots+n_{a-1}+q$ and $i=n_ap+q$. Then
the
ordering on the relabeled basis vectors is given by
$$
\epsilon_b(m)\le\epsilon_a(i)\Longleftrightarrow m\le  r_b(j).
\tag3.2.3
$$
Another way of saying this is: $\epsilon_b(r_b)$ is the largest basis
vector
of type $b$ that is smaller than $\epsilon_a(i)$ (or equal, in case
$b=a$).
So for instance we have:
$$
\aligned
    H_j &=\{\sum_{t=-\infty}^j c_t\epsilon_t\} \\
	&=\{\sum_{b=1}^k\sum_{s=-\infty}^{r_b} c_s^b \epsilon_b(s)\}
\endaligned
\tag3.2.4
$$
Combining the second relation of \thetag{3.2.2} and \thetag{3.2.3} we
find:
$$
\epsilon_{j-n} < \epsilon_b(r_b(j))\le\epsilon_j.
\tag3.2.5
$$
The ordering of the basis $\epsilon_a(i)$ determines which relabeled
elementary matrices are upper triangular:
$$
\Epsilon_{ba}^{mi} \text{ is upper triangular} \Longleftrightarrow
m\le
r_b(j),
\tag3.2.6
$$
if $\epsilon_j$ corresponds to $\epsilon_a(i)$.

\subsubhead {\bf 3.3} Pre-Heisenberg system of a
partition\endsubsubhead
Define now shift matrices $\Lambda^+_a$, $\Lambda^-_a$ in $\gllf$ by
$$
\Lambda^+_a  =\sum_{\ell \in \Bbb Z} \Epsilon_{aa}^{\ell\ell+1},\quad
\Lambda^-_a  =\sum_{\ell \in \Bbb Z} \Epsilon_{aa}^{\ell\ell-1}.
\tag3.3.1
$$
They act on the standard basis of $\Cinf$ by
$$
\Lambda^+_a  \epsilon_b(i) = \epsilon_a(i-1)\delta_{ab},\quad
\Lambda^-_a  \epsilon_b(i) = \epsilon_a(i+1)\delta_{ab},
\quad 1\le a\le k.
\tag 3.3.2
$$
Let
$$
\Heisnbar=\bigoplus_{a=1}^k
\left (
\left[\bigoplus_{k>0}\Bbb C (\Lambda^+_a)^k
\right]
\bigoplus
\left[\bigoplus_{k>0}\Bbb C (\Lambda^-_a)^k
\right]
\right)
\tag3.3.3
$$
be the commutative subalgebra of $\gllf$ generated by the shift
operators.
We refer to $\Heisnbar$ as the
{ \it pre-Heisenberg algebra}  of type $\nbar$, since in the
universal
central extension of
$\gllf$ the lift of $\Heisnbar$ is indeed a Heisenberg algebra, see
section 5.1.

When $\nbar$ is a partition in more than one part we have besides the
pre-Heisenberg algebra another ingredient in the theory: the
pre-translation group.
We need some definitions.
We write $H=\oplus_{a=1}^k H_a$ where $H_a$ is the subspace of $H$
that
can be written using only $\epsilon_a(i)$, $i\in \Bbb Z$. Let
$1_a:=\sum_{i\in \Bbb Z}\Epsilon^{ii}_{aa}$ be the projection
operator
$H\to
H_a$. Define  operators on $H$ by
$$
Q_a=\sum_{b\ne a} 1_b+ \Lambda_a^-, \quad a=1,2, \dots,k.
\tag 3.3.4
$$
Then $Q_a$ is invertible and we have
$$
Q_a\inv=\sum_{b\ne a} 1_b+ \Lambda_a^+, \quad a=1,2, \dots,k.
\tag 3.3.5
$$
Let $R=\oplus_{i=1}^{k-1} \Bbb Z\alpha_i$ be the root lattice of the
simple Lie algebra
$sl(k,\Bbb C)$, with $\alpha_i, i=1,2,\dots, k-1$, the simple roots.
We define a homomorphism  from the additive group $R$ to a
multiplicative
Abelian subgroup in $\Gllf$ by
$$
\alpha_i\mapsto T_{\alpha_i}:= Q_{i}^{\phantom {1}}Q_{i+1}\inv,\quad
1\le i \le k-1.
\tag3.3.6
$$
In
particular
$\alpha=\sum_{i=1}^k d_i\alpha_i$ gets mapped to
$T_\alpha:=\prod_{i=1}^k T_{\alpha_i}^{d_i}$.
The image of $R$ is called the  {\it pre-translation group} (of type
$\nbar$) and is denoted by $ T^{\nbar}$.

One sees immediately that elements of $\Heisnbar$ and $ T^{\nbar}$
commute. The pair $(\Heisnbar, T^{\nbar})$ will be called
the {\it pre-Heisenberg system of type $\nbar$}.
\subhead{\bf 4.}
Multicomponent KP equations
\endsubhead
\subsubhead {\bf 4.1} Time evolution\endsubsubhead
Let $n$ be a positive integer. We are going to associate
to every partition $\nbar$  of
$n$ a collection of ``continuous and discrete time flows'' on the
infinite
Grassmannian $Gr$, using the pre-Heisenberg system $(\Heisnbar,
T^\nbar)$ of the previous section.
Let $\Gamma^\nbar$ be the subgroup of elements of
$\Gllf$ of the form
$$
\vacwn(t,\alpha)= \exp( \sum_{i>0}\sum_{a=1}^k t_i^a
(\Lambda^+_a)^i)\cdot T_\alpha.
\tag 4.1.1
$$
Here $t=\{t^a_i\in \Bbb C\mid  1\le a\le k,i>0\}$ are  the
``continuous time parameters'' and  $\alpha\in R$, where $R$,
the root lattice of $sl(k,\Bbb C)$,
is thought of as  a ``discrete time lattice''.  We will often
identify the pair $(t,\alpha)$ with the
element $\vacwn(t,\alpha)\in \Gamma^\nbar$. The elements
\thetag{4.1.1}
satisfy, of course,
$$
\vacwn(t,\alpha+\beta)= \vacwn(t,\alpha)T_\beta,
\tag 4.1.2
$$
for all $\alpha,\beta\in R$.

We define the
action (``time flow of type $\nbar$'') of $\vacwn(t,\alpha)$ on
the Grassmannian in
the following way: for $W\in Gr$ we put
$$
W(t,\alpha)=\vacwn(t,\alpha)\inv\cdot W=\exp(-\sum
t_i^a\Lambda_a^i)\cdot
T_{-\alpha}\cdot W.
\tag4.1.3
$$
If
$W=g\cdot H_j$
then we have $W(t,\alpha)=g(t,\alpha) \cdot H_j$, where
$$
g(t,\alpha)=\vacwn(t,\alpha)\inv \cdot g.
\tag4.1.4
$$
Note that different choices of $n$ and $\nbar$ might give the same
flow
on the Grassmannian.
For example we obtain the same flow if we take for any positive
integer
$n$
the
{\it principal} partition $\nbar=n$ of $n$ into one part. This are
the
famous KP-flows.

We denote by $\Gamma^{j,\nbar}_W$ the collection of
points
$(t,\alpha)$
in $\Gamma^\nbar$ such that $W(t,\alpha)=g(t,\alpha)H_j$ belongs to
the
big
cell with respect to $H_j$, see subsection 2.2.

\subsubhead {\bf 4.2} Formal Laurent series and pseudo differential
operators\endsubsubhead
The multicomponent KP equation that we are going to introduce
consists of
equations for a $k\times k$ matrix function of a ``spectral
variable''
$z$,
which appear as follows.

Fix, as always,
a partition $\nbar$ of $n$ into $k$ parts. Denote by $e_a$, $1\le
a\le k$
the standard basis  vector of $\Bbb C^k$ with a 1 on the
$a^{th}$ place and 0 elsewhere. We think of
the $e_a$ as column vectors. Similarly denote by $E_{ab}$, $1\le a,b
\le k$ the elementary matrix in $gl(k,\Bbb C)$ with a 1 on the
$(a,b)^{th}$
place and 0 elsewhere. Let, as usual,
$\Bbb C[[z]]$ be the integral domain of formal power series in the
variable $z$ and let $\Bbb C((z))$ be its quotient field, the
field of formal Laurent series.
Denote then by $H^{(k)}=\oplus_{a=1}^k \Bbb C((z))e_a$ the space of
$k$-component formal Laurent series.
Let now $\jmath^{\nbar}:H\to H^{(k)}$ be the linear isomorphism given
by
$$
\jmath^{\nbar}(\epsilon_a(i))=z^{-i}e_a ,
\tag4.2.1
$$
For any linear map $A:H\to H$ we have an induced map
$A^{(k,\nbar)}:H^{(k)}\to
H^{(k)}$ given by $A^{(k,\nbar)}=\jmath^{\nbar}\circ
A\circ(\jmath^{\nbar})\inv$. When the partition $\nbar$ is clear from
the
context we write $\jmath$ and $A^{(k)}$.

For any $W\in Gr$ we will write  $W^{(k,\nbar)}$ (or simply
$W^{(k)}$,
if $\nbar$ is fixed)
for the image $\jmath^{\nbar}(W)\subset
H^{(k)}$. The image
of $H_j\subset H$ is
$$
H^{(k,\nbar)}_j=\oplus_{b=1}^k \Bbb C[[z]]z^{-r_b}e_b
	=\{\sum_{b=1}^k\sum_{j=-r_b}^{\infty} c_{bj}z^j e_b\},
\tag4.2.2
$$
with $r_b=r_b(j)$ defined in 3.2.1. In $H$ the subspace  $H_j$ is
related
to
the standard subspace $H_0$ by $H_j=\Lambda^{-j}H_0$, for $\Lambda$
the
shift
matrix $\sum_{i\in \Bbb Z}\Epsilon_{ii+1}$
in $\Gllf$. Similarly
$$
H^{(k)}_j=\diag(z^{-r_1},z^{-r_2}, \dots,z^{-r_k})H^{(k)}_0
\tag4.2.3
$$

On $H^{(k)}$ we have a natural action of the formal loop algebra
$gl(k,\Bbb
C((z)))$. Often, in practice, it happens that the image $A^{(k)}$ of
an
operator $A:H\to H$ ends up lying inside $gl(k,\Bbb C((z)))$. This is
not
the
case for $\Lambda^{(k)}$, in general, but for
example, for $a=1,2,\dots,k$ we have
$$
\aligned
(\Lambda_a^{+})^{(k)}&=z E_{aa},\\
(\Lambda_a^{-})^{(k)}&=
			z\inv E_{aa},\\
Q_a^{(k)}&=\diag(z^{-\delta_{ab}}),\\
T_{\alpha_i}^{(k)}&=
	\diag(z^{\delta_{i+1j}-\delta_{ij}}), \quad i=1,2,\dots,k-1 .
\endaligned
\tag4.2.4
$$
To check the first relation, we note that
$\Lambda^+_a\epsilon_b(i)=\delta_{ab}\epsilon_a(i-1)$, so the linear
transformation induced by $\jmath$ on $H^{(k)}$ maps
$\jmath(\epsilon_b(i))
=z^{-i}e_b$ to $\delta_{ab}z(z^{-i})e_a$, i.e., this induced map is
multiplication by the matrix $z E_{aa}$. The other relations are also
easily
checked.

In particular the group element $\vacwn$ of \thetag{4.1.1},
responsible for
the
time evolution on $Gr$,  corresponds to multiplication on
$H^{(k)}$ by
$$
\vacw(z;t,\alpha):=(\vacwn)^{(k)}=\exp(\sum_{i>0}\sum_{a=1}^k t^a_i
z^iE_{aa}) \cdot T_\alpha^{(k)},
\tag4.2.5
$$
where
$$
\align
T_\alpha^{(k)}&=\prod_{i=1}^{k-1}(T_i^{(k)})^{d_i},
\tag4.2.6a\\
\intertext{if $\alpha=\sum_{i=1}^{k-1} d_i \alpha_i$.
When we think of the root
lattice
as a sub lattice of $\oplus_{a=1}^k \Bbb Z \delta_a$, with
$\alpha_i=\delta_i
-\delta_{i+1}$ and $(\delta_a|\delta_b)=\delta_{ab}$, then
we find
}
T_\alpha^{(k)}&=\diag (z^{-( \alpha|\delta_1)},
z^{-( \alpha|\delta_2)},\dots,z^{-(
\alpha|\delta_k)}).\tag4.2.6b
\endalign
$$

Recall the evolution group $\Gamma^\nbar$ of subsection 4.1 and the
corresponding time flows on $Gr$. Of all the
generators  of one parameter subgroups of $\Gamma^\nbar$
we distinguish a particular one and call this the generator of the
$x$-flow: we define,
$$
\partial={\partial\over\partial x}:= \sum_{a=1}^k \partial^1_a, \quad
\partial^i_a:=\frac{\partial}{\partial{t^a_i}}.
\tag4.2.7
$$
(See \cite{FNR} for discussion of this process of singling out a
particular
combination of the times as ``$x$'').
We will often let this operator act from the right, in which case we
write
$\ddxl={{\overleftarrow\partial}\over\partial x}$. This  operator
acts on
$\vacw(z;t,\alpha)$ by
$$
    \vacw(z;t,\alpha)\cdot\ddxl =\vacw(z;t,\alpha) \cdot z.
\tag4.2.8
$$
We will also consider formally the inverse of $\ddxl$, defined by
$$
    \vacw(z;t,\alpha)\cdot\ddxl\inv =\vacw(z;t,\alpha) \cdot z\inv.
\tag4.2.9
$$
We will in the sequel have to consider $k$-component formal Laurent
series
of the form
$$
\vacw \cdot X(z), \quad X(z)=\sum_{a=1}^k \sum_{i=s}^m z^iX_a^ie_a,
\tag4.2.10
$$
so that the vector $X(z)$ is a Laurent polynomial in $z$.
We can, in this situation,
``trade in'' every
occurrence of a power of $z$ in $X(z)$ for the corresponding power of
$\ddxl$: if we write $X$ as in \thetag{4.2.10} and define
$$
\overleftarrow X=\sum_{a=1}^k\sum_{i=s}^{m}
\ddxl^i X_{a}^{i} e_{a},
\tag4.2.11
$$
then, clearly, $\vacw \cdot X=
\vacw \cdot \overleftarrow X$. This procedure introduces in the
theory the (non commutative) ring of matrix pseudo differential
operators,
which will play an important role in the sequel.

\subsubhead {\bf 4.3} The wave function\endsubsubhead
We will now use the concepts introduced in the previous sections
to define the wave function.

\proclaim{Definition 4.3.1}
Fix an integer $j$, an element $W=gH_j$ in $Gr$ and a partition
$\nbar$ of
$n$ into $k$ parts.

The  {\it wave function of type $(j,\nbar)$ of $W\in Gr$}
is the $k\times k$ matrix function, defined for
$(t,\alpha)\in\Gamma^{j,\nbar}_W$, obtained by juxtaposing
the $k$ columns with index
$r_1(j),r_2(j),\dots,r_k(j)$
(these numbers are defined in \thetag{3.2.1})
of a certain infinite matrix  and
applying to each column the isomorphism $\jmath^\nbar$
(defined in the previous section) to get a $k$-component formal
Laurent
series:
$$
\aligned
\Wwave(z;t,\alpha)&=\jmath^\nbar\left(\vacwn(z;t,\alpha)
\cdot g_-^j(t,\alpha)
\cdot(\epsilon_1(r_1)\epsilon_2(r_2)\dots
\epsilon_k(r_k))\right)\\
&=\vacw(z;t,\alpha) \cdot \jmath^\nbar(g_-^j(t,\alpha)
\cdot(\epsilon_1(r_1)\epsilon_2(r_2)\dots\epsilon_k(r_k))).
\endaligned
$$
\endproclaim
Mostly we will write $w_W(t,\alpha)$ or even $w_W$ for
$\Wwave(z;t,\alpha)$.
Note that the elements
$\epsilon_1(r_1)$, $\epsilon_2(r_2),\dots,\epsilon_k(r_k)$
all belong to $H_j$, so the columns
$g_-^j(t,\alpha) \cdot \epsilon_b(r_b)$ belong
to $W(t,\alpha)=\vacwn(t,\alpha)\inv \cdot W$, and hence the columns
of
the wave
function $w_W$ all belong to $W^{(k)}$. Note furthermore that the
columns
$g_-^j(t,\alpha) \cdot \epsilon_b(r_b)$ are of the form
$$
g_-^j(t,\alpha) \cdot\epsilon_b(r_b)= \epsilon_b(r_b) +
\sum_{c=1}^k\sum_{\ell>0}
g^{cb}_{r_c+\ell, r_b}(t,\alpha)\epsilon_c(r_c+\ell).
\tag4.3.1
$$
This is so because  $g_-^j \cdot\epsilon_b(r_b)=\epsilon_b(r_b) + X
\cdot\epsilon_b(r_b)$ with $X\in gl_{\infty-}^j$. Now
$X \cdot\epsilon_b(r_b)$ consists of a linear combination of vectors
$\epsilon_c(\ell)$ that are larger than $\epsilon_j=\epsilon_a(i)$ in
the
ordering of section 4.2, as one sees using the explicit form 2.2.3 of
$gl_{\infty-}^j$ and the inequality \thetag{4.2.4}.
Applying the isomorphism $\jmath$ to $H^{(k)}$ shows
then that the columns of the wave function are of the form
$$
\aligned
(w_W)_b(t,\alpha)&=\vacw(z;t,\alpha)
\cdot\jmath^\nbar(g_-^j(t,\alpha)
(\epsilon_b(r_b))),\\
&= \vacw(z;t,\alpha) \cdot
\left(z^{-r_b}e_b +
\sum_{c=1}^k\sum_{\ell>0}z^{-r_c-\ell}g^{cb}_{r_c+\ell,
r_b}(t,\alpha)
e_c\right),\\
&=\vacw(z;t,\alpha) \cdot\diag(z^{-r_1}, z^{-r_2},\dots,z^{-r_k})
\cdot\left(e_b +
\sum_{c=1}^k\sum_{\ell>0}z^{-\ell}g^{cb}_{r_c+\ell, r_b}(t,\alpha)
e_c\right).
\endaligned
\tag4.3.2
$$
The summation over $\ell$ in \thetag{4.3.1-2} is finite, since in our
case
$g_-^j$
contains only a finite number of non zero diagonals.

The point of introducing this wave function is that its columns form
a
basis for an important class of elements of $W^{(k)}$
over a ring of differential operators.
Indeed, fix a $(t,\alpha)\in \Gamma^{\nbar}$ and  define
$$
\wkfin(t,\alpha)=\{f\in W^{(k)}
    \mid f(z)=\vacw
    (z;t,\alpha) \cdot\left(\sum_{b=1}^k\sum_{\ell=s_b}^{m_b}
    z^\ell f_{b\ell} e_b\right)\}.
\tag 4.3.3
$$
It is clear that $\wkfin(t,\alpha)$ does depend trivially on
$\alpha\in R$
and we will delete $\alpha$ here. We will
often also, having
fixed $t$, suppress the  $t$ dependence.
The space $\wkfin$ is
the image of the space $W^{(k)}(t,\alpha)^{\text{fin}}$, the finite
order
part of  $W^{(k)}(t,\alpha)$, under the isomorphism given by
multiplication
by $\vacw(z;t,\alpha)$.  Note  that if
$(t,\alpha),(t,\alpha+\beta)\in
\Gamma_W^{j,\nbar}$ (so that $w_W(t,\alpha)$ and
$w_W(t,\alpha+\beta)$
exist)
the columns of $w_W(t,\alpha)$, $w_W(t,\alpha+\beta)$ and also of
$\partial(w_W)(t,\alpha)$, $\partial (w_W)(t,\alpha+\beta)$
belong to $\wkfin$.

We have the following generalization
of results of Drinfeld--Sokolov \cite{DS}, Segal--Wilson \cite{SW}:
\proclaim{ Proposition 4.3.2} Fix $(t,\alpha)\in \Gamma^{j,\nbar}_W$,
let
$W^{(k)}$
be the image of $W$ in the space of $k$-component functions $H^{(k)}$
and
define $\wkfin$ by \thetag{4.3.3}. Then $\wkfin$ is a free rank $k$
module
over the ring $\Bbb C[\ddxl]$, with basis
the columns of $w_W(t,\alpha)$. More
explicitly: there exists for every $f(z)\in \wkfin$ a unique
$k$-component
differential operator
$\overleftarrow P(f)=\sum_{b=1}^k \overleftarrow P_{b}(f)e_b$,
$P_{b}(f)\in
\Bbb C[\ddxl]$,
such that
$$
f=w_W(t,\alpha) \cdot \overleftarrow P(f).
\tag4.3.4
$$
\endproclaim
\demop
Let $\epsilon_j$ correspond to $\epsilon_a(i)$ and let $r_b$ the
numbers
associated with $j$ in section 4.2. Suppose that for $f(z)\in \wkfin$
as
in
\thetag{4.3.3} $f_{bm_b}\ne0$. Then we call $m_b+r_b$ {\it the
$b$-order}
of
$f(z)$ and $f_{bm_b}$ {\it the leading $b$-coefficient}. If all
$f_{b\ell}$
are zero the $b$-order is $-\infty$. (This ordering comes
from a refinement of the ordering \thetag{2.2.3} on $W(t,\alpha)$,
via
application of $\jmath$ and multiplication by $\vacw(z;t,\alpha)$).

For example,
the $b^{\text{th}}$ column of $w_W$ has $b$-order 0 and leading
$b$-coefficient is 1, whereas its $c$-order for $c\ne b$ is strictly
negative. If $f\in\wkfin$ then also $f \cdot \ddxl\in\wkfin$, with
the
$b$-order of $f \cdot \ddxl$ bigger by 1 than that of $f$
and with the leading $b$-coefficient
unchanged, for all $b=1,2,\dots,k$.

If $(t,\alpha)\in \Gamma^{j,\nbar}_W$, then we have
$W(t,\alpha)=g_-^j(t,\alpha) H_j$, so
that $W(t,\alpha)$ projects isomorphically to $H_j$. Applying
$\jmath$ we
see
that $W^{(k)}(t,\alpha)$ maps isomorphically to $H^{(k)}_j$. In
particular
if $f=\vacw \cdot X\in
\wkfin$, with $X=\sum_{b=1}^k\sum_{j=s_b}^{m_b}
    z^j f_{bj} e_b$ then $X= w_0\inv \cdot f\in W^{(k)}(t,\alpha)$
and
$X$ maps isomorphically to $H^{(k)}_j$ using the projection
$pr^{(k)}_{W,j}$.
Now if the $b$-order of $f$ were
negative for all $b$ then all $m_b$ would be smaller than $-r_b$ and,
taking explicit form \thetag{4.2.2} of $H_j^{(k)}$ into account,
$X$ would project to $0\in H^{(k)}_j$.
Since this projection is an isomorphism
$X$, and hence $f$, has to be zero. So
we see that for $f\in \wkfin$ at least one of the $b$-orders is
non negative, unless $f=0$.

Now, of course, the idea is, given $f\in \wkfin$, to reduce its
orders by
subtraction of  terms $(w_W)_b \cdot \overleftarrow P$, for suitable
differential operator
$\overleftarrow P$. More precisely,
let $m=m_b+r_b$ be the {\it total order} of f, i.e., the
maximum of the orders that occur, and let $\mu$ be the {\it
multiplicity}
of
$m$, i.e., the number of components $c$ for which the
$c$-order is equal to the total order. Then
$(w_W)_b \cdot \overleftarrow P$, with $\overleftarrow P
= (\ddxl)^{m}f_{bm_b}$, has the same $b$-order
$m=m_b+r_b$ and the same
leading $b$-coefficient as $f$ and the $c$-order, $c\ne b$, is at
most
$m-1$.
Subtracting we obtain an element $f-(w_W)_b \cdot \overleftarrow P$
of
$\wkfin$ of lower order in the $b$-component. If the multiplicity
$\mu$
was
1 the total order of $f-(w_W)_b \cdot \overleftarrow P$ is strictly
smaller than that of $f$. In case
the multiplicity is larger than $1$ the total order of $f-(w_W)_b
\cdot
\overleftarrow P$
will still be $m$, but the multiplicity is one
smaller than that of $f$. By repeating this process we reduce the
multiplicity and the total order and we can find  $k$ differential
operators
$\overleftarrow P_{a}$ such that $\tilde f:=f(z)-\sum_{a=1}^k(w_W)_a
\cdot
\overleftarrow P_a$ has  its $b$-order, for all $b$, less than $0$.
Then, as
we argued before, $\tilde f$
itself must be zero, so $f(z)=\sum_{a=1}^k(w_W)_a \cdot
\overleftarrow P_{a}$. It is easy to check that the differential
operators
$\overleftarrow P_{a}$ are unique.
\qede

\subsubhead {\bf 4.4} Differential difference multi-component
KP\endsubsubhead
In this subsection we derive the equations satisfied by the wave
function as
a function of the continuous and discrete time variables.

The wave function  $w_W(t,\alpha)$ of type $j,\nbar$,
defined whenever $(t,\alpha)\in \Gamma^{j,\nbar}_W$, can be written
as
$$
w_W(t,\alpha)=\vacw(z;t,\alpha) \cdot\overleftarrow w_W (t,\alpha),
\tag 4.4.1
$$
where $\overleftarrow w_W $ is a $k \times k$ matrix pseudo
differential
operator,
called the {\rm wave operator}, of the form
$$
\overleftarrow w_W =\text{diag}(z ^{-r_1},\dots,z^{-r_k})
\cdot (1_{k \times k} +
\sum_{i>0}\ddxl^{-i}w_i),
\tag4.4.2
$$
with the $w_i$ $k \times k$ matrices with entries in the ring of
functions $B$ and the numbers $r_b$ defined in   \thetag{3.2.1}.
To see this use the explicit form \thetag{4.3.2} for the columns of
the
wave
function and
use \thetag{4.2.9} to trade in negative powers of $z$ for powers of
$\ddxl\inv$. The wave operator will be used later on to define
resolvents.

Note that in \thetag{4.4.2}, according to our definitions, the
summation over
$i$
is finite. Also note that $\overleftarrow w_W $ is invertible as a
formal
pseudo differential operator (PDO), since
it starts out with an
invertible matrix and contains for the rest only negative powers of
$\ddxl$. Note finally that if we had used $g_-$, the component of the
ordinary Gauss decomposition (proposition 2.2.1) instead of $g_-^j$,
the
component in the Gauss
decomposition
adapted to $j$ (Lemma 2.2.2), to define the
wave function in \thetag {4.3.1},
the  first term in the expansion \thetag{4.4.2} would have been not
the
identity matrix but rather more complicated. The choice we make here
also
allows us to calculate the wave function in a rather straight forward
manner
in terms of the $\tau$-function, see Theorem 5.5.2.
In this sense the decomposition introduced in Lemma 2.2.2 is adapted
to
$j$.

There exist unique $k\times k$ matrix
pseudo differential
operators $\overleftarrow \Lambda_a$, $\overleftarrow T_{\alpha_i}$,
such
that
$$
\aligned
\partial^1_a \vacw&=\vacw \overleftarrow \Lambda_a,\\
\vacw(z;t,\alpha+\alpha_i)&=\vacw(z;t,\alpha)\overleftarrow
T_{\alpha_i}.
\endaligned
\tag{4.4.3}
$$
Explicitly we have
$$
\aligned
    \overleftarrow\Lambda_a &:=  \ddxl  E_{aa}\\
\overleftarrow T_{\alpha_i}&:=  \ddxl E_{ii} +\ddxl\inv E_{i+1i+1}+
\sum_{j\ne i,i+1}E_{jj}.
\endaligned
\tag4.4.4
$$
Now define
$$
\aligned
\psresa &:=\overleftarrow w_W \inv \cdot \overleftarrow \Lambda_a
\cdot
\overleftarrow w_W =  \ddxl
E_{aa}+[E_{aa},w_1]+ \Cal O(\ddxl\inv),\\
\pslatresai&:=\overleftarrow w_W \inv \cdot \overleftarrow
T_{\alpha_i}\inv\cdot \overleftarrow w_W =
\overleftarrow w_W \inv \cdot
\left( \ddxl\inv E_{ii}+\ddxl E_{i+1i+1}+\sum_{j\ne
i,i+1}E_{jj}\right)
\cdot \overleftarrow w_W .
\endaligned
\tag4.4.5
$$
We refer to $\psresa$ and $\pslatresai$ as the pseudo differential
resolvents and lattice resolvents
associated to $(W,j,\nbar)$ respectively. (See  \cite{GD} for
the concept of a resolvent. The lattice resolvent was introduced in
\cite
{BtK}). Note that in both $\psresa$ and $\pslatresai$ the first
diagonal
factor $\text{diag}(z^{-r_1},z^{-r_2},\dots,z^{-r_k})$
of $\overleftarrow{w}_W$ in \thetag{4.4.2} cancels, so that
resolvents and
lattice
resolvents have the same general form whichever $H_j$ cell or
partition
into $k$ parts we use.

The lattice resolvent
$\pslatresai$ is an invertible matrix pseudo differential
operator. We say that an invertible $k\times k$ matrix
pseudo differential operator $A$  is {\it in the big cell} if it
admits a
decomposition
$A=A_-A_+$, where $A_+$ is an invertible $k\times k$
matrix differential operator and
$A_-=1_{k\times k}+\Cal O(\ddxl\inv)$. Such a decomposition, if it
exists,
is
unique.
The resolvent $\psresa$ is in general  not invertible. In fact  $\Cal
R_a\Cal R_b=0$
for $a\ne b$. (When $k=1$
$\psresa$ is invertible, as a monic scalar PDO).
We will denote by $(\psresa^i)_+$ the differential operator part of
the
$i^{\text{th}}$ power of
$\psresa$ and by $(\psresa^i)_-$ the formal integral
operator $\psresa^i-(\psresa^i)_+$,
and similar for other possibly
non invertible matrix pseudo differential operators. So the notation
subscripts $\pm$ is not entirely unambiguous, but the meaning is
hopefully
clear from the context.

\proclaim{Proposition 4.4.2}
Let $W\in Gr_j$ and suppose that $(t,\alpha)$,
$(t,\alpha+\alpha_i)$
belong to $\Gamma^{j,\nbar}_W$. Then the lattice resolvent
$\pslatresai$
is
in the big   cell and we have:
$$
\aligned
    \partial^i_b w_W&= w_W \cdot(\Cal R^i_b)_+ ,\\
    w_W(t,\alpha+\alpha_i)&=w_W(t,\alpha)   \cdot(\pslatresai)_+\inv.
\
\endaligned
\tag4.4.6
$$
\endproclaim

\demo{Proof}
Until now we have defined the action  of
PDO\rq s with a finite number of negative powers of $\partial$ on
$w_0$ and the action of differential operators on expressions one
obtains
in
this way. We also need to define an action of arbitrary PDO's on
expressions of the form \thetag
{4.2.10}: we put
$$
\aligned
\vacw \cdot X(z) \cdot\ddxl E_{ab}&=\vacw \cdot
    (z+\partial) X(z)E_{ab},\\
\vacw \cdot X(z) \cdot
\ddxl\inv E_{ab}&=\vacw \cdot(z+\partial)\inv X(z)E_{ab},\\
	&=\vacw \cdot z\inv\sum_{i=0}^\infty (-z\inv\partial)^i X(z)
	E_{ab}.
\endaligned
\tag4.4.7
$$
Here we run into  a little trouble: $\vacw$ is a power series in $z$
and
there is a priori
no guarantee that the product in the last line of \thetag{4.4.7}
makes sense. The easiest way to circumvent this problem is not to try
to calculate this product and  instead interpret $\vacw$ as an
abstract
free
generator $v_0$ of a module $N$ of expressions $v_0 \cdot
\sum_{-\infty}^m
F_iz^i$ over the matrix PDO's with action given by
\thetag {4.4.7} (with $\vacw$ replaced by $v_0$). (cf. \cite{DS}).
In the obvious way we also
define differentiation with respect to the times $t^b_\ell$ on $N$.
We identify then $w_W=\vacw \cdot \overleftarrow w_W$ with the
element
$v_W=v_0 \cdot
\overleftarrow w_W$ and the proof of this Proposition
takes place in the module $N$.
It happens that for some elements
of $N$, such as $v_W \cdot(\Cal R^i_b)_+$, $v_W
\cdot(\pslatresai)_+\inv$,
one can give an interpretation as a formal Laurent
series; in particular we can interpret
$v_W \cdot(\Cal R^i_b)_+$, $v_W \cdot(\pslatresai)_+\inv$ as the
series
$w_W \cdot(\Cal R^i_b)_+$, $w_W \cdot(\pslatresai)_+\inv$.
This being understood we will in the sequel just write $\vacw$ for
$v_0$.

As we noted before, the columns of $w_W(t,\alpha)$  belong to
$W^{(k)}$ and in fact to $\wkfin$. The same is true of $\partial^i_b
w_W$
and
of $w_W(t,\alpha+\alpha_i)$.
So we can use
proposition 4.3.2 to conclude that   $\partial^i_b w_W$ and
$w_W(t,\alpha+\alpha_i)$ are of the form $w_W(t,\alpha) \cdot O$,
respectively
$w_W(t,\alpha) \cdot P$,
with $O,P$   $k\times k$ matrix
differential operators. On the other hand we have by \thetag{4.4.1}
and \thetag{4.4.3}:
$$
\aligned
\partial^i_b w_W&=\vacw \cdot(\overleftarrow \Lambda^i_b
\overleftarrow w_W +
\partial^i_b\overleftarrow w_W )\\
    &=w_W \cdot(\overleftarrow w_W \inv \overleftarrow\Lambda^i_b
    \overleftarrow w_W +
\overleftarrow w_W \inv \partial^i_b\overleftarrow w_W )\\
w_W(t,\alpha+\alpha_i)&=\vacw(z;t,\alpha)
		    \cdot(\overleftarrow  T_{\alpha_i} \overleftarrow
		    w_W(t,\alpha+\alpha_i))\\
		    &=w_W(t,\alpha) \cdot(\overleftarrow
			w_W \inv(t,\alpha) \cdot
		    \overleftarrow T_{\alpha_i}
		    \cdot\overleftarrow w_W(t,\alpha)
		    \cdot\overleftarrow w_W(t,\alpha)\inv
		    \cdot\overleftarrow w_W(t,\alpha+\alpha_i)).
\endaligned
\tag4.4.8
$$
Now $w_0$ and  hence also $w_W$
is a free generator for the action of
PDO\rq s, i.e., if for two PDO\rq  s $X,Y$ we have
$w_W \cdot X=w_W \cdot Y$ then $X=Y$. This implies
$$
\aligned
O&=\overleftarrow w_W \inv \overleftarrow \Lambda^i_b \overleftarrow
w_W
+\overleftarrow w_W \inv \partial^i_b\overleftarrow w_W =\Cal R_b^i+
\overleftarrow w_W \inv \partial^i_b\overleftarrow w_W ,\\
P&=\overleftarrow w_W\inv(t,\alpha) \cdot \overleftarrow T_{\alpha_i}
\cdot\overleftarrow w_W (t,\alpha)
\cdot\overleftarrow w_W (t,\alpha) \inv    \cdot\overleftarrow
    w_W(t,\alpha+\alpha_i)\\
	    &=(\pslatresai)\inv\cdot\overleftarrow w_W (t,\alpha)
\inv
		    \cdot\overleftarrow w_W (t,\alpha+\alpha_i).
\endaligned
\tag4.4.9
$$
Note that $\overleftarrow w_W \inv \partial^i_b\overleftarrow w_W $
is an
operator containing only
negative powers of $\ddxl$ while $O$ is a differential operator. This
implies
$$
\overleftarrow w_W \inv \partial^i_b\overleftarrow w_W  =-(\Cal
R^i_b)_-
\tag4.4.10
$$
Similarly
$\overleftarrow w_W (t,\alpha) \inv    \cdot\overleftarrow w_W
(t,\alpha+\alpha_i)$
is of the form $1_{k\times k}+\Cal O(\ddxl\inv)$ and $P=P(\alpha)$ is
an
invertible differential operator. By the same argument we see that
the
operator $P^\prime(\alpha)$ such that
$w_W(t,\alpha-\alpha_i)=w_W(t,\alpha)
\cdot P^\prime$ is an invertible differential operator. Since $w_W$
is a
free
generator we have $P^\prime(\alpha+\alpha_i)P(\alpha)=1$, i.e.,
the inverse of $P$ is also a differential operator. From this we see
that
$\pslatresai$ belongs
to the big cell and
$$
\overleftarrow w_W (t,\alpha) \inv    \cdot\overleftarrow w_W
(t,\alpha+\alpha_i) =
(\pslatresai)_-.
\tag4.4.11
$$
Combining \thetag{4.4.9}, \thetag{4.4.10} and \thetag{4.4.11} proves
the
Proposition.
\qede

\proclaim{Definition  4.4.3} Let $L$ be a matrix PDO of the form
$$L=\ddxl A +\Cal O(\ddxl^0)
\tag 4.4.12
$$
for $A$ a diagonal constant matrix with distinct
non zero eigenvalues $A_a$. Let $w(z)$ be a solution of
$$
w \cdot L=zA w,
\tag4.4.13
$$
and introduce the resolvent and lattice resolvent associated to L by
\thetag
{4.4.5} using $w$ for $w_W$.
Then the  $k$-component
differential-difference KP hierarchy is the
system of deformation equations for L:
$$
\aligned
\partial^i_b L &=[ L ,(\Cal R^i_b)_+],\\
L(t,\alpha+\alpha_i)&= (\pslatresai)_+ \cdot L(t,\alpha) \cdot
(\pslatresai)_+\inv.\\
\endaligned\tag4.4.14
$$
\endproclaim

Consider  the   $k\times k$ matrix pseudo differential operator
$$
\aligned
L_W(t,\alpha)&=\overleftarrow w_W (t,\alpha)\inv \cdot \ddxl A \cdot
\overleftarrow
    w_W(t,\alpha)=\sum_{a=1}^k A_a\psresa,\\
    &=\ddxl A +\Cal O(\ddxl^0)
\endaligned
\tag4.4.15
$$
Then $w_W L_W=zA w_W$ and one  finds, using that $L_W$, $\psresa$ and
$\pslatresai$
commute, that $L_W$ is a solution of the $k$-component
KP hierarchy, if $(t,\alpha)$, $(t,\alpha+\alpha_i)$ belong to
$\Gamma_W^{j,\nbar}$.

The compatibility equations for \thetag{4.4.6} are the ``zero
curvature
equations'' that are also useful:
$$
\align
    \partial^\ell_b (\psresa^m)_+ &=\partial^m_a (\Cal R_b^\ell)_+ -
			[(\psresa^m)_+,(\Cal
R_b^\ell)_+],\tag4.4.16a\\
    (\psresa(t,\alpha+\alpha_i)^\ell)_+&=
	-\partial^\ell_a(\pslatresai)_+(\pslatresai)_+
	\inv+(\pslatresai)_+ \cdot
	(\psresa(t,\alpha)^\ell)_+  \cdot
	(\pslatresai)_+\inv .\tag4.4.16b
\endalign
$$

\subsubhead {\bf 4.5} An example: the Davey-Stewartson-Toda
system\endsubsubhead
In this subsection we discuss a few of the equations that follow from
the equations \thetag{4.4.16}.

Our starting point is an element $W$ of the Grassmannian $Gr$ and the
choice
of a partition $\nbar$, defining a time flow $W\mapsto W(t,\alpha)$,
see
\thetag{4.1.3}. The simplest case is obtained by choosing
for any positive integer $n$ the principal partition $\nbar=n$ into
one
part. The resulting equations form, of course, the KP hierarchy,
discussed
extensively in the literature, (see e.g., \cite{SeW}),
with the discrete part \thetag{4.4.16b}
missing in this case.

The next simplest case occurs when we choose for any $n>0$ a
partition
$\nbar=(n_1,n_2)$ into two parts. Now there will be a doubly infinite
set
of continuous time parameters $(t_1^i,t_2^i)$, with $i>0$ while the
discrete
parameter $\alpha$ lives on the rank one root lattice
of $sl_2$ with generator $\alpha_1$: $\alpha=m\alpha_1, m\in \Bbb Z$.
We
will
indicate the dependence on the
discrete variable by a superscript:  we write $W^m(t)$ for
$W(t,m\alpha_1)$, etc. We introduce some new variables
$$
x={\frac 1 2}(t_1^1+t_1^2), \quad
\bar x={\frac 1 2}(t_1^1-t_1^2), \quad
\bar t={\frac 1 2}(t_2^1+t_2^2), \quad
t={\frac 1 2}(t_2^1-t_2^2),
\tag4.5.1a
$$
and the following differential operators:
$$
\partial=
    \partial_1^1+\partial_2^1,\quad
    \bar\partial=
\partial_1^1-\partial_2^1,\quad
\partial_{t}=
\partial_1^2-\partial_2^2.
\tag4.5.1b
$$
Let $h=\pmatrix1&0\\0&-1\endpmatrix$. We will need the following
resolvents:
$$
\aligned
R^m_{\bar x}(t)&=R_1^m(t)-R_2^m(t)= \overleftarrow w^m(t)\inv
    (\ddxl h)\overleftarrow w^m(t),\\
R^m_{t}(t)&=R_1^m(t)^2-R_2^m(t)^2= \overleftarrow w^m(t)\inv
    (\ddxl^2 h) \overleftarrow w^m(t).
\endaligned
\tag4.5.2
$$

The resolvents \thetag{4.5.2} are defined iff the corresponding
element $W^m(t)$ belongs to the big cell.
\proclaim{Lemma 4.5.1} Let $W^m(t)$ belong to the big cell.
Then the resolvents \thetag{4.5.2} have the expansion
$$
\align
R^m_{\bar x}&=\ddxl h+\pmatrix0&q^m(t)\\r^m(t)&0\endpmatrix +\Cal
O(\ddxl\inv),\tag4.5.3a\\
R^m_{t}&=\ddxl^2 h+\ddxl \pmatrix0&q^m(t)\\r^m(t)&0\endpmatrix +
\lambda^m(t)1_2+\mu^m(t)h +\\
&\qquad\qquad +\pmatrix 0&\frac12(\partial-\bar\partial)q^m(t)\\
    \frac12 (\partial+\bar\partial)r^m(t)&0\endpmatrix+
\Cal O(\ddxl\inv).\tag4.5.3b
\endalign
$$
Define $Q^m=\mu^m+\frac12{q^mr^m}$. Then we have the following
equations
for
$q^m,r^m,Q^m$:
$$
\align
\partial_t q^m&=-\frac12(\partial^2+\bar\partial^2){q^m}
	+(q^m)^2r^m-2q^mQ^m,
\tag4.5.4.a\\
\partial_t r^m&=\frac12(\partial^2+\bar\partial^2) {r^m}
	-(r^m)^2q^m+2r^mQ^m,
\tag4.5.4.b\\
(\partial^2-\bar\partial^2)Q^m&=\partial^2(q^mr^m).
\tag4.5.4.c\\
	    \endalign
$$
\endproclaim
The equations \thetag{4.5.4} form the Davey--Stewartson system
(\cite{DaS, SaA}).
\demop
We write for the wave operator and its inverse
$$
\aligned
\overleftarrow w^m(t)&=1_2+\ddxl\inv w^m_1+\ddxl^{-2}w^m_2+\dots,\\
\overleftarrow w^m(t)\inv&=1_2+\ddxl\inv
v^m_1+\ddxl^{-2}v^m_2+\dots,\\
\endaligned
\tag4.5.5
$$
where $w^m_i,v^m_i$ are $2\times 2$ matrices and where we have
ignored the
irrelevant diagonal factor $\diag(z^{-r_1},z^{-r_2})$
(see the remark after \thetag{4.4.5}). Then we
have
$$
v^m_1=-w^m_1,\quad v_2=-w^m_2+(w^m_1)^2.
\tag4.5.6
$$
If we write
$$
\aligned
R^m_{\bar x}&=\ddxl h+\gamma^m +\Cal O(\ddxl\inv),\\
R^m_{t}&=\ddxl^2 h+\ddxl \gamma^m + \delta^m+\Cal O(\ddxl\inv),
\endaligned
\tag4.5.7
$$
then
$$
\gamma^m=[h,w^m_1],\quad\delta^m=[h,w^m_2]+w^m_1[w^m_1,h]-2\partial
w^m_1
h.
\tag4.5.8
$$
This shows that $\gamma^m$ is off-diagonal, so that we can write
$$
\gamma^m(t)=\pmatrix 0&q^m(t)\\r^m(t)&0\endpmatrix,
\tag4.5.9
$$
and this proves \thetag{4.5.3.a}.
Next we consider the zero curvature equation
$$
[\bar \partial+(R^m_{\bar x})_+,\partial_t+(R^m_t)_+]=0.
\tag4.5.10
$$
Equating the coefficients of powers of $\ddxl$ to zero gives the
equations
$$
\align
[h,\delta^m]&=-\bar\partial\gamma^m-\partial\gamma^mh,\tag4.5.11.a\\
0&=\bar\partial\delta^m -\partial \delta^m.h-
	\partial_t\gamma^m+\partial^2\gamma^m h+
	\partial\gamma^m
\cdot\gamma^m+[\gamma^m,\delta^m].\tag4.5.11.b
\endalign
$$
The first equation shows that we can write
$$
\delta^m =\lambda^m(t)1_2+\mu^m(t)h +\pmatrix
0&\frac12(\partial-\bar\partial)q^m(t)\\
    \frac12 (\partial+\bar\partial)r^m(t)\endpmatrix
    \tag4.5.12
$$
which proves \thetag{4.5.3.b}. Substituting the Ans\"atze
\thetag{4.5.9, 4.5.12} in \thetag{4.5.11b } we obtain the system
$$
\aligned
2\bar\partial\lambda^m-2\partial\mu^m+\partial(q^mr^m)&=0,\\
-2\partial\lambda^m+2\bar\partial\mu^m+\bar\partial(q^mr^m)&=0,\\
-(\partial^2+\bar\partial^2)q^m/2-2\mu^m q^m-\partial_tq^m&=0,\\
(\partial^2+\bar\partial^2)r^m/2+2\mu^m r^m-\partial_tr^m&=0,\\
\endaligned
\tag4.5.13
$$
Now we eliminate the variable $\lambda^m$ by imposing on the first
two
equations of \thetag{4.5.14} the integrability condition
$\bar\partial\partial(\lambda^m)=\partial\bar\partial(\lambda^m)$.
This
gives
$$
\partial^2(2\mu^m-q^mr^m)=\bar\partial^2(2\mu^m+q^mr^m).
\tag4.5.14
$$
The last two equations of \thetag{4.5.13} combined with
\thetag{4.5.14}
and
the definition of $Q^m$
give then the Davey--Stewartson system \thetag{4.5.4a-c}.
\qede

To derive the discrete equations we need the lattice resolvent
$$
U^m(t)=\overleftarrow
w^m(t)\inv\pmatrix\ddxl\inv&0\\0&\ddxl\endpmatrix
    \overleftarrow w^m(t).
\tag4.5.15
$$
\proclaim{Lemma 4.5.2} Assume that $W^m(t)$ and $W^{m+1}(t)$ belong
to the big
cell. Then the lattice resolvent \thetag{4.5.15} admits a
factorization
$U^m=U^m_-U^m_+$ where $U^m_-=1_2+\Cal O(\ddxl\inv)$ and
$$
U^m_+=\pmatrix 0&2/{r^m}\\-{r^m}/2&\ddxl+\frac12
(\partial-\bar\partial)\log(r^m)\endpmatrix.
\tag4.5.16
$$
\endproclaim
\demop
The fact that the factorization exists follows from the proof of
Proposition
4.4.2. To calculate the positive factor write
$$
\aligned
U^m(t)&=\ddxl E_{22}+[E_{22},w^m_1]+\\
	&\qquad+\ddxl\inv\big\{
	E_{11}-\partial
	w^m_1E_{22}+[E_{22},w^m_{2}]+w^m_1[w_1,E_{22}]\big\}+
	\Cal O(\ddxl^{-2}),
\endaligned
\tag4.5.17
$$
where we have used the expansion \thetag{4.3.6}. Parametrize
$w^m_1=\pmatrix
a^m&q^m/2\\-r^m/2&b^m\endpmatrix$. Then we find
from \thetag{4.5.9}
$$
\delta^m=[h,w^m_2]+
\pmatrix-\frac12{q^mr^m}-2\partial a^m&\partial q^m-a^mq^m\\
	\partial r^m-b^mr^m     &\frac12{q^mr^m}+2\partial b^m
\endpmatrix.
\tag4.5.18
$$
Combining this with the explicit form \thetag{4.5.13} of $\delta^m$
gives
$$
[h,w^m_2]=\pmatrix0&-\frac12(\bar\partial+\partial){q^m}+a^mq^m\\
		-\frac12(\partial-\bar\partial){r^m}+b^mr^m&0
		\endpmatrix
\tag4.5.19
$$
Since $[E_{22},w_2^m]=-\frac12[h,w^m_2]$ we find
for the lattice resolvent the
expansion
$$
U^m=\ddxl E_{22}+\pmatrix 0&-\frac12{q^m}\\-\frac12{r^m}&0\endpmatrix
    +\ddxl\inv
    \pmatrix
    1+\frac14{q^mr^m}&\frac14(\bar\partial-\partial){q^m}\\
    \frac14(\partial-\bar\partial){r^m}&-\partial b^m-\frac14{q^mr^m}
    \endpmatrix+
    \Cal O(\ddxl^{-2}).
\tag4.5.20
$$
A simple calculation shows then that $U^m_+$ has to have the form
\thetag{4.5.16}.
\qede

A straightforward but tedious calculation now shows that
$$
-\bar \partial U^m_+ (U^m_+)\inv +U^m_+(R^m_{\bar
x})_+(U^m_+)\inv=\pmatrix
\ddxl&-4/r^m\\
\frac{r^m}4(\bar
\partial^2-\partial^2)\log(r^m)-\frac14q^m(r^m)^2&-\ddxl\endpmatrix
\tag4.5.21
$$
Hence, using \thetag{4.4.16b}, we find
$$
\align
q^{m+1}&=-4/r^m, \tag4.5.22a\\
r^{m+1}&=\frac{r^m}4
((\bar\partial^2-\partial^2)\log(r^m)-q^mr^m),\tag4.5.22.b\\
\endalign
$$
With $u^m=\log(r^m)$ the equations \thetag{4.5.23} imply
$$
\frac14
(\bar\partial^2-\partial^2)u^m=\exp(u^{m+1}-u^m)-\exp(u^m-u^{m-1}),
\tag4.5.23
$$
the 2-dimensional Toda lattice equation.
{}From the diagonal
components of the expression $-\partial_t U^m_+ (U^m_+)\inv
+U^m_+(R^m_{t})_+(U^m_+)\inv$
we find
$$
\aligned
\lambda^{m+1}&=\lambda^m+\partial\bar\partial\log(r^m),\\
\mu^{m+1}&=\mu^m+\frac12(\partial^2+\bar\partial^2)\log(r^m),
\endaligned
\tag4.5.24
$$
from which follows
$$
Q^{m+1}=Q^m+\partial^2\log(r^m).\tag4.5.23.c
$$

Note that, for fixed j, $W^m(t)$ will be outside the  $H_j$ cell for
all
$t$ except possibly  for a finite set of integers $m$. In fact, as
in \cite{BtK, Lemma 6.1.b}, one can prove that this set is,
in the case of the
polynomial Grassmannian we use, an uninterrupted
sequence $m_{\text{min}},m_{\text{min}}+1,m_{\text{min}}+2,
\dots,m_{\text{max}}$, so that we get from this construction
solutions of
the
{\it finite} 2-dimensional Toda lattice equation. To obtain solutions
of
the
infinite 2-dimensional Toda lattice  one has to choose a suitable
$W$ belonging to a bigger space, e.g., the Segal-Wilson Grassmannian.

\subhead{\bf 5.}
The calculation of the  wave function from $\semi$
\endsubhead

\subsubhead{\bf 5.1} Semi-infinite wedge space and Fermi-Bose
correspondence\endsubsubhead

In this subsection we collect some results from \cite{tKvdL} on the
Fermi-Bose
correspondence associated to a partition of $n$ that we will need
later, see also \cite{DJMK, Ka}.

The space $\Bbb C^\infty$ is a representation for the group
$Gl_\infty$, but
not
a highest weight representation. The fundamental highest weight
representations
of $\Glinf$ are contained in the {\it semi-infinite wedge space}
$\semi$, the collection of (finite) linear combinations of
semi-infinite
exterior products of elements of $\Bbb C^\infty$ of the form
$$
v_{0}\wedge v_{-1} \wedge v_{-2}\wedge\dots,
\tag5.1.1
$$
where $\{v_i\}_{i\le0}$ is an {\it admissible set}: a
set $\{v_i\}_{i\le0}$ of elements of $H$
is called admissible if there exist integers $k\in \Bbb Z, N\le 0$
such that
$v_i=\epsilon_{k+i}$ for all $i\le N$. Note that we might as well
assume
in the construction of a semi-infinite wedge that
the $v_i$ belong to $\Bbb
C^\infty$. (This is the reason for the notation $\semi$ instead of
$\Lambda^
{\infty\over 2} H$.)
For later constructions, however, we prefer to allow elements of
$H$ as members of an admissible set.
The integer $k$ is called the {\it charge} of
the wedge $	  v_{0}\wedge v_{-1} \wedge v_{-2}\wedge\dots$ and
the
semi-infinite wedge space decomposes in a direct sum of  subspaces of
fixed charge:
$$
\semi=\bigoplus_{k\in \Bbb Z}\Lambda^{\infty\over 2}_k\Cinf.
\tag5.1.2
$$
If in an admissible set $\{v_i\}_{i\le0}$  all the $v_i$ are of the
form
$v_i=\epsilon_{j_i}$ then the corresponding element
$$
\epsilon_{j_0}\wedge\epsilon_{j_{-1}}\wedge\epsilon_{j_{-2}}\dots
\tag 5.1.3
$$
of $\semi$ is called an {\it elementary wedge}.

The
action of $a\in\glinf$, $g\in\Glinf$ is given as usual by
$$
\aligned
    \rho(a) \cdot(v_{0}\wedge v_{-1} \wedge v_{-2}\wedge\dots)
    &=(a \cdot v_{0}\wedge v_{-1} \wedge v_{-2}\wedge\dots +\\
    &\quad\quad v_0 \wedge (a \cdot v_{-1}) \wedge
    v_{-2}\wedge\dots +
    \dots
    \\
    \rho(g)\cdot(v_{0}\wedge v_{-1} \wedge v_{-2}\wedge\dots)
    &=(g \cdot v_0)\wedge
    (g \cdot v_{-1}) \wedge(g \cdot v_{-2} )\wedge\dots.
\endaligned
\tag5.1.4
$$
This does not extend to  representations of the algebra $\gllf$ and
the
group $\Gllf$, because of the
infinities that occur (e.g., in the naive action of $\lambda
\sum_{i\in \Bbb Z}
\Epsilon_{ii}\in \Gllf$ on $\semi$, for $|\lambda|>1$).
	    However these infinities are relatively innocuous
and after ``renormalization'' one obtains  a projective
representation
$\hat
\rho $ of $\gllf$ and	 $\Gllf$,
or, what is the same thing, a representation of a central extension
of the algebra and
group. In terms of the algebra generators it reads
$$
\hat\rho(\Epsilon_{ij})=\rho(\Epsilon_{ij})-\delta_{ij}\theta_{0i},
\tag5.1.5
$$
where $\theta_{0i}=0$ if $i>0$ and 1 otherwise. This defines a non
trivial
central extension $\hatgllf=\gllf\oplus \Bbb C$ of the Lie algebra
$\gllf$.
The corresponding group will
be described in the next section.

Let the {\it fermion operators}
$\psi(i),\psi^*(i)$ be the linear operators on
$\semi$ that act on elementary wedges by
$$
\aligned
    \psi(i) \cdot(\epsilon_{i_0}\wedge \epsilon_{i_1} \wedge
    \epsilon_{i_2}\wedge\dots)
    &=\epsilon_i \wedge\epsilon_{i_0}\wedge
    \epsilon_{i_1} \wedge
    \epsilon_{i_2}\wedge\dots
    \\
    \psi^*(i) \cdot(\epsilon_{i_0}\wedge \epsilon_{i_1} \wedge
    \epsilon_{i_2}\wedge\dots)
    &=\sum_{k=0}^\infty(-1)^{k}\delta_{ii_k}
    \epsilon_{i_0}\wedge \epsilon_{i_1} \wedge
    \epsilon_{i_2}\wedge\dots \wedge\hat\epsilon_{i_k}\wedge\dots
\endaligned
\tag5.1.6
$$
Now fix an integer $n$ and a partition $\nbar$. Then we have, as in
section
{3.1}, a relabeling of the
basis vectors and a corresponding  relabeling of the fermion
operators:
the $k$-component fermions $\psi_a(i)$, $\psi^*_a(i)$, $i\in \Bbb
Z,1\le
a\le k$
correspond to $\psi(j),\psi^*(j)$ whenever $\epsilon_a(i)$
corresponds to
$\epsilon_j$. The anti-commutation relations for the relabeled
fermions
are
$$
\aligned
    \{\psi_b(i),\psi_c(\ell)\}&=\{\psi^*_b(i),\psi^*_c(\ell)\}=0,\\
    \{\psi_b(i),\psi^*_c(\ell)\}&=\delta_{bc}\delta_{i\ell}.
\endaligned
\tag5.1.7
$$
The action of  $\Epsilon_{bc}^{\ell m}$ on semi-infinite wedge space
is given
by
normal ordered fermion bilinears: if $v\in \semi$ then
$\hat \rho(\Epsilon_{bc}^{\ell m}) v=:\psi_b(\ell)\psi^*_c(m): v$,
where the {\it normal ordering} of fermions is defined by
$$
:\psi_b(\ell)\psi^*_c(m):\overset \text{def}\to =\left\{
\aligned\psi_b(\ell)\psi^*_c(m)\quad&\text{if\ }m>0\\
-\psi_c^*(m)\psi_b(\ell)\quad&\text{if\ }m\le 0
\endaligned\right..
\tag5.1.8
$$
Recall the operators
$(\Lambda_a^{\pm})^j,j\ne 0$ of \thetag{3.2.1}. They generate,
via the representation  $\hat \rho$, on $\semi$ a
Heisenberg algebra, the generators of
which we will denote by
$$
\alpha_a(\pm j):=\hat\rho((\Lambda_a^\pm)^j)=\sum_{\ell\in \Bbb Z}
:\psi_a(\ell)\psi(\ell\pm j):,  \quad
j>0,a=1,\dots,k.
\tag5.1.9
$$
It is natural to introduce
the operator
$$
\alpha_a(0):=\sum_{\ell\in \Bbb Z}:\psi_a(\ell)\psi^*_a(\ell):,
\quad a=1,\dots,k.
\tag5.1.10
$$
One checks
that
$$
[\alpha_a(j),\alpha_b(\ell)]=j\delta_{j+\ell,0}\delta_{ab},\quad
j,\ell
\in \Bbb
Z,1\le a,b\le k.
\tag5.1.11
$$
Furthermore we need linear invertible operators $\hat Q_a$, $1\le
a\le k$,
on
$\semi$ that
satisfy the following defining relations:
$$
\aligned
    \hat Q_a \cdot \bold{v}_0&=\psi_a(1) \cdot \bold{v}_0,\\
    \hat Q_a\psi_b(i)\hat Q_a\inv&=\psi_b(i+\delta_{ab}),\\
    \hat Q_a\psi^*_b(i)\hat Q_a\inv&=\psi^*_b(i+\delta_{ab}).
\endaligned
\tag5.1.12
$$
In \thetag{5.1.12} $\bold{v}_0$ is a distinguished element   of
$\semi$,
the $0^{\text{th}}$ vacuum. In general one defines the
{\it $j^{\text{th}}$ vacuum} by:
$$
\bold{v}_j=\epsilon_j \wedge \epsilon_{j-1} \wedge \epsilon_{j-2}
\wedge\dots.
\tag5.1.13
$$
We have, for $r_a$ the numbers \thetag {3.2.1} associated to $j$,
$$
\alpha_a(0) \cdot \bold{v}_j=r_a \bold{v}_j,\quad a=1,\dots,k.
\tag5.1.14
$$
The operators $\hat Q_a$ are called {\it fermionic translation
operators}. They satisfy the following relations
(\cite{tKvdL}):
$$
\{\hat Q_a,\hat Q_b\}=0,\quad
\text{ for } a\ne b
\tag5.1.15
$$
Introduce next fermionic fields, formal power series with operator
coefficients:
$$
\aligned
\psi_a(z)&:=\sum_{\ell\in \Bbb Z}\psi_a(\ell)z^{\ell},\\
\psi^*_a(z)&:=\sum_{\ell\in \Bbb Z}\psi^*_a(\ell)z^{-\ell},\quad
a=1,\dots,k.\\
\endaligned
\tag5.1.16
$$
Now we can express the fermion fields completely in terms of the
Heisenberg
generators $\alpha_a(k)$ and the fermionic translation operators
$\hat
Q_a$.
The {\it parity operator} on $\semi$ is defined  by
$$
\chi=(-1)^{\sum_b \alpha_b(0)}.
\tag5.1.17
$$
The parity operator acts as $(-1)^k$ on the charge $k$ sector
$\Lambda^{\infty\over 2}_k\Cinf$of $\semi$.
It follows from Theorem 1.3 of
\cite{tKvdL} that we have the following {\it
bosonization formulae:}
$$
\aligned
    \psi_a(z)&=\chi \hat
    Q_a(-z)^{(\alpha_a(0)+1)}
    \exp\left(-\sum_{\ell<0}{1\over\ell}
	z^{ -\ell}\alpha_a(\ell)\right)
    \exp\left(-\sum_{\ell>0}{1\over\ell}
	z^{-\ell}\alpha_a(\ell)\right),
\\
    \psi^*_a(z)&=\hat Q_a\inv\chi (-z)^{-\alpha_a(0)}
    \exp\left( \sum_{\ell<0}{1\over\ell}
	z^{-\ell}\alpha_a(\ell)\right)
    \exp\left( \sum_{\ell>0}{1\over\ell}
	z^{-\ell}\alpha_a(\ell)\right), \quad a=1,\dots,k.
\endaligned
\tag5.1.18
$$
So the only way the various bosonizations of $k$-component fermions
are
distinguished is through the zero-modes $\alpha_a(0)$, which, by
\thetag{5.1.14}, are able to detect which partition we are using.
One can use this ``bosonized'' form for the fermions to express the
whole
representation of the Lie algebra $\glinf$ in terms of the Heisenberg
operators $\alpha_a(i)$ and the fermionic translation operators,
\cite
{tKvdL}. We won't
need this in the sequel.

\subsubhead{\bf 5.2} Central extension of $\Gllf$ and
	group action on $\semi$\endsubsubhead

We have now defined an action on $\semi$ of a central extension
of $\gllf$,
described by an exact sequence
$$
0\to \Bbb C \to\hat{gl}^{\text{\it lf}}_\infty\overset \pi_*
\to\to\gllf\to
0.\tag5.2.1
$$
In this subsection we sketch the construction of the corresponding
group and
its action on $\semi$,
following the approach of \cite{SeW, PrS}, to which we refer for more
details.

The reason that the usual action of $g\in \Gllf$ on $v_{0}\wedge
v_{-1} \wedge
v_{-2}\wedge\dots$ by $gv_{0}\wedge gv_{-1} \wedge
gv_{-2}\wedge\dots$ does not work is that the set $\{gv_i\}_{i\le0}$
is,
in
general, not admissible, even if $\{v_i\}_{i\le0}$ is. Now an
admissible
set
$\{v_i\}_{i\le0}$ in $\Cinf$, if the $v_i$ are linearly independent,
is
the
basis for the finite order part $W^{\text{fin}}$ for some (unique)
$W\in Gr$. In the same way $\{gv_i\}_{i\le0}$ is a basis for
$(gW)^{\text{fin}}$, (not necessarily admissible).
The idea is now
to replace the possibly non admissible basis $\{g.v_i\}$ by an
admissible
one, say $\{w_i\}$, and to replace the wedge $v_0\wedge
v_{-1}\wedge\dots$
by
$w_0\wedge  w_{-1}\wedge\dots$.
Because of the ambiguity in the choice of this
admissible basis, we obtain in this way
a projective representation of $Gl^{lf}_\infty$,
or, equivalently, a representation of a central extension of
this group. Below we will make this precise.

Denote the group of invertible matrices of size
$\Bbb Z_{\le0}\times \Bbb Z_{\le0}$
with a finite number of nonzero lower triangular diagonals by:
$$
Gl(H_0)^{\text{lf}}=\{a=\sum_{i,j\le0}a_{ij}\Epsilon_{ij}\mid
a_{ij}=0
\text{ if } i-j\gg 0, a\text{ invertible}\}.
\tag5.2.2
$$
A subgroup of $Gl(H_0)^{\text{lf}}$ is
$$
\Cal T =\{t\in Gl(H_0)^{\text{lf}}\mid t=1+\text{finr}\}.
\tag5.2.3
$$
Here and in the sequel ``$\text{finr}$''
denotes a finite rank matrix. A finite rank matrix in
$Gl(H_0)^{\text{lf}}$
is one with only a finite number of
nonzero columns.

Every $W\in Gr$ has an admissible basis for
$W^{\text{fin}}$, for instance the canonical basis of \thetag{2.2.4}.
It will be convenient to think of an admissible set  as a
matrix $\underline v=(\dots v_{-2}v_{-1}v_0)$ of size $\Bbb Z\times
\Bbb
Z_{i\le0}$, with
the $v_i$ as columns. On such matrices we have a right action of
$Gl(H_0)^{\text{lf}}$. It is
easy to see  that if $\underline v$ and $\underline
v^\prime$ are admissible bases of $W^{\text{fin}}$
then $\underline v^\prime=\underline v \cdot t$ for $t\in \Cal T$.
We will use frequently that $\underline v$ and $\underline v^\prime$
are bases for the finite order part $W^{\text{fin}}$ of the same $W$
iff
the corresponding wedges $\bold v$ and $\bold v^\prime$ differ by
a constant.  Moreover
if $\underline {\tilde v}$ is any basis for $W^{\text{fin}}$ then we
can
find
an $a\in Gl(H_0)^{\text{lf}}$ such that $\underline {\tilde v} \cdot
a\inv$
is the canonical basis. In particular if, for admissible $\underline
v$,
$g \cdot\underline v=\{gv_i\}_{i\le0}$ is not admissible, then we can
find
an
$a\in Gl(H_0)^{\text{lf}}$ such that $g \cdot \underline v \cdot
a\inv$ is
admissible.

This leads to the introduction of
$$
\Cal E=\{(g,a)\mid g\in \Glnlf, a\in Gl(H_0)^{\text{lf}},
		g_{--}-a=\text{finr}\}.
\tag5.2.4
$$
Here we write $g\in \Gllf$ in block form with respect to the
decomposition
$H=H_0\oplus H_0^\perp$:
$$
g=\pmatrix g_{--}&g_{-+}\\g_{+-}&g_{++}\endpmatrix.
\tag5.2.5
$$
The subgroup of $g\in \Gllf$ such that $g_{--}:H_0\to H_0$
is a Fredholm operator of index $0$, is denoted by $\Glnlf$ and
is the called the identity component of
$\Gllf$. One checks that $(g_1,a_1)
\cdot(g_2,a_2) =(g_1g_2,a_1a_2)$ gives $\Cal E$ a group structure.
Suppose
$(g,a)\in \Glnlf\times Gl(H_0)^{\text{lf}}$ and let $\underline v$ be
an
admissible basis corresponding to a wedge of charge $0$. Then
$g\underline
v a\inv$ is
admissible iff $(g,a)\in\Cal E$.

The inclusion $t\in \Cal T\mapsto (1,t)\in \Epsilon$ gives us an
exact
sequence:
$$
1\to \Cal T\to \Cal E\to \Glnlf\to 1.
\tag5.2.6
$$
Note that every $t\in \Cal T$ has a determinant.
\proclaim{Lemma 5.2.1}Let
$$
\aligned
\Cal T_1&=\{t\in \Cal T\mid \det(t)=1\},\\
\Cal E_1&=\{(1,t)\in \Cal E\mid t\in \Cal T_1\}.
\endaligned
\tag 5.2.7
$$
Then $\Cal T_1$ is normal in $Gl(H_0)^{\text{lf}}$ and
$\Cal E_1$ is normal in $\Cal E$.
\endproclaim
\demop
We must check that for all $a\in Gl(H_0)^{\text{lf}}$ $ata\inv$
belongs to
$\Cal T_1$ if $t$ does. Since $t=1+f$, $f=\text{finr}$ we have
$ata\inv=1+afa\inv\in \Cal T$, so it remains to check that
$\det(ata\inv)=1$.
Since $a$ might not have a determinant we need a little
argument for this. It runs as follows: write $t=1+f$ as above
and choose a basis $\{v_1,v_2,\dots\}$ of $H_0$ such that for $k>k_0$
all
basis elements $v_k$  belong to  the kernel of $f$ and put
$V=\oplus_{i=1}^kv_k$. Then $V$ is a finite
dimensional subspace of $H_0$ and we get by restriction and
projection
a map $t|_V:V\to V$. One checks that $\det(t|V)$ is
independent of the choices made here, so we can define
$\det(t)=\det(t|_V)$. Define then $V^\prime=a \cdot V$,
$t^\prime=ata\inv$,
so that  $ \det(t^\prime)=\det(t^\prime|_{V^\prime})$. Since
$a|_V:V\to
V^\prime$ is an isomorphism of finite dimensional vector spaces we
see
that
$\det(t)=\det(t^\prime)$, as we wanted to prove.

For the second part we must
check, for all $(g,a)$ in $\Cal E$ and $(1,t)$ in $\Cal E_1$,   that
$(g,a) \cdot(1,t)
\cdot(g\inv,a\inv)=(1,ata\inv)$ belongs to $\Cal E_1$, but this
follows
from
the first part.
\qede

Taking quotients we obtain from \thetag{5.2.6} the exact sequence:
$$
1\to \Cal T/\Cal T_1\simeq \Bbb C^*\to \Cal E/\Cal E_1\to \Glnlf\to
1.
\tag5.2.8
$$
This gives us a central extension $\hatGlnlf:=\Cal E/\Cal E_1$
of the identity component of $\Gllf$.

To get a central extension of the whole group $\Gllf$ we need the
shift
automorphism given by
$$
\sigma(g)=\Lambda \cdot g \cdot \Lambda\inv.
\tag5.2.9
$$
The semi-direct product $\Glnlf\ltimes_\sigma \Bbb Z$ (with
multiplication
$(g,k) \cdot(h,l)=(g\sigma^k(h),k+l)$) is isomorphic to $\Gllf$ by
the map
$(g,k)\mapsto g \cdot\Lambda^k$. The shift automorphism $\sigma$
lifts to
an automorphism $\hat \sigma$ of $\Cal E/\Cal E_1$ as follows: an
element of $\Cal E/\Cal
E_1$ can be written as $(g, a\Cal T_1)$, with $g\in \Glnlf$, $a\in
GL(H_0)^{
\text{lf}}$, such that $g_{--}-a=\text{finr}$. Then let  $\hat
\sigma(
(g, a\Cal T_1)):=(\sigma(g),\bar\sigma(a)\Cal T_1)$ where $\bar
\sigma(a)$
is
obtained by adding to $a$ a row and column of zeroes
and a diagonal 1:
$$
\bar \sigma(a)=\pmatrix a&0\\0&1\endpmatrix\in
Gl(H_0)^{\text{lf}}.
\tag5.2.10
$$
One checks that this is independent of the representation of the
coset
$a\Cal
T_1$. To see that this indeed defines an automorphism note that the
fiber
of
the projection $\Cal E/\Cal E_1 \to \Glnlf$ over $g$ is $\Bbb C^*$
and
that
$\hat \sigma$ defines a homomorphism from the fiber over $g$ to the
fiber
over  $\sigma(g)$ with kernel 1. Therefore this homomorphism has to
be an
isomorphism  and hence $\hat \sigma$ is an automorphism.

Next one defines $\hatGllf:= \hatGlnlf\ltimes_{\hat \sigma}\Bbb Z$
and we
get an exact sequence
$$
1\to \Bbb C^*\to \hatGllf\to \Gllf\to 1,
\tag5.2.11
$$
and this is
the central extension of $\Gllf$ we were looking for. One checks that
this exact sequence corresponds to the Lie algebra extension
\thetag{5.2.1}.

Next we have to define an action of $\hatGllf$ on $\semi$. It
suffices to establish an action on wedges $v_0\wedge v_{-1}\wedge
v_{-2}\wedge\dots$, since as soon that is known  we can
extend to all of $\semi$ by linearity. We first discuss wedges in
some
more
detail.

To a non zero wedge $\bold{v}=v_0\wedge v_{-1}\wedge
v_{-2}\wedge\dots$
we can associate (non uniquely) a linear independent admissible set
$\{v_i\}_{i\le0}$.  If $ \bold{v}$ is a
wedge of charge $k$ then the corresponding matrix $\underline v$
is of the form
$$
\underline v=\Lambda^{-k}\pmatrix v_-\\v_+\endpmatrix,
\quad v_-=1+\text{finr},
\quad v_+=\text{finr}.
\tag5.2.12
$$
Here the subscripts $\pm$ refer to the decomposition of a  $\Bbb
Z\times
\Bbb
Z_{\le0}$ matrix into blocks induced by the decomposition
$H=H_0\oplus
H_0^\perp$, cf. \thetag{5.2.5}.
Denote by $\Cal A$ the collection of all matrices of the form
\thetag{5.2.12} with linearly independent columns. On $\Cal A$ we
have an
action from the right of the group $\Cal T_1$ of \thetag{5.2.7}.
We can identify, in a
bijective manner, a non trivial wedge $\bold{v}$ with an orbit of
$\Cal T_1$ in $\Cal A$ via $\bold{v}\leftrightarrow \underline v
\Cal T_1$. In order to define
an action of $\hatGllf$  on $\semi$ it therefore
suffices to define an action on
the orbit space $\Cal A/\Cal T_1$.

Let $\hat Q\in \hatGllf=((e,e\Cal T_1),1)$ be the canonical lift of
the shift
matrix $\Lambda\inv=\sum\Epsilon_{i+1i}\in \Gllf$.
Just as any element of $\Gllf$ can be written as $g \cdot
\Lambda^{-k}$,
with
$g\in \Glnlf$ we can write an element of $\hatGllf$ as $(g,a\Cal T_1)
\cdot
\hat
Q^k$, $(g,a\Cal T_1)\in \hatGlnlf$. The action of $\hat Q$
reads
in terms of elements of $\Cal A/\Cal T_1$:
$$
\hat Q^j \cdot\left \{\Lambda^{-k}\pmatrix v_-\\v_+\endpmatrix\Cal
T_1\right\}=
\Lambda^{-(k+j)}\pmatrix v_-\\v_+\endpmatrix\Cal T_1.
\tag5.2.13
$$
Next we define the action of $(g^\prime,a^\prime\Cal
T_1)\in\hatGlnlf$. We
should have
$$
(g^\prime,a^\prime\Cal T_1) \cdot
\left \{
    \Lambda^{-k}\pmatrix v_-\\v_+\endpmatrix
    \Cal T_1
\right\}
=
\hat Q^k \cdot
\left (
    \hat Q^{-k}(g^\prime,a^\prime\Cal T_1)\hat Q^k
\right) \cdot \pmatrix
    v_-\\v_+\endpmatrix\Cal T_1
\tag5.2.14
$$
Now note that $\hat Q^{-k}(g^\prime,a^\prime\Cal T_1)\hat Q^k$ is an
element
of
$\hatGlnlf$, so it is of the form $(g,a\Cal T_1)$. To complete the
definition
of the $\hatGllf$ action we put
$$
(g,a\Cal T_1) \cdot \pmatrix v_-\\v_+\endpmatrix\Cal T_1=g
\cdot\pmatrix
v_-\\v_+\endpmatrix \cdot a\inv \Cal T_1.
\tag5.2.15
$$
We leave to the reader the easy verification that these definitions
make
sense,
i.e., are independent of the choice of representatives for the coset
$a\Cal
T_1$
and for the orbit $ \pmatrix v_-\\v_+\endpmatrix\Cal T_1$, and that
the
right hand side of \thetag{5.2.15} indeed belongs to $\Cal A/\Cal
T_1$.

\subsubhead{\bf 5.3} The projection $\pi:\hatGllf\to\Gllf$ and the
translation
group
\endsubsubhead

In the previous subsection we have constructed the central extension
$\hatGllf$ of $\Gllf$. With the help of the fermions that act on
$\semi$
we
will describe the projection $\pi:\hatGllf\to\Gllf$ very explicitly.
Using
this we show that the fermionic translation operator $\hat Q_a$
\thetag{5.1.10} that occurs in the bosonization formula
\thetag{5.1.16}
belongs to the group $\hatGllf$ and projects to the shift operator
$Q_a$
of
\thetag{3.2.4}. The fermionic translation operators $\hat Q_a$ are
the
ingredients for the lift of the translation group
$T^\nbar\subset\Gllf$ to
the central extension $\hatGllf$.

\proclaim{Proposition 5.3.1} Let $g=\sum g_{ij}\Epsilon_{ij}\in
\Gllf$ and let
$\hat g$ be any lift of $g$ in $\hatGllf$. Then, for all $i\in \Bbb
Z$, we
have, as operators on $\semi$,
$$
\hat g \cdot\psi(i) \cdot\hat g\inv=\sum_j g_{ji}\psi(j).
$$
\endproclaim
\demop
We write as before $(g,a\Cal T_1) \cdot \hat Q^k$ for an element of
$\hatGllf$,
with $\hat Q$ the canonical lift of the shift matrix $\Lambda\inv$,
see
\thetag
{5.2.13}. It is clear
that
$$
\hat Q \cdot\psi(i) \cdot\hat Q\inv=\psi(i+1),\quad i\in \Bbb Z.
\tag5.3.1
$$
Hence the proposition holds for elements of $\hatGllf$ of the form
$\hat
g=\hat
Q^k$, $k\in \Bbb Z$. It remains to prove the theorem for $(g,a\Cal
T_1)\in
\hatGlnlf$.

If we represent a wedge $\bold{v}\in \semi$ by an orbit $\underline
v\Cal T_1$
then the action \thetag{5.1.6} of $\psi(i)$ amounts to adding the
vector
$\epsilon_i$ on the right to $\underline v$:
$$
\psi(i) \cdot \underline v\Cal T_1=(\underline v\mid \epsilon_i)\Cal
T_1.
\tag5.3.2
$$
This is independent of the choice of $\underline v$: If $t\in \Cal
T_1$ then
$(\underline vt\mid \epsilon_i)\Cal T_1=(\underline v\mid \epsilon_i)
\pmatrix t&0\\0&1\endpmatrix\Cal T_1=(\underline v\mid\epsilon_i)\Cal
T_1$.
The proposition then states that the endomorphism $\hat g
\cdot\psi(i)
\cdot\hat g\inv$
acts on an orbit $\underline v\Cal T_1$ by adding on the right the
column
vector
$g \cdot \epsilon_i$. This is a small calculation:
$$
\aligned
(g,a\Cal T_1) \cdot\psi(i) \cdot(g\inv,a\inv\Cal T_1) \cdot\underline
v\Cal
T_1
&=(g,a\Cal T_1) \cdot\psi(i) \cdot (g\inv\underline v a)\Cal T_1\\
&=(g,a\Cal T_1) \cdot(g\inv\underline v a\mid\epsilon_i)\Cal T_1\\
&=\hat Q \cdot\left(\hat Q\inv(g,a\Cal T_1)\hat Q\right) \cdot
    \hat Q\inv\cdot(g\inv\underline v a\mid\epsilon_i)\Cal T_1\\
&=\hat Q \cdot(\Lambda g\Lambda\inv,\bar \sigma(a)\Cal T_1)  \cdot
	(\Lambda g\inv\underline v a\mid \Lambda \epsilon_i)\Cal
T_1\\
&=\hat Q \cdot(\Lambda \underline v a
    \mid \Lambda g \cdot\epsilon_i)\bar \sigma(a)\inv\Cal T_1\\
&= \hat Q \cdot( \Lambda\underline v
    \mid \Lambda g \cdot\epsilon_i)\Cal T_1\\
&=(\underline v\mid g \cdot\epsilon_i)\Cal T_1.
\endaligned
\tag5.3.3
$$
\qede

On $\semi$ there exists a unique positive definite
Hermitian form $\langle\, .\mid .\,\rangle$ such that the elementary
wedges
\thetag{5.1.3}
are orthonormal. With respect to this form the adjoint of $\psi(i)$
is
$\psi^*(i)$ and  the adjoint of $\hat\rho(\Epsilon_{ij})$ is
$\hat\rho(\Epsilon_{ij}^\dagger)=\hat\rho(\Epsilon_{ji})$.
To discuss the adjoint action
of $\hatGllf$ on $\psi^*(i)$ we need a concrete description of this
Hermitian form.

\proclaim{Lemma 5.3.2} Let $\bold{v}$, $\bold{w}$ be wedges in
$\semi$ with
charge $k$, $l$ respectively. Let $\underline v\Cal T_1$ and
$\underline
w\Cal T_1$ be the corresponding orbits in $\Cal A/\Cal T_1$. Then
$$
\langle \bold{v}\mid\bold{w}\rangle=\delta_{kl}\det(\underline
v^\dagger\underline w)
\tag5.3.4
$$
\endproclaim
\demop
Let us define for the moment by
$H(\bold{v},\bold{w})=\delta_{kl}\det(\underline
v^\dagger\underline w)$ a Hermitian form  on $\semi$.
To show that $H$ coincides with the standard
Hermitian form we must check that the elementary wedges
are orthonormal for $H$. To this end let, for $i\in \Bbb Z$
and $\lambda\in \Bbb
C$, $A_i(\lambda)=(\exp(\lambda
\Epsilon_{ii}),1\Cal T_1)\in \hatGlnlf$. Then we have
$$
H(A_i(\lambda)\bold{v},\bold{w})=\delta_{kl}\det((A_i(\lambda)
\underline
v)^\dagger\underline w)=\delta_{kl}\det(\underline
v^\dagger A_i(\lambda^*)\underline
w)=H(\bold{v},A_i(\lambda^*)\bold{w})
\tag5.3.5
$$
For an elementary wedge
$\bold{v}=\epsilon_{i_0}\wedge\epsilon_{i_{-1}}\wedge
\epsilon_{i_{-2}}\wedge\dots$, with $i_0>i_{-1}>i_{-2}>\dots$, we
have
$$
A_i(\lambda) \cdot \bold{v}=\left\{
    \aligned
    e^{\lambda}\bold{v} &\quad \text{ if }
    i\in\{i_0,i_{-1},i_{-2},\dots\}, \\
    \bold{v}&\quad \text{ if } i\notin\{i_0,i_{-1},i_{-2},\dots\}.
    \endaligned
\right.
\tag5.3.6
$$
Let $\bold{w}=\epsilon_{j_0}\wedge\epsilon_{j_{-1}}\wedge
\epsilon_{j_{-2}}\wedge\dots$, with $j_0>j_{-1}>j_{-2}>\dots$,
be another elementary wedge and let
$i=\max(i_0,j_0)$. If $i_0>j_0$ we have
$$
\exp(\lambda)H(\bold{v},\bold{w})=H(A_i(\lambda^*)\bold{v},\bold{w})=
H(\bold{v},A_i(\lambda)\bold{w})=H(\bold{v},\bold{w})
\tag5.3.7
$$
and hence $H(\bold{v},\bold{w})=0$, whereas when $i_0<j_0$ we have
$$
\exp(\lambda)H(\bold{v},\bold{w})=H(\bold{v},A_i(\lambda)\bold{w})=
H(A_i(\lambda^*)\bold{v},\bold{w})=H(\bold{v},\bold{w})
\tag5.3.8
$$
and also in this case $H(\bold{v},\bold{w})=0$. Repeating this
argument
for all other pairs $(i_{-1},j_{-1})$, $(i_{-2},j_{-2})$, $\dots$, we
find
that $H(\bold{v},\bold{w})\ne0$ implies $\bold{v}=\bold{w}$. Finally
let
$\bold{v}=\epsilon_{i_0}\wedge\epsilon_{i_{-1}}\wedge
\dots\wedge\epsilon_{i_{-N-1}}\wedge\epsilon_{-N-k}\wedge
\epsilon_{-N-k-1}\wedge\dots$ be an elementary wedge of charge $k$,
with
the
first $-N$ exterior factors different from those of
the $k^{\text{th}}$ vacuum. Then
$H(\bold{v},\bold{v})=\det(M^\dagger M)$, where $M$ is the $\Bbb
Z\times
N$
matrix with columns $\epsilon_{i_{-N-1}},
\dots,\epsilon_{i_{-1}},\epsilon_{i_0}$. It is clear that $M^\dagger
M=1$,
proving that $H$ makes the elementary wedges orthonormal.
\qede

If $g\in \Gllf$ then the necessary and sufficient condition for the
Hermitian
conjugate matrix $g^\dagger$ also to belong to $\Gllf$ is that
$g$ contains only a finite number of nonzero upper triangular
diagonals.
If this condition is satisfied we say that $g$ has {\it finite
width}.
Suppose now that $g$ has finite width and let $\hat g=
(g,a\Cal T_1)$ be a lift of
$g$. Then $a\in Gl(H_0)^{\text{lf}}$ automatically also
has finite width (and $
a^\dagger\in Gl(H_0)^{\text{lf}}$). We calculate the adjoint of $\hat
g$
with
respect to the Hermitian form of $\semi$:
$$
\aligned
\langle \hat g\bold{v}\mid\bold{w}\rangle&=\det((g\underline v
a\inv)^\dagger
\underline w)\\
&=\det((a\inv)^{\dagger}\underline v^\dagger g^\dagger\underline w)\\
&=\det(\underline v^\dagger g^\dagger\underline w
(a\inv)^{\dagger})\\
&=\langle\bold{v}\mid\hat g^\dagger \bold{w}\rangle,
\endaligned
\tag5.3.9
$$
with
$$\hat g^\dagger:=(g^\dagger, a^\dagger\Cal T_1)\in \hatGlnlf.
\tag5.3.10
$$
So we see that the adjoint of $\hat g$ with respect to the standard
Hermitian
form of $\semi$ is a lift of the Hermitian conjugate matrix
$g^\dagger$.

Let now $\hat g$ be the lift of a finite width element. Then we have
the
following analogue of the Proposition {5.3.1}
$$
\hat g \cdot\psi^*(i) \cdot\hat g\inv=\sum_j (g\inv)_{ij}\psi^*(j).
\tag5.3.11
$$
Indeed
$$
\aligned
\hat g \cdot\psi^*(i) \cdot\hat g\inv
&=((\hat g\inv)^\dagger \cdot\psi(i) \cdot\hat g^\dagger)^\dagger\\
&=(\sum_j (g\inv)^\dagger_{ji}\psi(j))^\dagger\\
&=\sum_j (g\inv)_{ij}\psi^*(j)
\endaligned
\tag5.3.12
$$
We have derived here \thetag{5.3.10} under the assumption that $g$
has
finite
width, the only situation in which we will use this formula. We leave
it
to
the reader to prove this result for general $g$.

Recall now the elements $Q_a$ of $\Gllf$ defined in \thetag{3.2.4}.
These act
on the relabeled basis introduced in section 3.1 by
$$
Q_a \cdot\epsilon_b(j)=\epsilon_b(j+\delta_{ab}), \quad
a,b=1,2,\dots,k,
\quad j\in \Bbb Z.
\tag5.3.13
$$
Denote by $\hat Q_a^\prime$ any lift of $Q_a$ to $\hatGllf$. Then by
proposition 5.3.1 and equation \thetag{5.3.10}
we find
$$\aligned
    \hat Q^\prime_a\psi_b(i)(\hat Q_a^\prime) \inv
	&=\psi_b(i+\delta_{ab}),\\
    \hat Q^\prime_a\psi^*_b(i)(\hat Q^\prime_a) \inv
	&=\psi^*_b(i+\delta_{ab}).
\endaligned
\tag5.3.14
$$
Comparing this with \thetag{5.1.10} we see that $\hat Q^\prime_a$ has
the
same adjoint action on the fermions as the fermionic translation
operators
introduced in section 5.1, that had a priori no relation to
$\hatGllf$.
To compare the action of $\hat Q^\prime_a$ and $\hat Q_a$ on $\semi$
we
need
a lemma.

\proclaim {Lemma 5.3.3} Let $\hat Q^\prime_a$ be a lift of $Q_a$.
Then
$$
\hat Q_a^\prime \cdot \bold{v}_0=\nu \psi_a(1) \bold{v}_0,
\quad \nu\in \Bbb C^*
\tag5.3.15
$$
\endproclaim
\demop The $0^{\text{th}}$ vacuum
$\bold{v}_0=\epsilon_0\wedge\epsilon_{-1}\wedge\epsilon_{-2}\wedge
\dots$
corresponds  to the orbit $\underline v_0\Cal T_1$, where we can
choose
$  \underline v_0=(\dots \epsilon_{-2}\epsilon_{-1}\epsilon_0)$. The
columns
for
this matrix $\underline v_0$ form an admissible basis for
$$
H_0^{\text{fin}}=\bigoplus_{j\le0} \Bbb C\epsilon_j=\bigoplus_{b=1}^k
\bigoplus_{i\le0} \Bbb C\epsilon_b(i).
\tag5.3.16
$$
Now if $\hat Q_a^\prime $ is any lift of $Q_a$ then the wedge $\hat
Q_a^\prime
\cdot \bold{v}_0$ corresponds to the orbit $\underline v^\prime\Cal
T_1$,
with the
columns of $\underline v^\prime$ an admissible basis for $(Q_a \cdot
H_0)^{
\text{fin}}$. Since $Q_a=\sum_{b\ne a}1_b+\Lambda_a^-$, and
$\Lambda_a^-
\epsilon_b(i)=\epsilon_b(i+\delta_{ab})$, we see that
$$
(Q_a \cdot H_0)^{\text{fin}}=\Bbb C\epsilon_a(1)\oplus
\bigoplus_{b=1}^k \bigoplus_{i\le0} \Bbb C
\epsilon_b(i).
\tag5.3.17
$$
Hence a particular simple admissible basis for  $(Q_a \cdot
H_0)^{\text{fin}}$
is $\{\epsilon_a(1)\}\cup \{\epsilon_b(i)\mid b=1,2,\dots,k;i\le0\}$,
corresponding to the orbit $(\underline v_0\mid \epsilon_a(1))\Cal
T_1$
and to the wedge
$\psi_a(1)\bold{v}_0$. Any two admissible bases of $(Q_a \cdot
H_0)^{\text{fin}}$ correspond to wedges that differ by at most a non
zero
factor.
\qede

This lemma, combined with \thetag{5.3.13} and \thetag{5.1.10}, shows
that for
any
lift $\hat Q_a^\prime$ of $Q_a$ we have $\hat Q_a=\nu \hat
Q_a^\prime$ for
some
$\nu\in \Bbb C^*$. Hence

\proclaim{Proposition 5.3.4} Let $\hat Q_a$ be the fermionic
translation
operators
on $\semi$ defined in \thetag{5.1.10}. Then $\hat Q_a\in \hatGllf$
and
$\hat
Q_a$
projects to $Q_a\in\Gllf$.
\endproclaim

Next we turn to the translation operators $T_\alpha$ responsible for
the
discrete time evolution on the Grassmannian. Since in the central
extension
the $\hat Q_a$ no longer commute (as the $ Q_a$ do) we have to make a
choice here. We think of the root lattice $R=\oplus_{i=1}^{k-1}
\Bbb Z \alpha_i$ as a
sublattice of the rank $k$ lattice $\oplus_{i=1}^{k} \Bbb Z\delta_i$
through
the identification $\alpha_i=\delta_i-\delta_{i+1}$, $i=1,2,\dots,
k$. The
$\delta_j$ are orthogonal for a symmetric bilinear form $( \cdot|
\cdot)$.
We can then write uniquely for any $\alpha\in R$
$$
\alpha=
\sum_{j=1}^\ell p_{i_j}\delta_{i_j}+\sum_{j=\ell+1}^k
n_{i_j}\delta_{i_j},
\tag 5.3.18
$$
with the $p_{i_j}$ non negative and the $n_{i_j}$ strictly negative.
We make then
a choice of the ordering in the two sets of subscripts:
$$
i_1<i_2<\dots i_\ell;\quad i_{\ell+1}>\dots>i_k.
\tag 5.3.19
$$
Let $\hat F=\langle \hat Q_i,i=1,2,\dots,k\rangle$ be the group
generated
by
the fermionic translation operators. We then define, using the
ordering
\thetag{5.3.19}, a map $R\to \hat F$, $\alpha\mapsto \hat T_\alpha$
by
$$
\hat T_\alpha:=
\hat Q_{i_1}^{p_{i_1}}\hat Q_{i_2}^{p_{i_2}}
\dots \hat Q_{i_\ell}^{p_{i_\ell}}\hat
Q_{{i_{\ell+1}}}^{n_{i_{\ell+1}}}
\dots \hat Q_{i_k}^{n_{i_k}}.
\tag5.3.20
$$
Clearly $\hat T_\alpha$ is then a lift of the translation operator
$T_\alpha$ (defined in
subsection 3.2) and the group $\hat{T}^\nbar$ generated by the
$\hat{T}_{\alpha_i}$
is a central extension of the Abelian group $
T^\nbar$ by $\Bbb Z_2$,
defined by a cocycle $\epsilon:R\times R\to \Bbb Z_2$  given by
$$
\hat T_\alpha\hat T_\beta=\epsilon(\alpha,\beta)\hat T_{\alpha+\beta}
\tag5.3.21
$$
So $\epsilon$ satisfies the cocycle properties
$\epsilon(\alpha,\beta)
\epsilon(\alpha+\beta,\gamma)=\epsilon(\alpha,\beta+\gamma)
\epsilon(\beta,\gamma)$ and
$\epsilon(\alpha,0)=\epsilon(0,\alpha)=1$.
The choice \thetag{5.3.19} was made to ensure simple properties of
the
cocycle $\epsilon$, as described in the following Lemma.
\proclaim{Lemma 5.3.5} The cocycle $\epsilon$ satisfies
for all $\alpha,\beta\in R$:
\roster
\item $\epsilon(\alpha,-\alpha)=1$,
\item
$\epsilon(\alpha,\beta)\epsilon(\beta,\alpha)=(-1)^{(\alpha|\beta)}$,
\item $\epsilon(-\alpha,-\beta)=(-1)^{(\alpha\mid
\beta)}\epsilon(\alpha,\beta)$.
\endroster
\endproclaim
The proof of the Lemma consists of  simple computations using the
fact
that
the $\hat Q_i$ satisfy the anti-commutation relations \thetag{5.1.15}
and
the
fact that the sum of the coefficients $p_{i_j},n_{i_j}$ is zero.
Note that part \therosteritem{1} of the Lemma implies the useful
relation $\hat T_\alpha^{-1}=\hat T_{-\alpha}$ for all $\alpha \in
R$.

\subsubhead{\bf5.4} Group decomposition and
$\tau$-functions\endsubsubhead

In this section we will study the Gauss decomposition in $\Gllf$
by means of the action of the central extension $\hatGllf$ on
$\semi$. In fact
we will need a slight refinement: if $g\in\Gllf$ with Gauss
decomposition
$g=g_-Pg_+$ then we can write (uniquely) $g_+=h \cdot g_+^s$, with
$h$ a
diagonal
matrix and $g_+^s$ a upper triangular matrix with ones on the
diagonal
to get a decomposition
$$
g=g_-Phg^s_+.
\tag5.4.1
$$
Note that in the Gauss decomposition the factors $g_-,g_+$ are in
general not
uniquely defined. However, the element $h$ here {\it is} uniquely
determined;
later on we will see that its entries are essentially
the $\tau$-functions of $W=gH_j$ (when $W$ is in the $H_j$ cell), to
be defined in Definition 5.4.3 below.
For any lift $\hat g$ of $g$ we can find elements $\hat g_{-}$, $
\hat P$,
$\hat
h$ and $\hat g^s_+$ projecting to the corresponding elements without
a hat to
get
$$
\hat g=\hat g_-\hat P\hat h\hat g^s_+.
\tag5.4.2
$$
This decomposition is however not unique: each of the factors can be
multiplied
by a non zero complex
constant as long as the product of these factors is one. The
following lemma
fixes a normalization.
\proclaim{Lemma 5.4.1} Let $g=g_-Phg^s_+$ be a fixed factorization
\thetag
{5.4.1}. Assume that $PH_j=H_j$.
Then,
for any lift $\hat g$ of $g$ there is a unique factorization as in
\thetag{5.4.2}
such that
\roster
\item $\hat g^s_+\bold{v}_j=\bold{v}_j$.
\item $\hat P\bold{v}_j=\bold{v}_j$.
\item $\langle \bold{v}_j\mid \hat g_-\bold{v}_j\rangle=1$
\endroster
\endproclaim
\demop
For the first part note that for any lift $\hat g^s_+$ of $g^s_+$ the
wedge
$\hat g^s_+ \bold{v}_j$ corresponds to an
admissible basis
for $g^s_+H_j=H_j$, since $g^s_+$ is upper triangular. As before we
use
that
two
wedges corresponding to two admissible bases of the same point of
$Gr$
differ
by
a non zero factor, so that $\hat g^s_+ \bold{v}_j=\nu \bold{v}_j$. By
changing
the lift we can make $\nu=1$. The proof for the second part is the
same.
For
the third part we note that $g_-$ has finite width, and $g_-^\dagger$
is
an
upper
triangular element of $\Gllf$ that by \therosteritem{1} has a unique
lift
$\hat g_-^\dagger\in \hatGllf$
such that $\hat g_-^\dagger\bold{v}_j=\bold{v}_j$. Then the
adjoint of $\hat g_-^\dagger$ is the unique lift of $g_-$ that
satisfies
\therosteritem3.
\qede

To decide whether or not $PH_j=H_j$ one can use the following Lemma.

\proclaim{Lemma 5.4.2} Let $\hat P$ be any lift of a permutation
matrix $P$.
Then
$$	\langle\bold{v}_j\mid \hat P\bold{v}_j\rangle\ne
0\Longleftrightarrow
PH_j=H_j
$$
\endproclaim
\demop
The wedge $\hat P \bold{v}_j$ corresponds to some admissible
basis for $PH_j$. Since $P$ is a permutation matrix it is clear that
$PH_j^{\text{fin}}$
has a basis consisting of basis
vectors $\{\epsilon_{\sigma_P(i)}\}_{i\le j}$, where $\sigma_P$ is
the
permutation associated to $P$. Let $S^j_P=\{\sigma_P(i)\mid i\le j\}$
and
order the elements of $S^j_P$ as $i_0>i_{-1}>i_{-2}>\dots$. Then
$\{v_{i_m}\}_{m\le0}$ is an admissible basis of $PH_j^{\text{fin}}$
and we have
$$
\hat P \bold{v}_j =
\nu\epsilon_{i_0}\wedge\epsilon_{i_{-1}}
    \wedge\epsilon_{i_{-2}}\wedge\dots,\quad \nu\in \Bbb C^*
\tag5.4.3
$$
Since elementary wedges are orthogonal we have
$$
\aligned
\langle \bold{v}_j\mid \hat P\bold{v}_j\rangle\ne0
&\Longleftrightarrow
\hat P \bold{v}_j = \nu \bold{v}_j,\quad \nu\in \Bbb C^*\\
&\Longleftrightarrow
\sigma_P\text{ restricts to a permutation of }\{i\mid i\le j\}\\
&\Longleftrightarrow
PH_j=H_j
\endaligned
\tag5.4.4
$$
\qede
Consider next the following lift to  $\hat \Gllf$
of $w_0^{\underline n}(t,\alpha)$:
$$
    \hatvacmbf(t,\alpha)=\exp(\sum_{i>0}\sum_{a=1}^kt^i_a\alpha_a(i))
    \cdot
\hat T_{\alpha}\in
\hatGllf.
\tag5.4.5
$$
Let $\hat
g\in\hatGllf$ be any lift of $g\in \Gllf$. Then we define
$$
\hat g(t,\alpha)= \hatvacmbf(t,\alpha)\inv\hat g.
\tag5.4.6
$$

\proclaim{Definition 5.4.3}  Fix an integer $j$, a positive integer
$n$, a
partition $\nbar$ of
$n$
and let   $W=gH_j\in Gr$. Fix a lift $\hat g$ of $g$. The {\it
$\tau$-function of type $j,\nbar$ of $W$} is the function on
$\Gamma^\nbar$
given as ``vacuum expectation value'':
$$
\tauwjn(t,\alpha):=\langle v_j\mid  \hat g(t,\alpha)\cdot
v_j\rangle.
\tag5.4.7
$$
\endproclaim
Note that if $\hat g^\prime$ is another element of
$\hat\Gllf$ projecting to $g$ the
$\tau$-function calculated with
$\hat g^\prime$ will differ from \thetag{5.4.7} by a
constant. This will be irrelevant in the sequel. Also note that the
$\tau$-function is defined for all values of $(t,\alpha)$, also when
$W(t,\alpha)$ does not belong to the big cell. In fact the
$\tau$-function
of type $j,\nbar$ determines whether or not $W(t,\alpha)$ belongs to
the   $H_j$ cell:

\proclaim{Proposition 5.4.4} Let $j$, $\nbar$, $W$ be as above. Let
$\Gamma^\nbar
$ be the group of evolutions of type $\nbar$ on $Gr$ and let
$\Gamma^{j,\nbar}_W$  be the subset of $(t,\alpha)\in\Gamma^\nbar$
such
that
$pr:W(t,\alpha)=g(t,\alpha)H_j\to H_j$ { is an isomorphism.} (see
section 2.2). Then for all
$(t,\alpha)\in\Gamma^\nbar$
the following two statements are equivalent:
\roster
\item
    $\tauwjn(t,\alpha)\ne 0$,
\item
    $(t,\alpha)\in\Gamma^{j,\nbar}_W$.
\endroster
\endproclaim
\demo{Proof} We use the Gauss decomposition of
$\hat g(t,\alpha) \in \hat\Gllf$. We suppress for
simplicity the reference to $(t,\alpha)$ and write $\hat g=\hat
g_-\hat
P\hat
h \hat g^s_+$, with the normalization as in Lemma 5.4.1. Then,
writing
$\tau$
for $\tauwjn(t,\alpha)$:
$$
\aligned
    \tau&=\langle \bold{v}_j\mid \hat g_- \cdot \hat P\cdot \hat h
\cdot
    \hat g_+^s \cdot \bold{v}_j\rangle\\
    &= \langle (\hat g_-)^\dagger \cdot \bold{v}_j\mid   \hat P \cdot
    \hat h
    \cdot \bold{v}_j\rangle\\
    &=\exp(\lambda_j)\langle \bold{v}_j\mid \hat P \cdot
    \bold{v}_j\rangle.
\endaligned\tag5.4.8
$$
Here we have used Lemma 5.4.1, and the fact
that $\bold{v}_j$ is an eigenvector of elements of lifts of diagonal
matrices: $\hat h \cdot \bold{v}_j
=\exp(\lambda_j)\bold{v}_j$, for some $\lambda_j$ (depending on
$(t,\alpha)$). Using
Lemma 5.4.2 we see from \thetag {5.4.7}
that $\tauwjn(t,\alpha)\ne 0$ iff
$PH_j=H_j$. But $PH_j=H_j$ iff $W(t,\alpha)=g_-(t,\alpha)H_j$. Now
$W(t,\alpha)=g_-(t,\alpha)H_j$,
for $g_-(t,\alpha)$ strictly lower triangular,
is equivalent to $W(t,\alpha)$ belonging to the  $H_j$ cell.
\qede
Note that from the proof of this proposition it follows that for
$(t,\alpha)\in
\Gamma^{j,\nbar}_W$ we have, if $\hat g(t,\alpha)=\hat g_- \cdot\hat
P
\hat  h\cdot \hat g_+^s$,
$$
\hat P\hat h g_+^s\cdot \bold{v}_j=\tauwjn(t,\alpha)  \bold{v}_j,
\tag 5.4.9
$$
a  fact that will be used in the calculation of the wave function in
terms
of
$\tau$-functions.

\subsubhead{\bf5.5 } Relation between wave function and
$\tau$-function\endsubsubhead
In the construction of   the
wave function of a point $W$ of the Grassmannian the lower triangular
part
$g_-^j$
of the
Gauss decomposition adapted to $H_j$ occurs (see
definition 4.4.1).
It is a rather trivial observation that one can
calculate matrix elements of $g_-^j$ by lifting it  to
the central extension $\Gllf$ and using the semi-infinite wedge space
(see the introduction for the finite dimensional situation). First
note
that there is a unique lift of $g_-^j$ such that
$\langle \bold{v}_j\mid \hat g_-^j\bold{v}_j\rangle=1$. The proof is
as in
Lemma 5.4.1.  To find the coefficient of the
matrix $\Epsilon_{pq} $ in $g_-^j$ we observe that $\Epsilon_{pq}$
is represented on $\semi$ by the operator
$\psi_p\psi_q^*$ (if $p\ne q$). The following Lemma is then, maybe,
not
too
surprising.

\proclaim{ Lemma 5.5.1} Fix an integer $j$ and a positive integer $n$
and a
partition $\nbar$ of $n$. Let $g\in\Gllf$ and let $g(t,\alpha)$ be
given
by
\thetag{4.1.4}. Assume that $g(t,\alpha)$ belongs to the  $H_j$ cell,
see
section 2.2. Let
$g_-^j(t,\alpha)\in \Gllf$ be the minus component of the
Gauss decomposition of $g(t,\alpha)$ of Lemma 2.2.2 adapted to $j$.
Write
$ g_-^j(t,\alpha)=\sum g_{pq}(t,\alpha)\Epsilon_{pq}$.
Let $\hat g(t,\alpha)$ be an arbitrary lift of $g(t,\alpha)$ and lift
$ g_-^j(t,\alpha)$ to
the unique element $\hat g_-^j(t,\alpha)\in \hat \Gllf$ that
satisfies
$\langle \bold{v}_j\mid \hat g_-^j(t,\alpha)\bold{v}_j\rangle=1$.
Then  for all $p\in \Bbb Z,q\le j$ we have
$$
\align
g_{pq}(t,\alpha)&=\langle (\psi_p\psi_q^*) \cdot \bold{v}_j
\mid \hat g_-^j(t,\alpha)\cdot \bold{v}_j\rangle,\tag5.5.1.a\\
&=\langle (\psi_p\psi_q^*) \cdot \bold{v}_j
\mid \hat g(t,\alpha)\cdot \bold{v}_j\rangle/\tauwjn(t,\alpha)
    ,\tag5.5.1.b
\endalign
$$
where $\tauwjn(t,\alpha)$ is the $\tau$-function of type $(j,\nbar)$
of
$W=
gH_j$.
\endproclaim

\demo{Proof}
For simplicity we will mostly suppress the dependence on
$(t,\alpha)$.
If $p\ne q$, and $p\le j$,
then, by definition of $g_-^j$, we have $g_{pq}=0$, but also
$(\psi_p\psi_q^*) \cdot \bold{v}_j=0$, so in that case
\thetag{5.5.1a} is
true.
In case $p=q\le j$ we have $g_{pp}=1$, $(\psi_p\psi_p^*) \cdot
\bold{v}_j=\bold{v}_j$ and  \thetag{5.5.1a} holds because
of the normalization of the lift $\hat g_-^j$.

Assume next  that $ p> j$. Then we use Lemma 5.3.1 to calculate the
right hand side of \thetag{5.5.1a}. The elementary wedge
$(\psi_p\psi_q^*)\cdot \bold{v}_j= \epsilon_{j} \wedge \epsilon_{j-1}
\wedge\dots \wedge \epsilon_{q+1} \wedge\epsilon_p
\wedge \epsilon_{q-1} \wedge\dots$ corresponds to the orbit
$\underline
v^{p,q}_j\Cal T_1$, where $\underline
v^{p,q}_j$ is the $\Bbb Z\times \Bbb Z_{\le0}$
matrix with
columns
$(\dots\epsilon_{q-1}\epsilon_p\epsilon_{q+1}\dots\epsilon_j)$.
The wedge $\hat g_-^j\bold{v}_j$ corresponds to an orbit $\underline
v\Cal T_1$,
where we can take $\underline v$ to be
$$
\underline v=g^{j}_-\Lambda^{-j}\pmatrix1\\0\endpmatrix a\inv,
\tag5.5.2
$$
for $a$ the identity matrix. With this choice $\underline v$ is
a $\Bbb Z\times \Bbb Z_{\le0}$ matrix
of the form
$(\dots v_{-2}v_{-1}v_0)$ with for all $i\le0$
$$
v_i=g_-^j \cdot \epsilon_{j+i}.
\tag5.5.3
$$
Therefore
$$
\aligned
\langle (\psi_p\psi_q^*) \cdot \bold{v}_j
\mid \hat g_-^j\cdot \bold{v}_j\rangle&=\det(
\underline v^{p,q}_j)^\dagger\underline v),\\
	&=\det
\pmatrix g_{pq}&g_{pq-1}&\dots &g_{pj}\\
	    0    & 1	&\dots	&0     \\
	    \vdots&\vdots &\ddots&\vdots\\
	    0&  0 &   \dots     &1
\endpmatrix=g_{pq},
\endaligned
\tag5.5.4
$$
proving \thetag{5.5.1.a} also in this case.

Now $g_-^j(t,\alpha)$ is a factor in the decomposition
$$
g(t,\alpha)=g_-^j \cdot f^j \cdot P \cdot g_+=g_-^j
\cdot f^j \cdot P \cdot h \cdot g^s_+,
\tag5.5.5
$$
where $f^j$ is defined in \thetag{2.2.4} and we have decomposed $g_+$
in
a
diagonal part $h$ and an upper triangular part $g^s_+$ with ones on
the
diagonal. Since
$g_-^j \cdot f^j$ is the factor $g_-$ in the Gauss decomposition and
$(t,\alpha)\in \Gamma_W^{j,\nbar}$ we can apply Lemma 5.4.1 to get
for any
lift $\hat g(t,\alpha)$ a unique factorization
$$
\hat g(t,\alpha)=\hat g_-^j \cdot\hat f^j \cdot\hat P
\cdot\hat h \cdot\hat g^s_+,
\tag5.5.6
$$
such that
$$
\hat P\bold{v}_j=\bold{v}_j,\quad \hat g^s_+\bold{v}_j=\bold{v}_j,
\quad\langle\bold{v}_j\mid\hat g_-^j \cdot
\hat f^j \bold{v}_j\rangle=1.
\tag5.5.7
$$
Now $f^jH_j=H_j$ so we have $\hat f^j\bold{v}_j=\nu \bold{v}_j$, with
$\nu\in \Bbb C^*$. But because of the normalization of $\hat g_-^j$
we
must
have in fact $\nu=1$ and $\hat f^j\bold{v}_j=\bold{v}_j$. This
implies
that
$$
\aligned
\hat g_-^j\cdot \bold{v}_j&=\hat g(t,\alpha) \cdot(\hat g^s_+)\inv
\cdot\hat h\inv \cdot\hat P\inv \cdot(\hat f^j)\inv\cdot \bold{v}_j\\
&=\hat g(t,\alpha)\cdot \bold{v}_j/\tauwjn(t,\alpha),
\endaligned
\tag5.5.8
$$
using \thetag{5.4.8}.
\qede

Note that only because we have assumed that $g_-^j$ is the minus part
of the
Gauss decomposition adapted to $j$ we are able to calculate its
matrix
elements $g_{pq}$ in a simple way in the semi-infinite wedge
representation: if we had used the ordinary Gauss decomposition of
\thetag
{2.2.2} the finite matrix in
\thetag{5.5.4} would not have had zeroes below the diagonal.

In Lemma 5.5.1 we calculated the coefficients of $g^{j}_-$  in an
expansion
in terms of the standard basis $\Epsilon_{pq}$. After the relabeling
of
type
$\nbar$, so that $g^{j}_-=\sum g^{bc}_{rs}\Epsilon_{bc}^{rs}$,
we find of course  for the new coefficients an expression in terms of
the
relabeled fermions: if $\epsilon_c(s)$ corresponds to $\epsilon_q$,
with
$q\le j$, then
$$
g^{bc}_{rs}=\langle (\psi_b(r)\psi_c^*(s)) \cdot \bold{v}_j
\mid \hat g(t,\alpha)\cdot \bold{v}_j\rangle/\tauwjn(t,\alpha),
\tag5.5.9
$$
for all $1\le b\le k$, $r\in \Bbb Z$.  In particular
the columns of type $c$ of $g^{j}_-$ that occur in the definition of
the
wave function of type $j,\nbar$ have $s=r_c$ so that the
coefficients that occur in the wave function are of the form
$g^{bc}_{rr_c}$ for $c=1,2,\dots,k$. This will be used in the next
Theorem,

\proclaim{Theorem 5.5.2} Fix $g\in\Gllf$, a lift of $g$ to $\hat g\in
\hatGllf$,
an integer $j$,
a positive integer $n$ and a
partition $\nbar$ of $n$ into $k$ parts and let   $W=gH_j\in Gr$. Let
$w_W^{j,\nbar}(z;t,\alpha)$, $\tauwjn(t,\alpha)$ be  the wave
function and
the $\tau$-function associated to these data.
Then, if we write $w^{j,\nbar}_W(z;t,\alpha)=\sum_{b,c=1}^k
w_{bc}E_{bc}$,
we
have for all $(t,\alpha)\in \Gamma^{j,\nbar}_W$:
$$
w_{bc}=\langle
\psi_c^*(r_c)\bold v_j\mid \hat w^\nbar_0(t,\alpha)\inv
\psi_b^*(z) \hat g \rangle/\tauwjn(t,\alpha).
\tag5.5.10
$$
More explicitly we have
$$
w_{bc}(z;t,\alpha)=
(-1)^{r_b-r_c}z^{-(\alpha|\delta_b)}z^{-r_b}
(-z)^{\delta_{bc}-1}\epsilon(\alpha,\beta)
\exp(\sum_{\ell>0}z^\ell t_\ell^b)
\frac{\tauwjn(t\langle b\rangle,\alpha+\beta)}{\tauwjn(t,\alpha)} ,
\tag5.5.11
$$
where
$$
\tauwjn(t\langle b\rangle,\alpha+\beta)=
\exp(-\sum_{\ell>0}
\frac{z^{-\ell}}{\ell}\frac{\partial}{\partial t^b_\ell})
\tauwjn(t,\alpha+\beta),
\tag5.5.12
$$
where $\beta=\delta_b-\delta_c\in R$ is  the root corresponding
to the root vector $E_{bc}\in sl(k,\Bbb C)$,
and $\epsilon $ is the cocycle defined in \thetag{5.3.21}.
\endproclaim
\demo{Proof} For simplicity we suppress the reference to $(t,\alpha)$
and
mostly also to $z$ in the
proof.
Now by definition
of the wave function we have for the $c^{\text{th}}$ column
$$
\aligned
    w_{bc} &= e_b^t \cdot w_0(z;t,\alpha) \cdot\jmath^\nbar(g^j_-
\cdot
	\epsilon_c(r_c)),\\
    &= \exp(\sum_{i>0}t_i^bz^i)z^{<\alpha,\delta_b>}
    \sum_{r\in \Bbb Z} g^{bc}_{rr_c}z^{-r} e_b.
\endaligned
\tag5.5.13
$$
Hence by
\thetag{5.5.9} we have
$$
\aligned
\sum_{r\in Z}g^{bc}_{rr_c}z^{-r} &=\sum_{r\in \Bbb Z}z^{-r}
\langle \psi_b(r)\psi^*_c(r_c) \cdot \bold{v}_j \mid
    \hat g(t,\alpha) \cdot
    \bold{v}_j\rangle/
    \tauwjn(t,\alpha),\\
&= \langle
\sum_{r\in \Bbb Z}z^{r}
\psi_b(r)\psi^*_c(r_c) \cdot \bold{v}_j \mid
    \hat g(t,\alpha) \cdot
    \bold{v}_j\rangle/
    \tauwjn(t,\alpha),\\
&=\langle
    \psi_b(z)\psi^*_c(r_c) \cdot \bold{v}_j \mid
    \hat g(t,\alpha) \cdot
    \bold{v}_j\rangle/
    \tauwjn(t,\alpha),\\
    &=\langle
    \psi^*_c(r_c) \cdot \bold{v}_j \mid
    \psi^*_b(z)\hat g(t,\alpha) \cdot
    \bold{v}_j\rangle/
    \tauwjn(t,\alpha).\\
\endaligned
\tag5.5.14
$$
Here we have implicitly extended the Hermitian form $\langle
\cdot\mid
\cdot
\rangle $ on $\semi$ to a map $\langle \cdot\mid \cdot
\rangle_1:\semi[[z,z\inv]]\times \semi\to \Bbb C[[z,z\inv]]$ such
that
$\langle z \bold u\mid \bold v \rangle_1=z\inv\langle \bold u\mid
\bold v
\rangle $. There is a similar map $\langle \cdot\mid \cdot
\rangle_2:\semi\times \semi[[z,z\inv]]\to \Bbb C[[z,z\inv]]$ such
that
$\langle  \bold u\mid z\bold v \rangle_2=z\langle \bold u\mid \bold v
\rangle $ and we have
$$
\langle \psi_b(z) \bold u\mid \bold v \rangle_1=\langle \bold u\mid
\psi^*_b(z)\bold v\rangle_2.
\tag5.5.15
$$
In \thetag{5.5.14} we have dropped the subscripts $1,2$, as we will
continue
to do in the sequel. Now, $\hat g(t,\alpha)=\hat
w^\nbar_0(t,\alpha)\inv
\hat g$ and
$$
\psi_b^*(z)\hat w^\nbar_0(t,\alpha)\inv=\exp(\sum_{i>0}
-t^b_iz^i)z^{\langle \alpha,\delta_b\rangle}\hat
    w^\nbar_0(t,\alpha)\inv
\psi_b^*(z),
\tag5.5.16
$$
as follows from (see \thetag{5.1.12,5.3.20}),
$$
\aligned
\hat T_\alpha \psi_b^*(z) \hat T_\alpha\inv&= z^{\langle
\alpha,\delta_b\rangle}\psi_b^*(z),\\
[\alpha_a(i),\psi_b^*(z)]&=-z^i\delta_{ab}\psi_b^*(z).
\endaligned
\tag5.5.17
$$
Using this, and the explicit form \thetag{4.3.5-6} for $w_0(z)$,
in \thetag{5.5.14} gives  part \thetag{5.5.10} of the theorem.

We next continue with the calculation of $w_{bc}^\prime(z):=\langle
    \psi^*_c(r_c) \cdot \bold{v}_j \mid
    \psi^*_b(z)\hat g(t,\alpha) \cdot
    \bold{v}_j\rangle$.
The fermion operator $\psi^*_c(r_c)$ is the coefficient of $z^{-r_c}$
in
the field $\psi^*(z)$, so using \thetag{5.1.14}, \thetag{5.1.17} and
the bosonization formula
\thetag{5.1.18} we get
$$
\psi^*_c(r_c) \cdot \bold{v}_j=\hat Q_c\inv(-1)^{j-r_c}\cdot \bold
v_j
\tag5.5.18
$$
This gives, also using the fact that the fermionic translation
operators $\hat Q_c$ are unitary,
$$
\aligned
w_{bc}^\prime(z)&=(-1)^{j-r_c}\langle
	\bold{v}_j \mid
	\hat Q_c \psi^*_b(z)\hat g(t,\alpha) \cdot
    \bold{v}_j\rangle.\\
\endaligned
\tag5.5.19
$$
In the fermion field $\psi^*_b(z)$ the operator $(-z)^{-\alpha_b(0)}$
occurs. We need to move this to the left to let it act on $\bold
v_j$.
We have in general
$$
    [\alpha_a(0),\hat Q_b]=\delta_{ab}\hat Q_b,
\tag5.5.20
$$
so
$$
\hat Q_c\hat Q_b\inv(-z)^{-\alpha_b(0)}=\hat
Q_c(-z)^{-(\alpha_b(0)+1)}\hat
Q_b\inv=(-z)^{-(\alpha_b(0)+1)+\delta_{bc}}\hat Q_c\hat Q_b\inv.
\tag5.5.21
$$
This gives
$$
\aligned
w_{bc}^\prime(z)
&=(-1)^{j-r_c}\langle
	(-z)^{(\alpha_b(0)+1)-\delta_{bc}}\bold{v}_j \mid
	\hat Q_c \hat Q_b\inv \chi\exp\left(
    \sum_{\ell<0}{1\over\ell}
    z^{-\ell}\alpha_b(\ell)\right) \cdot,\\
&\qquad\qquad \cdot\exp\left( \sum_{\ell>0}{1\over\ell}
    z^{-\ell}\alpha_b(\ell)\right)\hat g(t,\alpha) \cdot
    \bold{v}_j\rangle,\\
&=(-1)^{-r_c}(-z)^{\delta_{bc}-r_b-1}
    \langle \exp\left( \sum_{\ell>0}{-1\over\ell}
    z^{-\ell}\alpha_b(\ell)\right) \bold{v}_j
    \mid\cdot\\
    &\qquad\qquad \cdot\hat Q_c \hat Q_b\inv
    \hat g(t\langle b\rangle,\alpha) \cdot
    \bold{v}_j\rangle,\\
&=(z)^{-r_b}(-1)^{-r_b-r_c}(-z)^{\delta_{bc}-1}
    \langle  \bold{v}_j
    \mid \hat Q_c \hat  Q_b\inv
    \hat g(t\langle b\rangle,\alpha) \cdot
    \bold{v}_j\rangle,\\
\endaligned
\tag5.5.22
$$
where $ t\langle b\rangle$ is as in \thetag{5.5.12} and we have used
that $\alpha_b(\ell)\bold v_j=0$ if $\ell>0$.

Finally we have by \thetag{5.3.20} and Lemma 5.3.5:
$$
\aligned
\hat Q_c \hat Q_b\inv \hat T_\alpha\inv
&=\hat T_\beta\inv\hat T_\alpha\inv,\\
&=\hat T_{-\beta}\hat T_{-\alpha},\\
&=\epsilon(-\beta,-\alpha)\hat
T_{-(\alpha+\beta)},\\
&=\epsilon(\alpha,\beta)\hat T_{\alpha+\beta}\inv.
\endaligned
\tag5.5.23
$$
Putting this all together gives
then the rest of the theorem.
\qede

Note that in approach of, say, \cite{DJKM} the formula
\thetag{5.5.10} would
be the {\it definition} of the wave function, while in the approach
of
\cite{Di3} the $\tau$-function would be defined by a formula like
\thetag{5.5.11}.

The relation between the wave function and the $\tau$-function given
by
Theorem
5.5.2 allows us also to express the coefficients of resolvents,
lattice
resolvents, etc. in terms of $\tau$-functions. For instance for the
example
of the
Davey-Stewartson-Toda system discussed in section 4.6 we find that
the
first
coefficient $w_1^m$ of the expansion \thetag{4.6.6} of the wave
operator
is
given by
$$
w_1^m=\pmatrix -\partial^1_1\log(\tau^m)&-\tau^{m-1}/\tau^m\\
    -\tau^{m+1}/\tau^m& -\partial^1_2\log(\tau^m)
    \endpmatrix.
\tag5.5.24
$$
(For simplicity we use the $j=0$ part of the Grassmannian and write
$\tau^m(t)$ for $\tau_W^{0,\nbar}(t,m\alpha_1)$, see section 4.6, and
suppress the time dependence).
{}From \thetag{4.6.9--10} we then see that
$$
\aligned
q^m&=-2\tau^{m-1}/\tau^m,\\
r^m&=2\tau^{m+1}/\tau^m.
\endaligned
\tag5.5.25
$$
A small calculation using the diagonal part of $\delta^m$, see
\thetag{4.6.9},
then proves that the expression of the variable $Q^m$ in terms of
$\tau$-functions is given by
$$
Q^m=\partial^2\log(\tau^m).
\tag5.5.26
$$.
\subhead{\bf 6.}
Multicomponent KP equations in bilinear form and Pl\"ucker equations
\endsubhead

\subsubhead{\bf6.0 } Introduction\endsubsubhead
In the previous section we have found a relation between the wave
function
of a
point of the (polynomial) Grassmannian and $\tau$-functions, matrix
elements
of
the
fundamental representations of $\glinf$. These wave functions are
solutions
of
the linear equations \thetag{4.4.6}. In this section we will
reformulate the linear equations \thetag{4.4.6} in terms of bilinear
equations
of Hirota type: any (formal) solution  of the linear equations is at
the same
time
a solution to the bilinear equations (Proposition 6.1.1). This allows
us to
make
the connection with the Pl\"ucker equations of
the embedding of $Gr$ in $\Bbb P\semi$, which are equations for the
$\tau$-function.
So we get, just as in the
one component case, three
equivalent descriptions of the multi component KP hierarchy: the wave
functions
of the multi component KP hierarchy satisfy both the linear equations
and the
bilinear equations and they
can be expressed in terms of $\tau$-functions that satisfy there own,
equivalent, system of equations.

\subsubhead{\bf6.1 } Dual Grassmannian and Pl\"ucker
equations\endsubsubhead
The equations \thetag{4.4.6} for the wave function $w_W(t,\alpha)$
can be
formulated
in terms of a bilinear equation involving also the so called dual
wave
function
$w_W^*(t,\alpha)$. This bilinear equation amounts to an orthogonality
relation
between $W\in Gr$ and $W^*$, an element of the dual Grassmannian
$Gr^*$,
as was explained in \cite{HP} in an analytic context for the
KP-hierarchy.
So
we
start out this subsection by briefly discussing the dualization of
all our constructions. Then we
derive, for completeness sake,
the bilinear equations, show that the bilinear equations for the wave
function and its dual are equivalent to the differential difference
linear
equations \thetag{4.4.6} and give the connection to the Pl\"ucker
equations
of the embedding $Gr_j\to \Bbb P\Lambda^{\frac\infty2}_j\Bbb
C^\infty$.

The starting point of the theory developed  until now was the space
$H$ of
infinite
column vectors, see \thetag{2.1.2}. The dual notion is
$$
H^*=\{\sum_{i=m}^\infty c_i\epsilon^*_i\mid c_i\in \Bbb C,
		    m\in \Bbb Z\},
\tag6.1.1
$$
where $\epsilon^*_i$ is the linear function on $H$ given by $\langle
\epsilon^*_i, \epsilon_j\rangle=\delta_{ij}, i,j\in \Bbb Z$.
We can think of $H^*$ as consisting  of
infinite row vectors and the natural bilinear pairing $H^*\times H\to
\Bbb
C$
is
then just matrix multiplication between a row and column vector.

On $H^*$ we have the action of $g\in\Gllf$ given by
$$
g \cdot\epsilon_i^*=\epsilon_i^* g\inv=\sum_{k=m}^\infty
\epsilon^*_k(g\inv
)_{ik}.
\tag6.1.2
$$
Define, cf. \thetag{2.2.1},
$$
H^*_j=\{\sum_{k=j+1}^{\infty}c_k \epsilon^*_k\}\subset H^*.
\tag 6.1.3
$$
Then $Gr^*$ is the collection of $W^*\subset H^*$ of the form $g
\cdot
H^*_j
=H_j^* g\inv$ for some $g\in \Gllf$, $j\in \Bbb Z$. There is a
natural
1-1 correspondence between points of $Gr$ and $Gr^*$:  $W=g H_j$
corresponds to
$W^*=g \cdot H^*_j$.  One checks that such $W$ and $W^*$ are
orthogonal
with respect to the pairing $\langle ,\rangle$, since $H_j$ and
$H_j^*$
are.

We define also, given a partition $\nbar$, a relabeling on $H^*$: put
$\epsilon^*_j=\epsilon^*_a(i)$ if $\epsilon_j=\epsilon_a(i)$ in $H$.
Then
we
have
a map to $k$-component row vector Laurent series, cf. \thetag{4.3.1}:
$$
\jmath^{*\nbar}:H^*\to H^{*(k)},\quad
    \jmath^{\nbar}(\epsilon^*_a(i))=z^{i-1}e^*_a ,
\tag6.1.4
$$
where $e^*_a, a=1,2,\dots,k$ is a basis for the $k$-component row
vectors
and
$H^{*(k)}=\oplus \Bbb C((z))e_a^*$.
The bilinear pairing between $H^*$ and $H$ then translates into the
residue
pairing on $H^{*(k)}\times H^{(k)}$:
$$
(f^{*},g)=\resz (f^{*}(z)g(z)),\quad f^* \in H^{*(k)},g\in
H^{(k)},
\tag6.1.5
$$
where $\resz$
is the coefficient of $z\inv$ in a formal power series.

On the dual Grassmannian we have the time evolution of type $\nbar$
given by
$W^*
\mapsto W^*(t,\alpha):=\vacwn(t,\alpha)\inv \cdot W^*=W^*
\vacwn(t,\alpha)$,
with $\vacwn(t,\alpha)$ given by \thetag{4.1.1}. An element $W$ of
$Gr$
belongs to the  $H_j$ cell iff the corresponding element $W^*$ of the
dual
Grassmannian belongs to the  $H_j^*$ cell. In particular the set
$\Gamma^{j,\nbar}_W$ of elements $(t,\alpha)$ of the evolution group
such
that $
W(t,\alpha)$ belongs to the  $H_j$ cell is equal to
$\Gamma^{j,\nbar}_{W^*}$.

The wave function $w_W$ of type $j,\nbar$ was defined using the
columns $g^j_-
\cdot \epsilon_b(r_b(j))$, with $r_b(j)$ given by \thetag{3.2.1}. The
analogous
dual numbers are $r_b^*(j)=r_b(j)+1$, with the following
interpretation:
$\epsilon_b^*(r_b)$ is the basis vector of type $b$ occurring
in $H^*_j$ with smallest argument. Note that
$$
\jmath^{*\nbar}(\epsilon_b^*(r^*_b))=z^{r^*_b-1}e_b^*=z^{r_b}e^*_b.
\tag6.1.6
$$
Now we define the dual wave function using the rows of $(g_-^j)\inv$
and
the
numbers $r_b^*$: for $(t,\alpha)\in \Gamma^{j,\nbar}_W $
$$
w_W^*(t,\alpha):=\jmath^{*\nbar}\left
((\epsilon^*_1(r^*_1)\epsilon^*_2(^*r_2)\dots
\epsilon^*_k(r^*_k))^t (g^j_-)\inv\right) \cdot w_0(t,\alpha)\inv
\tag6.1.7
$$
The dual wave function can be written in terms of the dual wave
operator,
acting now from  the left (cf. \thetag{4.5.1}):
$$
w_W^*(t,\alpha)=\overrightarrow w_W(t,\alpha) \cdot
\vacw(t,\alpha)\inv,
\quad \overrightarrow w_W=(1_{k\times k}+\sum_{i>0}w^*_i
\overrightarrow \partial^{-i})\diag(z^{r_1},z^{r_2},\dots,z^{r_k}).
\tag6.1.8
$$
Denote by $W^{*(k)}$ the image of $W^*$ in   $H^{*(k)}$. Then the
rows of
$w^*_W(t,\alpha)$ belong to $W^{*(k)}$ for all
$(t,\alpha)\in \Gamma^{j,\nbar}_W $, just as the columns of
$w_W(t,\alpha)$
belong to $W^{(k)}$. Because $W^{*(k)}$ and $W^{(k)}$ are orthogonal
for the
residue pairing we find the {\it  bilinear identity}:
$$
\resz( w_W^*(t,\alpha)w_W(t^\prime,\alpha^\prime)) =0,\quad
(t,\alpha),(t^\prime,\alpha^\prime)\in \Gamma^{j,\nbar}_W.
\tag6.1.9
$$
Both the linear equations \thetag{4.5.6} for the wave function
and the bilinear equations
\thetag{6.1.9} are derived in apparently different ways
from the geometry of the Grassmannian. In fact these equations are
equivalent. Indeed, consider two matrix PDO's acting
from the left and right:
$$
\overrightarrow P=\sum p_i\partial^i,\quad
\overleftarrow Q=\sum \ddxl^i q_i, \quad p_i,q_i\in Mat_n(\Bbb C).
\tag6.1.10
$$
Then we define the dual operators to act in the opposite direction:
$$
(\overrightarrow P)^*:=\sum p_i(-\ddxl)^i,\quad
(\overleftarrow Q)^*:=\sum (-\partial)^i q_i.
\tag6.1.11
$$

\proclaim {Proposition 6.1.1} Let
$$
\alignedat2
w(z;t,\alpha)&=\vacw(z;t,\alpha) \cdot \overleftarrow w,
&\quad \overleftarrow w&=\text{diag}(z ^{-r_1},\dots,z^{-r_k})
    \cdot (1_{k \times k} +\sum_{i>0}\ddxl^{-i}w_i), \\
w^*(z;t,\alpha)&=\overrightarrow w\cdot \vacw(z;t,\alpha)\inv ,
&\quad \overrightarrow w&=(1_{k \times k} +
    \sum_{i>0}w^*_i\overrightarrow \partial^{-i}) \cdot\text{diag}(z
^{r_
    1},\dots,z^{r_k}) ,
\endalignedat
$$
with $w_i $, and $w^*_i$ matrices of size $k$. If $w$, $w^*$ are both
defined and infinitely differentiable
for $(t,\alpha)$ and $(t^\prime,\alpha+\gamma) $, then
the bilinear equations
$$
\resz(w^*(z;t,\alpha)w(z;t^\prime,\alpha+\gamma))=0
\tag6.1.12
$$
are equivalent to the equations
$$
\align
\overrightarrow w&=(\overleftarrow w\inv)^*,\tag6.1.13a\\
\partial^i_b w&= w \cdot(\Cal R^i_b)_+ ,\tag6.1.13b\\
    w(z;t,\alpha+\gamma)&=w(z;t,\alpha)   \cdot(\Cal U_\gamma)_+\inv,
    \tag6.1.13c
\endalign
$$
where $\Cal R^i_b=\overleftarrow w\inv \cdot\overleftarrow
\Lambda^i_b \cdot\overleftarrow w$, $\Cal U_\gamma=\overleftarrow
w\inv \cdot\overleftarrow
T_\gamma\inv \cdot\overleftarrow w$.
\endproclaim

We omit the proof, which is rather similar to the one for the KP
hierarchy,
cf.
\cite{Di2}.

In Theorem 5.5.2 we found an expression for the wave function in
terms of
matrix elements of the
semi-infinite wedge space. There is, as one might expect,
a similar expression for the dual wave function in terms of some dual
semi-infinite wedge space. However the situation is simpler than
that.
Note that  $w^*_W$ consists essentially of  rows of $g_-\inv$, and
since
$g_-=1+X$, with $X\in \glinfmjn$, we have $g_-\inv=1-X$, because
$X^2=0$.
Hence we can calculate the matrix elements of $g_-\inv$ and thus
$w^*_W$
in terms of the standard semi-infinite wedge space. Writing
$w^*_W=\sum_{bc}^k w^*_{bc}E_{bc}$ we find by essentially the same
calculation as in Theorem 5.5.2 that
$$
w^*_{bc}=z^{-1}\langle
\psi_b^*(r^*_b)\bold v_j\mid\hat w^\nbar_0(t,\alpha)\inv \psi_c(z)
\hat g
\bold v_j
\rangle/\tauwjn(t,\alpha).
\tag6.1.21
$$

Consider the orbit $\Cal O_j$ of the group $\hatGlnlf$ through the
vacuum
$\bold
v_j$ in $\semi$. The projectivization of $\Cal O_j$ can be identified
with the component $Gr_j$ of $Gr$.
It is well known (\cite{KP}) that the points of $\Cal O_j$
are
characterized as follows: $\tau_j\in \Lambda^{\frac\infty2}_j \Bbb
C^\infty$
belongs to $\Cal O_j$ iff
$$
\sum_{i\in \Bbb Z} \psi(i)\tau_j\otimes \psi^*(i)\tau_j=0.
\tag6.1.22
$$
So the equation \thetag{6.1.22} might be called the Pl\"ucker
equation for
the embedding of $Gr_j$ in $\Bbb P\Lambda_j^{\infty\over 2} \Bbb
C^\infty$.
Note that if we use the bosonization formulae \thetag{5.1.18} we
obtain
differential-difference equations for the $\tau$-function. We refrain
from
writing down explicit equations.

Using the  relabeled fermionic fields of \thetag{5.1.16} we can write
this
as
$$
Res_z[ {z\inv}\sum_{c=1}^k \psi_c(z)\tau_j \otimes\psi^*_c(z)\tau_j
]=0.
\tag6.1.23
$$
Multiplying this by $ \hat w_0^\nbar(t,\alpha)  \otimes
\hat w_0^\nbar(t^\prime,\alpha^\prime)/
\tauwjn(t,\alpha)\tauwjn(t^\prime,\alpha^\prime)$ and taking the
inner
product with the element
$ \langle \psi_b(r^*_b)\bold v_j\mid\otimes \langle
\psi^*_d(r_d)\bold v_j\mid$ gives
$$
Res_z [\sum_{c=1}^k
\frac{\langle \psi_b(r^*_b)\bold
v_j\mid \hat w_0^\nbar(t,\alpha)\psi_c(z)\tau_j}{
\tauwjn(t,\alpha)}
\frac{\langle \psi^*_d(r_d)\bold
v_j\mid \hat w_0^\nbar(t^\prime,\alpha^\prime)
	\psi^*_c(z)\tau_j}{
	\tauwjn(t^\prime,\alpha^\prime)}]=0,
\tag6.1.24
$$
which is, using \thetag{5.5.10, 6.1.21},
the bilinear identity \thetag{6.1.9}. In other words the Pl\"ucker
equation
for the embedding of $Gr_j$ in $\Bbb P\Lambda_j^{\infty\over 2} \Bbb
C^\infty$ is nothing but the bilinear equation
for a point $W$ of the component $Gr_j$.

\subhead{\bf 7.}
Reduction to loop groups and KdV type equations
\endsubhead
\subsubhead {\bf 7.0} Introduction  \endsubsubhead
In the previous sections we have found for every $W$ in the
Grassmannian and
for
every choice of partition of $n$ into $k$ parts a solution of the
differential-difference $k$-component KP hierarchy.

In the classical
1-component KP case one can for every integer $n$ impose constraints
on
solutions of the KP to obtain solutions of the $n$-KdV hierarchy,
related
to
$\tilde Gl^{\text{\it lf}} (n,\Bbb C)$,
the loop group of $Gl(n,\Bbb C)$. This procedure is called
$n$-reduction
and amounts to considering $\tilde Gl^{\text{\it lf}} (n,\Bbb C)$
in a natural way as a subgroup of $\Gllf$. This gives rise to a
subspace
$Gr^n$ of $Gr$, called the $n$-periodic Grassmannian. The embedding
of
$\tilde Gl^{\text{\it lf}} (n,\Bbb C)$ in $\Gllf$ and $Gr^n$ are
discussed
in section 7.1.

In the general case that we
are studying we
can  similarly
impose for every partition $\nbar$ of $n$ into $k$ parts constraints
on
the $k$-component KP to obtain solutions of what one might call the
$\nbar$-reduced KP-hierarchy. This is discussed in section 7.2.

In section 7.3 the $\nbar$-reduced KP hierarchy is rewritten in terms
of
$n\times n$ matrices depending on a variable $\lambda$, so as to make
the
relation with the loop group  $\tilde Gl^{\text{\it lf}} (n,\Bbb C)$
and the $n$-periodic Grassmannian explicit. The $\nbar$-reduced KP
hierarchy
rewritten in this way will be called the $\nbar$-KdV hierarchy.
Every conjugacy class in the
symmetric group is determined by a partition $\nbar$ and
determines a gradation of $\tilde gl^{\text{\it lf}}
(n,\Bbb C)$. In \cite{Wi2} Wilson proposes to construct for every
gradation
of $\tilde gl^{\text{\it lf}} (n,\Bbb C)$ a hierarchy of differential
equations. We show that our $\nbar$-KdV hierarchies are
essentially the equations Wilson had in mind for the gradation
corresponding
to $\nbar$. (He was dealing with the modified equations,
related to the infinite $n$-periodic flag manifold in a similar way
as our
$\nbar$-KdV hierarchies are
related to the $n$-periodic Grassmannian. Also\ he
didn't discuss the discrete part of the hierarchies.)
\subsubhead {\bf 7.1} $n$-periodicity  \endsubsubhead
Recall from section 4.4 the map $\jmath^\nbar$ from the
space $H$ of
infinite column vectors to $H^{(k)}$, the space of $k$-component
formal
Laurent series in the variable $z$, constructed using
the partition $\nbar$. Here
we will consider the space  $H^{(n)}$ of $n$-component formal Laurent
series in the variable $\lambda$: let $\tilde e_i$, $1\le i\le n$ be
a
basis
for  $\Bbb C^n$, define
$$
H^{(n)}:=\oplus_{i=1}^{n}\Bbb C((\lambda))\tilde e_i,
\tag7.1.1
$$
and introduce the isomorphism $H\to H^{(n)}$ by
$$
\epsilon_j\mapsto \lambda^{-p-1}\tilde e_q,
\tag7.1.2
$$
where $j=np+q, 1\le q\le n$.  This corresponds to the construction of
$H^{(k)}$ of section 4.4 for the partition of $n$ into $n$ parts.
Let $\tilde gl^{\text{\it lf}} (n,\Bbb C)$ be the collection of
formal
loops
of the form $\sum_m^\infty \lambda^iA_i$ with $A_i\in gl(n,\Bbb C)$
and
$m$ some integer. 	This forms in the natural way a Lie algebra
and we can
define a (Lie) algebra
homomorphism
$\tilde gl^{\text{\it lf}} (n,\Bbb C)\to \gllf$  by
$$
\lambda^p \tilde E_{ij}\mapsto \sum_{\ell \in \Bbb Z}
			    \Epsilon_{n(\ell-p) +i,n\ell+j},
\tag7.1.3
$$
where $\tilde E_{ij}$ is the element of the natural
basis of $n\times n$ matrices with as only nonzero entry
a 1 on the $ij^{\text{th}}$ place. (In general we will use a tilde $\
\tilde{}\ $
to indicate that an
object is related to the loop algebra, loop group or anything ``of
size
n'').

The image of this homomorphism consists of the {\it $n$-periodic}
elements,
i.e., those $X$ in $\gllf$ such that
$$
X=\Lambda^n X\Lambda^{-n},
\tag7.1.4
$$
where $\Lambda=\sum\Epsilon_{ii+1}$ is the shift matrix in $\Gllf$.
If we expand $X$ as $X=\sum X_{ij}\Epsilon_{ij}$ then the condition
\thetag
{7.1.4} implies for the coefficients   $X_{i+n,j+n}=X_{ij}$ for all
$i,j\in
\Bbb Z$. In the same way we get a surjective
homomorphism from the formal loop group
$\tilde Gl^{\text{\it lf}} (n,\Bbb C)$ (consisting of the invertible
elements of
$\tilde gl^{\text{\it lf}} (n,\Bbb C)$)  to  the subset of
$n$-periodic elements of $\Gllf$.

Since $H^{(k)}$ and $H^{(n)}$ are both isomorphic to $H$ there is an
isomorphism
$\jmath_{n,k}:H^{(k)}\to H^{(n)}$ given explicitly by
$$
e_az^{-i}\mapsto \lambda^{-p-1}\tilde e_t,
\tag7.1.5
$$
where $i=n_a p+q$, $1\le q\le n_a$ and $t=n_1+\dots+n_{a-1}+q$. Using
this
isomorphism we can translate, of course, the action of $k \times k$
matrices
on
$H^{(k)}$ into an action on $H^{(n)}$. In particular the diagonal
matrix
$zE_{aa}$
acting on  $H^{(k)}$ corresponds to the action of the $n \times n$
block
diagonal matrix
$$
\Cal P_a=
\diag (0_{n_1},\dots ,0_{n_{a-1}},P_{n_a},0_{n_{a+1}},\dots ,
0_{n_k}).
\tag7.1.6
$$
where $0_{n_c}$ is the zero matrix of size $n_c\times n_c$ and
$P_{n_a}$ is the $n_a\times n_a$ matrix
$$
P_{n_a}=
\pmatrix&0      &1      &0       &\ldots   &0     &0      \cr
	&0      &0      &1       &\ldots   &0     &0      \cr
	&\vdots &\vdots &\vdots  &\ddots   &\vdots&\vdots \cr
	&0      &0      &0       &\ldots   &1     &0      \cr
	&0      &0      &0       &\ldots   &0     &1      \cr
	&\lambda&0      &0       &\ldots   &0     &0      \cr
    \endpmatrix.
\eqno(7.1.7)
$$
Since we have seen in subsection 4.4 that the generator $\Lambda^+_a$
of
the
pre-Heisenberg
algebra $\Heisnbar$ of type $\nbar$ corresponds in $H^{(k)}$ to
$zE_{aa}$ we find that the action of the positive part of the
pre-Heisenberg algebra in $H^{(n)}$
is generated by the elements $\Cal P_a$. The action of the
pre-translation
group is similarly generated by the elements $ \tilde T_{\alpha_i}=
\Cal P_{n_i}\inv +\Cal P_{n_{i+1}} +\sum_{j\ne i,i+1} 1_{n_j}$, $1\le
i\le k$.

We have seen that the shift matrix $\Lambda$ does not correspond
under
$\jmath^\nbar$ to an element of the formal loop group of size $k$,
leading
to
the introduction of the numbers $r_b$ of \thetag{4.3.1}. Now, when we
use
the
map \thetag{7.1.2}, the situation
is much simpler: the element $P_n$ of $\tilde Gl(n,\Bbb C)$ is the
image
of
the shift matrix $\Lambda$.
This means in particular that, if we denote by $H_j^{(n)}$ the
image of $H_j$ in $H^{(n)}$, we have, cf. \thetag{4.3.3},
$$
H_j^{(n)}=P_n^{-j}H^{(n)}_0=P_n^{-j}\oplus_{i=1}^n \Bbb
C[[\lambda]]\tilde e_i.
\tag7.1.8
$$

Let $g\in\Gllf$ be $n$-periodic and let $W=gH_j$. Since $\Lambda^nH_j
=H_{j-n}\subset
H_j$ we have $\Lambda^n W=\Lambda^n g H_j=g \Lambda^n H_j\subset W$.
Conversely,
elements of $Gr$ satisfying $\Lambda^n W\subset W$ can be  obtained
from
$H_j$ by
an $n$-periodic group element (see \cite{PrS}) and are called also
$n$-periodic.
The collection of $n$-periodic elements in $Gr$ is denoted by
$Gr^{n}$,
and
$Gr^n$ is called the $n$-periodic Grassmannian.

\subsubhead {\bf 7.2} The $\nbar$ reduced $k$-component KP
hierarchy\endsubsubhead

The infinite shift matrix $\Lambda^n\in \Gllf$ corresponds to
multiplication
by
$\diag(z^{n_1},\dots,z^{n_k})$ in $H^{(k)}$ and to multiplication by
$\lambda=\lambda 1_{n\times n}$ in $H^{(n)}$. For simplicity we also
write
often
$\lambda$ for $\diag(z^{n_1},\dots,z^{n_k})$ in $H^{(k)}$.
So if $W\in Gr^n$ we have for the
image $W^{(k)}$ in $H^{(k)}$ the relation
$\lambda W^{(k)}\subset W^{(k)}$. In particular for the wave
function, the
columns of which belong  to $W^{(k)}$ when
$(t,\alpha)\in \Gamma^{j,\nbar}_W$, we have
$$
\lambda w_W(t,\alpha)\subset W^{(k)},
\tag7.2.1
$$
where we say that a matrix belongs to $W^{(k)}$ if its columns do.
Now by
Proposition 4.4.2 this means that there is for every
positive integer $\ell$ a unique $k\times k$ matrix
differential operator $M^{(\ell)}$, such that
$$
\lambda^\ell w_W(t,\alpha)=w_W \cdot M^{(\ell)}(t,\alpha),
\tag7.2.2
$$
Just as $w_W$ also $M^{(\ell)}$ is defined for  $(t,\alpha)\in
\Gamma^{j,\nbar}_W$.

One sees that the resolvent \thetag{4.5.5} satisfies
$$
w_W \cdot\psresa=zE_{aa}w_W,
\tag7.2.3
$$
so we have, using that $\lambda=\diag(z^{n_1},\dots,z^{n_k})$,
$$
w_W \cdot( M^{(\ell)}-\sum_{a=1}^k\psresa^{\ell n_a})=0.
\tag7.2.4
$$
By the proof of Proposition 4.4.2 this means that
$$
M^{(\ell)}= \sum_{a=1}^k\psresa^{\ell n_a}, \quad \ell>0.
\tag7.2.5
$$
So in the linear combination of resolvents on the right hand side of
\thetag
{7.2.5} the negative powers of $\ddxl$ that might occur cancel to
give a differential operator.

Introduce now the derivation
$$
\partial^{\ell,\nbar}=\sum_{a=1}^k\partial_{t^a_{\ell n_a}},\quad
\ell>0.
\tag7.2.6
$$
Then we have
$$
\aligned
\partial^{\ell,\nbar}w_W&=w_W \cdot \big (\sum_{a=1}^k \Cal R_a^{\ell
	n_a}\big)_+,\\
&=w_W \cdot M^{(\ell)}.
\endaligned
\tag7.2.7
$$
On the other hand, writing $w_W=w_0 \cdot\overleftarrow w_W$, with
$\overleftarrow w_W$ the
wave operator, we have
$$
\aligned
\partial^{\ell,\nbar}w_W&=w_0 \cdot
    \big (\sum_{a=1}^k z^{\ell n_a}E_{aa}\big)
    \cdot\overleftarrow w_W+ w_0 \cdot \partial^{\ell,\nbar}
    \overleftarrow w_W,\\
    &=w_W \cdot M^{(\ell)}+w_0 \cdot
\partial^{\ell,\nbar}\overleftarrow
    w_W.
\endaligned
\tag7.2.8
$$
Comparing \thetag{7.2.7} and \thetag{7.2.8} we find that
the wave operator $\overleftarrow w_W$ and hence all resolvents
$\Cal R_a$, lattice resolvents $\Cal U_{\alpha_i}$ (\thetag{4.7.5})
and
the
multi-component KP operator $L$ of \thetag{4.5.11} are independent of
the
variables corresponding to $\partial^{\ell,\nbar}$, $\ell>0$ (in the
$n$-periodic case). Conversely
if $W\in Gr$ produces a solution of the multi-component KP hierarchy
that
is
independent of these variables it belongs to the $n$-periodic
Grassmannian.

\proclaim{Definition  7.2.1} The $\nbar$-reduced $k$-component
differential-difference
KP hierarchy is the
system of deformation equations for the $k\times k$ matrix PDO $L$ of
the form $L=A\ddxl +\Cal O(\ddxl^0)$ given by
$$
\aligned
\partial^i_b L &=[ L ,(\Cal R^i_b)_+],\\
L(t,\alpha+\alpha_i)&= (\pslatresai)_+ \cdot L(t,\alpha) \cdot
(\pslatresai)_+\inv.\\
\endaligned\tag7.2.9a
$$
together with the condition
$$
\partial^{\ell,\nbar} L=0,\quad \ell>0.
\tag7.2.9b
$$
\endproclaim

So we get for every $n$-periodic element $W\in Gr^n$ a
solution $L_W=w_W\inv A\ddxl w_W$
of the $\nbar$-reduced $k$-component KP hierarchy.

\subsubhead {\bf 7.3} The $\nbar$-KdV hierarchy\endsubsubhead

In this section we want to give an alternative description of the
$\nbar$-reduced
$k$-component KP hierarchy. In the case $k=1$ there are (at least)
two
other descriptions available of the $\nbar$-reduced hierarchies:
the scalar Lax equation approach, involving $n^{\text{th}}$ order
scalar
differential operators, and the approach using first order matrix
differential operators (see, e.g., \cite{DS}). We discuss both
methods in
the
general case of an arbitrary partition. It will turn out that the
scalar
Lax
operator method does not generalize in a satisfactory way.

The $k \times k$  matrix
differential operator $M^{(\ell)}$ introduced in \thetag{7.2.2}
satisfies the same equations as the
pseudo-differential operator $L$ \thetag{4.5.11}:
$$
\aligned
\partial^a_i M^{(\ell)} &=[ M^{(\ell)},(\psresa^i)_+],\\
M^{(\ell)}(t,\alpha+\alpha_i)&= (\pslatresai)_+ \cdot
	M^{(\ell)}(t,\alpha) \cdot (\pslatresai)_+\inv.\\
\endaligned
\tag7.3.1
$$
If we consider the partition of $n$ into 1 part (the principal
partition),
then
we have for $M:=M^{(1)}$ the relation $M=L^n$ and
$L$ is the $n^{\text{th}}$ root of the operator $M$. (We take $A=1$
here).
As is well known in this case the equations
\thetag{7.3.1} form the generalized $n$-KdV hierarchy (the discrete
part
is
now, of course, absent) and  this hierarchy is   equivalent
to the $n$-reduced 1-component KP hierarchy, cf., \cite{SeW}. So the
matrix
$M$ seems to be the natural generalization of the Lax operator in the
principal case.

However, in general the equations \thetag{7.3.1} are just a
consequence of the
equations \thetag{7.2.9} and not equivalent to them,
since  the operator $M$ contains less information than $L$. In
fact in the extreme case where $\nbar$ is the partition of $n$
into $n$ parts (the homogeneous case) $M$ is just $\ddxl 1_{n\times
n}$.
This
is so because then the {\it differential} operator $M$ is the sum of
the
pseudo
differential operators
$\psresa$ that are of the form $\ddxl E_{aa} +[ E_{aa},w_1]+
\Cal O(\ddxl\inv)$. So the scalar Lax operator formulation of
$n$-reduced
KP
does not seem to generalize simply to the general case, cf.,
\cite{Di1}. Therefore we now sketch how the
$n$-reduced $k$-component KP hierarchy \thetag{7.2.9} fits
in the framework of $n\times n$ matrix differential operators.

Recall the construction of the $k$-component KP hierarchy: we started
with
a $W$ in the Grassmannian, mapped it to $W^{(k)}$ in $H^{(k)}$ and
considered
the natural flow $W^{(k)}\mapsto w_0(t,\alpha)\inv \cdot  W^{(k)}$.
Then
we
considered the subspace $W^{(k)}_{\text{fin}}$ of elements of
$W^{(k)}$ of
the form  $w_0(t,\alpha)$ times a finite order (in $z$) $k$-component
vector.
The space $W^{(k)}_{\text{fin}}$ was stable under the action of
$\ddxl$
and
we proved in Proposition 4.4.2 that $W^{(k)}_{\text{fin}}$ was in
fact a
free rank $k$ module over $\Bbb C[\ddxl]$, with basis the columns of
the
wave function $w_W$. This lead in Proposition 4.5.2 to linear
equations
for
the wave function and this in turn produced the $k$-component KP
hierarchy.

Now, in the $n$-periodic case, $W^{(k)}_{\text{fin}}$ is not only
invariant
under the action of $\ddxl$ but also under the action of $\lambda=
\diag(z^{n_1},z^{n_2},\dots,z^{n_k})$, i.e., $W^{(k)}_{\text{fin}}$
is a $\Bbb C[\lambda]$ module. It turns out that
$W^{(k)}_{\text{fin}}$
is  free of rank $n$ and we can give an explicit basis. Then
the time evolution of this basis will as before lead to linear
equations
and we will obtain in very much the same way as before a
collection of differential difference equations, the $\nbar$-KdV
hierarchy.
In the $k$-component KP case the objects one deals with are $k\times
k$
matrices over $\ddxl$ (since $W^{(k)}_{\text{fin}}$ is rank $k$
over $\Bbb C[\ddxl]$) whereas in the $\nbar$-KdV case we deal with
$n\times n$ matrices over $\lambda$ (since now we think of
$W^{(k)}_{\text{fin}}$ as a rank $n$ module over $\Bbb C[\lambda]$).

Now we turn to some of the details of the construction.
Fix an integer j and consider an $n$-periodic element $g$ of $\Gllf$
and
the
corresponding element $W=gH_j$ of $Gr^n$. We embed $W$ in $H^{(n)}$
using
the map \thetag{7.1.2}, so that we now deal with $n$-component
vectors
depending on $\lambda$.

We say that an element $\tilde g\in \tilde Gl^{\text{\it lf}} (n,\Bbb
C)$ is
in
the $\tilde H_j$ cell if
$$
\tilde g=
\tilde g_-^{j} \cdot\tilde g_+^{j},
\tag7.3.4
$$
where
$$
\aligned
    \tilde g_-^j&=P_n^{-j}\left (\tilde
    1_{n\times n}+\sum_{i<0}\lambda^{i}\tilde
    g_i \right) P_n^j, \\
    \tilde g_+^j&=P_n^{-j}\left (
    \sum_{i\ge 0}\lambda^{i}\tilde g_i\right )P_n^j,\quad \tilde
g_i\in
    gl(n,\Bbb C), \tilde g_0\in Gl(n,\Bbb C).
\endaligned
\tag7.3.5
$$
This is called the $j$-Birkhoff decomposition of $\tilde g$.
An element $\tilde g$ is in the $\tilde H_j$cell
iff the corresponding
(under the map \thetag{7.1.3})
$n$-periodic element of $\Gllf$ is in the $H_j$ cell.
The image of $\tilde g_-^j$ (resp. $\tilde g_+^j$)
in $\Gllf$  does not
quite coincide with the minus component $g_-^j$ (resp. $g_+^j$)
of the Gauss decomposition of $g$ adapted to $H_j$, since
$g_-^j$ ($g_+^j$) is not $n$-periodic, in general.
So in case $g$ is $n$-periodic we have two natural decompositions
adapted
to $H_j$. However the
columns $j-n+1,j-n+2,\dots,j$ of the $n$-periodic elements
corresponding
to
$\tilde g_-^j$ and $\tilde g_+^j$
are equal to the same columns of
the original factors $g_-^j$ and $g_+^j$ of the Gauss decomposition
of $g$
adapted to $H_j$ (the other columns will differ, in general).
Since this are the only columns that play a r\^ole it is irrelevant
which decomposition of $g$ we use.
Besides the Birkhoff decomposition in the loop group we will need in
the
sequel also the following corresponding decomposition
of the formal loop algebra:
$$
\tilde gl(n,\Bbb C)=\tilde gl(n,\Bbb C)_-^j\oplus \tilde gl(n,\Bbb
C)_+^j,
\tag7.3.6
$$
with (cf. \thetag{2.2.3} and \thetag{7.1.8}):
$$
\aligned
    \tilde gl(n,\Bbb C)_-^j
	&=P_n^{-j}\left(gl(n,\Bbb C)\otimes \lambda\inv
	\Bbb C[\lambda\inv]\right ) P_n^j,\\
    \tilde gl(n,\Bbb C)_+^j&=P_n^{-j}\left(gl(n,\Bbb C)\otimes \Bbb
	C[[\lambda]]\right ) P_n^j.
\endaligned
\tag7.3.7
$$

Fix also a partition $\nbar$ of $n$ and consider the time flow from
the
corresponding Heisenberg algebra $\Heisnbar$ on $H^{(n)}$generated by
$$
    \tvacmbf (t,\alpha)=\exp(\sum_{i>0}\sum_{a=1}^k t^a_i
    \Cal P_a^i)\tilde T_\alpha,
\tag7.3.8
$$
The {\it  $n\times n$ loop group wave function of type $j,\nbar$},
associated
to
$W$ and defined for $(t,\alpha)\in \Gamma^{j,\nbar}_W$, reads:
$$
\tilde w_W(t,\alpha)=\tvacmbf (t,\alpha) \cdot \tilde
g^j  _-(t,\alpha),
\tag7.3.9
$$
where $\tilde g_-^j$ is the minus component in the Birkhoff
decomposition
for
$\tilde H_j$ of $\tilde g(t,\alpha)=\tilde w_0^{\nbar}
(t,\alpha)\inv \tilde g$. We can also
describe $\tilde g_-^j$ by noting
that the $i^{\text{th}}$
column $\tilde g^j_- \cdot \tilde e_i$ is the unique
element of $W^{(n)}(t,\alpha)$ of the form $P_n^{-j}(\tilde e_{n_i}
+\Cal O(
\lambda\inv))$. So the loop group wave function consists essentially
of $n$
columns of $g_-(t,\alpha)$, whereas the wave function contains only
$k$
columns.

The point of the introduction of the loop group wave function is that
its columns form a basis for $\tilde W^{(k)}_{\text{fin}}$, the image
of $\wkfin$ in $H^{(n)}$.
\proclaim{ Proposition 7.3.2} Let $W$ be $n$-periodic.
Fix $(t,\alpha)\in \Gamma^{j,\nbar}_W$. Then $\wkfin$ is a free rank
$n$
module
over the ring $\Bbb C[\lambda]$, with basis
the columns of $\tilde  w_W(t,\alpha)$.
\endproclaim

The proof of this Proposition is very much the same as that of
Proposition 4.3.2 and is omitted.

Next we define  $n\times n$ loop resolvents and lattice resolvents by
$$
\aligned
    \tpsresa &:=  \tWmbf(t,\alpha)
    \inv \cdot \Cal P_a \cdot  \tWmbf (t,\alpha)\\
    \tpslatresai &:=   \tWmbf
    (t,\alpha) \inv \cdot \tilde
    T_{\alpha_i}\inv \cdot \tWmbf (t,\alpha)
\endaligned
\tag7.3.10
$$
So the resolvent $\tpsresa$ belongs to the loop algebra and the
lattice
resolvent $\tpslatresai$ to the loop group. We use the convention
that
subscripts on $\tpsresa$
will refer to the Lie algebra  decomposition 7.3.7 for $j=0$,
while subscripts on
$\tpslatresai$, when it is in the $H_0$ cell, refer to the
Birkhoff decomposition of type 0.

Then the analogue of Proposition 4.4.2 is

\proclaim{Lemma 7.3.3}
Let $W\in Gr_j$ and suppose that $(t,\alpha)$,
$(t,\alpha+\alpha_i)$
belong to $\Gamma^{j,\nbar}_W$. Then the loop
lattice resolvent $\tpslatresai$ is
in the  $H_0$  cell and we have:
$$
\aligned
    \partial_{t^b_i} \tilde  w_W(t,\alpha) &= \tilde  w_W(t,\alpha)
    \cdot(\tilde{\Cal R}^i_b)_+ \\
    \tilde  w_W(t,\alpha+\alpha_i)&=\tilde  w_W(t,\alpha)
    \cdot(\tilde {\Cal U}_{\alpha_i})_+\inv \
\endaligned
\tag7.3.11
$$
\endproclaim

The proof is the same as that of Proposition 4.4.2.

Note that the derivations with respect to the times \thetag{7.2.6}
act trivially
on the $\tilde w_W$, i.e., $\partial^{\ell,\nbar}\tilde w_W=
\tilde w_W \cdot\sum \Cal P_a^{\ell n_a}=\lambda^\ell \tilde w_W$.

The analogue of the operator $L_W$ that solves the $k$-component KP
hierarchy
is the  resolvent $\tilde L=\tilde w_w\inv \cdot
\sum A_a\Cal P_a \cdot\tilde w_W$. We could now
define the $\nbar$-reduced KdV hierarchy in terms of deformation
equations
for $\tilde L$. However in our situation it turns out that all
information
of $\tilde L$ is already contained in the in the positive part
$\tilde
L_+$
and  it is convenient to formulate the theory in terms of this.

To proceed we need a little digression on finite order automorphisms
and
twisted
loop algebras associated to a partition $\nbar$.  (For more details
and
proofs see \cite{tKvdL}).
We associate to every partition $\nbar$ a diagonal matrix in
$gl(n,\Bbb
C)$:
$$
H_{\nbar}=\sum_{a=1}^k H_a,
\quad H_a=\frac1{2n_a}\sum_{j=1}^{n_a}(n_a-2j+1) E_{aa}^{jj}.
\tag7.3.12
$$
We use $H_{\nbar}$ to define an automorphism of $gl(n,\Bbb C)$:
$$
\sigma_{\nbar}=\exp(2\pi i ad(H_{\nbar})):gl(n,\Bbb C)\to gl(n,\Bbb
C).
\tag7.3.13
$$
If $N^\prime $ is the least common multiple of the parts of the
partition
$\nbar$ then the order $N$ of $\sigma_{\nbar} $ is equal to
$2N^\prime$ in
case $N^\prime(\frac1{n_a}+\frac1{n_b})$ is odd for some pair of
parts
$n_a$, $n_b$ and $N$ is equal to  $N^\prime$ otherwise. Then
$\sigma_{\nbar} $
defines a $\Bbb Z/N \Bbb Z$ grading of
$gl(n,\Bbb C)$:
$$
\aligned
gl(n,\Bbb C)&=\oplus_{i=0}^{N-1}\frak g_i,\\
\frak g_i&=\{x\in gl(n,\Bbb C)\mid \sigma_{\nbar} (x)=\omega^i x\},
\quad \omega=\exp(2\pi i/N).
\endaligned
\tag7.3.14
$$
Next we define the twisted loop algebra:
$$
L(\frak g,\sigma_{\nbar} ):=\{\sum_{i=m}^\infty
\mu^i y_{\bar \imath}\mid y_{\bar \imath}\in
\frak g_{\bar \imath}\}.
\tag7.3.15
$$
Write
$\mu=\exp(i\theta/N)$ and put $\Phi_{H_{\nbar}}=\exp(-i\theta
ad(H_{\nbar}))$.
Then $\Phi_{H_{\nbar}}:L(\frak g,\sigma_{\nbar} )\to \tilde gl(n,\Bbb
C)$
is an isomorphism between the twisted loop algebra
$L(\frak g,\sigma_{\nbar} )$ and  the untwisted loop algebra $\tilde
gl(n,\Bbb C)$, with	$\lambda=\mu^N=\exp(i\theta)$ as loop
variable.

Define a Cartan subalgebra $\frak h_{\nbar}$ of $gl(n,\Bbb C)$
associated to
$\nbar$ with basis $E_a^i$, $a=1,2,\dots,k$ and $i=1,2,\dots,n_a$,
where
$E_a=E_{aa}^{n_a1}+\sum_{j=1}^{n_a-1}E_{aa}^{jj+1}$. Under
$\sigma_{\nbar} $
the Cartan subalgebra $\frak h_{\nbar}$ is mapped
to itself and so  the decomposition
\thetag{7.3.14} induces a decomposition of $\frak h_{\nbar}$. We can
then
define as in \thetag{7.3.15} the twisted loop algebra $L(\frak
h_{\nbar},
\sigma_{\nbar} )$. The image of $L(\frak h_{\nbar},
\sigma_{\nbar} )$ under the isomorphism $\Phi_{H_{\nbar}}$ is
precisely
the
Heisenberg algebra $\tilde \Cal H_{\nbar}$ generated by $\Cal P_{a}$
and
$\Cal Q_{a}:=\Cal P_{a}\inv$. In particular the element
$\mu^{N/n_a}E_a$
corresponds to $\Cal P_a$.
For different $\nbar$ one obtains distinct Heisenberg
algebras in $\tilde gl(n,\Bbb C)$ and one proves also that {\it all\
}
Heisenberg algebras of $\tilde
gl(n,\Bbb C)$ (up to isomorphism) are obtained in this way.

The decomposition
$gl(n,\Bbb C)=\frak h_{\nbar}\oplus \frak h_{\nbar}^{\perp}$, where
$\frak h_{\nbar}^{\perp}$ is the direct sum of the root spaces with
respect
to $\frak h_{\nbar}$, induces a decomposition
$$
L(\frak g,\sigma_{\nbar} )=L(\frak h_{\nbar},\sigma_{\nbar} )\oplus
L(\frak h_{\nbar}^\perp,\sigma_{\nbar} ).
\tag7.3.16
$$
We also need the twisted loop group $L(\frak G,\sigma_{\nbar} )$, the
collection of units in $L(\frak g,\sigma_{\nbar} )$. The image of
$L(\frak G,\sigma_{\nbar} )$ under the isomorphism $\Phi_{H_{\nbar}}$
is the untwisted loop group $\tilde Gl(n,\Bbb C)$. If $\tilde g
\in \tilde gl(n,\Bbb C)$, or $\tilde g\in \tilde Gl(n,\Bbb C)$,
then we write $\tilde
g^{\sigma_{\nbar}}$ for the inverse
image in the twisted loop algebra or twisted loop group.
\proclaim{Lemma 7.3.4}
\roster
\item
If $\tilde g_-=1_n+\Cal O(\lambda\inv)$ then $\tilde
g_-^{\sigma_{\nbar}}=
1_n+\Cal O(\mu\inv)$.
\item
    Let $\tilde g_-$ as above. Then there exist two formal
loops $u^{\sigma_{\nbar}}=\sum_{i<0}\mu^{i}u_{i}$,
$v^{\sigma_{\nbar}}=\sum_{i<0}\mu^{i}v_{i}$,
with $u_{i}\in (\frak h_{\nbar})_{\bar \imath}$,
$v_{i}\in (\frak
h_{\nbar}^\perp)_{\bar \imath}$, such that
$$
\tilde g_-^{\sigma_{\nbar}}=\exp(u^{\sigma_{\nbar}})
\exp(v^{\sigma_{\nbar}}).
$$
\endroster
\endproclaim

\demop
The element $H_{\nbar}$ induces a $\Bbb Z$ grading of $\tilde
gl(n,\Bbb C)$ as
follows: if $y\in \frak g_{\bar m}$, then
$$
[H_{\nbar},y]=s y, \quad s=\frac{\bar m}{N}+\ell\ ,
\tag7.3.17
$$
for some $\ell\in \Bbb Z$, $\bar m=0,1,\dots,N-1$. We put then
$$
\text{deg}(\lambda^r y)= (r+s)N=(r+\ell)N + \bar m.
\tag 7.3.18
$$
This is the grading that corresponds to the grading by powers of
$\mu$
in $L(\frak g, \sigma_{\nbar})$ under the standard isomorphism:
we have: $\Phi_{H_{\nbar}}\inv(
\lambda^r y)=\mu^{(r+\ell)N + \bar m}y$. One calculates
that the eigenvalues of $H_{\nbar}$ are of the form
$$
{q\over n_b}-{p\over
n_a} +{1\over 2n_a} -{1\over 2n_b},
\tag7.3.19
$$
where $1\le p\le n_a$ and
$1\le q\le n_b$. From this one sees that
the absolute value of the eigenvalues of $H_{\nbar}$  is strictly
less
than
1, and that the only possibilities in \thetag{7.3.18} are $m=0,l=0$
or
$0<m\le N-1$ and  $l=0,-1$.
This means that a homogeneous element
of the form $\lambda^{-i}y$, $i>0$, has strictly negative degree and
maps to
an element of $L(\frak g,\sigma_{\nbar})$ containing a strictly
negative
power of
$\mu$. This proves part \therosteritem{1}. The proof of part
\therosteritem {2} is similar to that of Lemma 5.1 in \cite{BtK}.
\qede

We return to the formulation of the $\nbar$ reduced KP hierarchy in
terms of
$n\times n$ matrix operators.
Let $\partial_y=\sum A_a \partial_{t^a_1}$ and $\Cal P_y=\sum A_a
\Cal P_a$.
Then one considers the operator
$$
\aligned
D_y(t,\alpha)&=\overleftarrow\partial_y -\tilde  w_W(t,\alpha)\inv
\partial_y(\tilde  w_W(t,\alpha))=\overleftarrow\partial_y -\tilde
L(t,\alpha)_+,\\
&=\ddxl_y-\Cal P_y - q(t,\alpha),
\endaligned
\tag7.3.20
$$
for some matrix  $q(t,\alpha)$.
The components of $q(t,\alpha)$ are
then considered the fundamental fields of the
theory. Note that $\Cal P_y+q$ is the polynomial part (in $\lambda$)
of
an element in the adjoint
orbit through $\Cal P_y$ of the group of elements of the form $g=
1+\Cal O(\lambda\inv)$, and hence $q$ is constant in $\lambda$.
In particular the degree \thetag
{7.3.18} of the homogeneous
components of $q$ is strictly less than the maximum of $N\over n_a$,
$a=1,
\dots,k$.
The most general form for $q$ is obtained if one chooses the
constants $A_a$
such that $\Cal P_y=
\sum A_a\Cal P_a$ is regular, i.e., such that the centralizer of it
is
just $\tilde H_{\nbar}$.  However
the theory would work also if $\Cal P_y$ is not regular: then some of
the
components of $q$ would be zero.
\proclaim{Lemma 7.3.5} Let $D_y(t,\alpha)$ as above and
denote by $D_y^{\sigma_{\nbar}}$ the inverse image of this operator
in the
twisted
loop algebra $L(\frak g,\sigma_{\nbar})$:
$D_y^{\sigma_{\nbar}}=
\overleftarrow\partial_y -\sum A_a\mu^{N\over
{n_a}}E_a-q^{\sigma_{\nbar}}$.
Then there exists a unique
formal power series $v^{\sigma_{\nbar}}=\sum_{i<0}\mu^{i}v_{i}$,
with  $v_{i}\in (\frak
h_{\nbar}^\perp)_{\bar \imath}$, such that
$$
\exp(ad(v^{\sigma_{\nbar}}))
(D_y^{\sigma_{\nbar}})=\overleftarrow
\partial_y -\sum A_a\mu^{N\over {n_a}}
E_a + \sum_{i<\max(N/n_a)}
k_i
\mu^{i}, k_i
\in (\frak h_{\nbar})_{\bar {\imath}}.
\tag7.3.21
$$
The components of $v_i$ and
$k_i$ are $y$-differential polynomials in the components of
$q^{\sigma_{\nbar}}$
and hence in the components of $q(t,\alpha)$.
\endproclaim

\demop
First note that trying to construct $v^{\sigma_{\nbar}}$ directly
from \thetag{7.3.21} by expanding in powers of $\mu$ seems not to
lead
in the usual way (cf. \cite{DS}) to
a simple recursion scheme for the $v_i$
because of the inhomogeneity of $\Cal P_y$. Therefore we break up
$D_y$
in pieces corresponding to the homogeneous components of $\Cal P_y$.

Indeed, let $\Cal P_y +q(t,\alpha)=[Ad(\tilde g_-)(\Cal P_y)]_+$  for
$\tilde g_-=1+\Cal O(\lambda\inv)$. Define $\Cal P_a+q_a=[Ad(\tilde
g_-)(\Cal
P_a)]_+$ and $D_a=\overleftarrow
\partial_{t^a_1}-\Cal P_a-q_a$. We have $D_y=\sum_{a=1}^k A_aD_a$,
and
$q=\sum_{a=1}^k A_aq_a$. We will write as in Lemma 7.3.4
\allowbreak $\tilde g_-^{\sigma_{\nbar}}=\exp(u^{\sigma_{\nbar}})
\exp(v^{\sigma_{\nbar}})$. Define $\tilde w=\tilde w_0 \tilde g_-$,
so that $\tilde w ^{\sigma_{\nbar}}=\tilde w_0
\exp(u^{\sigma_{\nbar}})
\exp(v^{\sigma_{\nbar}})$ and we have
$$
\tilde w ^{\sigma_{\nbar}}D_a^{\sigma_{\nbar}}=0.
\tag7.3.22
$$
{}From this one easily sees that for all $a$:
$$
\exp(ad(v^{\sigma_{\nbar}}))
(D_a^{\sigma_{\nbar}})=\overleftarrow
\partial_{t^a_1} -\mu^{N\over {n_a}}
E_a + \sum_{i<N/n_a}
k^{(a)}_i
\mu^{i}, \quad k^{(a)}_i
\in (\frak h_{\nbar})_{\bar {\imath}},
\tag7.3.23
$$
and hence also that
$\exp(ad(v^{\sigma_{\nbar}}))(D_y^{\sigma_{\nbar}})$ is of  the same
form,
proving  therefore the existence of
a $v^{\sigma_{\nbar}}$ such that \thetag{7.3.21} holds, cf.
\cite{BtK}. Now
fix an $a$ between 1 and $k$. Comparing
the coefficient
of $\mu^{N/n_a -j}$ on both sides of \thetag{7.3.23}
we obtain  an equation of the form
$$
\aligned
[v_{-j}, E_{a}] +
k^{(a)}_{N/n_a-j}&=
\text{ polynomial in } q_a^{\sigma_{\nbar}} \text{ and in}\\
&v_{-l}, -l>-j,\text{ and ${t^a_1}$
derivatives. }
\endaligned
\tag7.3.24
$$
Using the partition $\nbar$ we decompose  $v_{-j}$ in blocks: write
$v_{-j}=\sum_{a,b=1}^k v_{-j,ab}$, where the matrix $v_{-j,ab}$ is
zero
outside the block with index $ab$ of
size
$n_a\times n_b$. Then the commutator in the left hand side of
\thetag{7.3.24}
becomes $\sum_{b=1}^k
A_a\allowmathbreak[v_{-j,ba}C_{n_a}-C_{n_a}v_{-j,ab}]$, where
$C_p=E_{p1}+\sum
E_{ii+1} $ is a cyclic matrix of size $p$. Now $C_p$ acts from the
left (right) as an invertible linear
transformation on the space of $p\times q$ ($q\times p$)  matrices
for any
$q$.
Furthermore the adjoint action of $C_p$ is an invertible linear
transformation on the orthogonal complement of the linear span of the
powers in $C_p$ in the space of $p\times p $ matrices.
Using these facts one can express the blocks  $v_{-j,ab}$,
$v_{-j,ba}$ for
$b=1,\dots,k$
and  $k^{(a)}_{N/n_a-j}$ uniquely in terms of $q^{\sigma_{\nbar}}_a$
and
the $v_{-l}$, $-l>-j$, and
their $t^a_1$ derivatives. Letting now also $a$ run from 1 to $k$ we
express
{\it all} blocks of $v_{-j}$ in terms of these variables.
Note that to find $v_{-j,ab}$
according to this method one can consider the
diagonalization \thetag{7.3.23} of $D_a$ or of $D_b$. These must give
the
same result, since the consistency of the procedure it ensured by the
existence of at least one solution $v^{\sigma_{\nbar}}$.
Induction leads now to the conclusion that $v^{\sigma_{\nbar}}$
and $\sum k_i\mu^i$ (with $k_i=\sum A_ak_i^{(a)}$) are
polynomials in $q_a^{\sigma_{\nbar}}$
and $t^a_1$ derivatives, $a=1,\dots,k$. Next we use the fact (which
follows from \thetag {7.3.22}) that
$$
\aligned
0=[D_y,D_a]&= [\overleftarrow\partial_y-\Cal P_y -q,\overleftarrow
    \partial_{t^a_1}-\Cal P_a -q_a],\\
\endaligned
\tag7.3.25
$$
This gives us an expression for $\partial_{t^a_1}(q)$ in
terms of derivatives of $q_a$ with respect to $y$. Since the
projection
$q\mapsto q_a$ is just differentiation with respect $A_a$ we find
that we can express all $t^a_1$ derivatives  of $q_a$  in terms of
$y$ derivatives and the Lemma follows.
\qede

{}From the proof of the last Lemma it follows that all
resolvents, being of the form $R(p)=\tilde w\inv \cdot p \cdot \tilde
w=\tilde g_-\inv \cdot p \cdot \tilde
g_-=\exp(-ad(v))(p)$, are differential polynomials of the fundamental
field $q(t,\alpha)$. This makes the following definition reasonable:
\proclaim{Definition 7.3.6}
The $\nbar$-KdV hierarchy is the collection of deformation equations
of the
operator $D_y$ (of the form \thetag{7.3.20} with
$\Cal P_y +q(t,\alpha)=[Ad(\tilde g_-)(\Cal P_y)]_+$ for some $\tilde
g_-=1+\Cal
O(\lambda\inv)$) given by
$$
\aligned
\partial_{t^b_i} D_y &=[(\tilde \Cal R_b^i)_+, D_y],\\
D_y(t,\alpha+\alpha_i)&=(\tilde {\Cal U}_{\alpha_i})_+(D_y(t,\alpha))
(\tilde {\Cal U}_{\alpha_i})_+\inv.
\endaligned
\tag7.3.26
$$
Here $(\Cal R_b^i)_+$ and $(\tilde {\Cal U}_{\alpha_i})_+$ are the
positive
parts
of the resolvents \thetag{7.3.10}, using $\tilde  w=\tilde w_0 \tilde
g_-$
instead of $\tilde w_W$.
\endproclaim

Note that we have made here the simple choice $\partial_y=\sum
A_a\partial_{t^a_1}$
to determine   the ``spatial variable'' $y$
in our system of equations \thetag{7.3.6}. It will be clear from the
construction
that we can make much more general choices for this spatial variable
and still
get a reasonable set of equations, cf. \cite{FNR}.

As an example of a $\nbar$-KdV hierarchy consider the 2-reduction of
the
Davey-Stewartson-Toda system of section 4.6, for the partition
$2=1+1$. The resulting system is the  AKNS-Toda system:
the field $Q^m$, being the second $x$ derivative of the
$\tau$-function, see
\thetag{5.5.32}, is in the
2-periodic case identically zero, since the $\tau$-function is now
independent of
$x$. Furthermore, in the resolvents and lattice resolvents
the operator $\ddxl$ can be effectively replaced by the variable $z$;
this is
because
$$
w_W\ddxl=w_0 \cdot \overleftarrow w_W \ddxl
=zw_0 \cdot  \overleftarrow  w_W+w_0\partial( \overleftarrow  w_W)
= w_W,
\tag7.3.27
$$
since also the wave operator is $x$ independent in the $2$-periodic
case.
This gives the theory described in \cite{BtK}.

\enddocument
\bye